\documentclass[twocolumn]{aastex63}
\usepackage{amssymb}	
\usepackage{amsmath}	
\usepackage{array}
\usepackage{bm}
\usepackage{CJK}
\usepackage{graphicx}	
\usepackage{hyperref}
\usepackage{xifthen}
\usepackage{xspace}
\usepackage{natbib}
\newcommand{\aizw}[1]{%
  {\color{black} #1}%
}

\newcommand{\aizwrev}[1]{%
  {\color{black} #1}%
}

\begin{document}
\begin{CJK*}{UTF8}{gbsn}
\title[Revealing asymmetry in DSHARP disks]{Axisymmetric Modeling of DSHARP Dusty Disks: Asymmetric Structures and Inner-Disk Dispersal}

\author[0000-0001-8877-4497]{Masataka Aizawa}
\affiliation{College of Science, Ibaraki University, 2-1-1 Bunkyo, Mito, 310-8512, Ibaraki, Japan }
\affiliation{Tsung-Dao Lee Institute, Shanghai Jiao Tong University, Shengrong Road 520, 201210 Shanghai, P. R. China }
\affiliation{RIKEN Cluster for Pioneering Research, 2-1 Hirosawa, Wako, Saitama 351-0198, Japan }

\author{Takayuki Muto}
\affiliation{Division of Liberal Arts, Kogakuin University, 1-24-2 Nishi-Shinjuku, Shinjuku-ku, Tokyo 163-8677, Japan }

\author[0000-0002-3001-0897]{Munetake Momose}
\affiliation{College of Science, Ibaraki University, 2-1-1 Bunkyo, Mito, 310-8512, Ibaraki, Japan }
\affiliation{NAOJ Chile, National Astronomical Observatory of Japan, Los Abedules 3085 Oficina 701, Vitacura 7630414, Santiago, Chile}

\begin{abstract}
High-resolution observations of Class II protoplanetary disks frequently reveal annular structures that may indicate the presence of embedded planets. In this study, we model brightness profiles and geometries for \aizwrev{16} dusty disks in the DSHARP observations to identify asymmetric substructures, including possible planet-induced signatures. We find no compelling evidence for circumplanetary emission in these systems. We identify a possible one-armed spiral in Elias 27; while previous studies report an $m=2$ spiral, the morphology of the newly identified spiral agrees with a spiral for a possible protoplanet. Although non-detection of circumplanetary emission is consistent with low expected luminosities, the absence of additional dust spirals except for Elias 27 \aizwrev{may constrain properties of potential embedded planets given their theoretical detectability}. The analyses further suggest that spiral amplitudes and phases correlated with gap and ring locations in WaOph 6 and IM Lup, appearing as \aizwrev{deflections} of spirals in images. \aizw{Five} disks exhibit strong residual asymmetries attributable to the vertical extent of their dust layers. We find that the brightness temperature of inner disks ($1$-$5$ au) declines with stellar age on a $\sim 3$ Myr timescale, while the total flux shows no clear decreasing trend. This trend is consistent with the presence of pressure bumps that retain large dust grains at outer radii, while allowing the inner disks to disperse. The universality of disk dispersal timescale at millimeter wavelengths, observed in both the extended DSHARP disks and the compact disks from previous demographic surveys, may constrain timing of planet formation, including habitable planets. 
\end{abstract}

\section{Introduction}

High-resolution observations by the Atacama Large Millimeter/submillimeter Array (ALMA) uncover diverse substructures in protoplanetary disks \citep[e.g.,][]{bae2023}. Notably, the unprecedented angular resolution allows us to resolve rings and gap structures in the dust continuum emission of these disks \citep{2015ApJ...808L...3A,andrews2018, huang_ring_2018, bae2023}. One of the leading scenarios explaining these rings and gaps is disk-planet interaction, in which planets carve out density gaps in the disk \citep{lin1979,goldreich1980}. Nevertheless, other physical mechanisms—such as snowline effects \citep{zhang2015}, sintering \citep{okuzumi2016}, secular gravitational instability \citep{takahashi2014}, and non-ideal MHD effects \citep{flock2015}—can also produce similar ring and gap structures, so the origins of these annular features remains uncertain. For example, based on the observed gap locations, \cite{marel2019} conclude that the snow line model is unlikely to be the common origin of systems with multiple rings and that the gap locations do not exhibit systematic trends indicative of planetary resonances.

To test the planetary hypothesis for the observed annular structures, independent observational evidence for the presence of planets is thus  essential. Possible signatures include planetary thermal emission, circumplanetary disk (CPD) emission, spirals, and kinematic disturbances. More than a dozen of kinematic signatures attributed to embedded planets have been reported \citep[e.g.,][]{pinte2018,pinte2020}; for a recent review see also \citet{pinte2023}. On the other hand, \citet{speedie_obs_2022} search for planetary spirals in 10 disks with kinematic signatures of embedded planets but did not find any conclusive cases. \cite{uyama2025} also search for emission from planets with kinematic signals in HD 163296 with JWST/NIRCam coronagraphic imaging, but do not \aizw{detect the putative planets at the predicted locations. Instead, they identify a point-source candidate at a different position, although its nature remains uncertain. }

Currently, observational evidence of planetary or CPD emission in disks remains limited to a few systems: PDS~70 \citep{keppler2018,wagner2018,benisty2021}, AB~Aur \citep{currie2022}, HD~169142 \citep{reggiani2014,hammond2023}, MWC~758 \citep{wagner2023}, and \aizwrev{WISPIT 2} \citep{close2025,van2025}. \cite{bae2022} also report a possible detection of CPD emission from \aizwrev{$^{13}\mathrm{CO}\,(J=2\mbox{--}1)$} emission in AS 209. \aizw{These detections are predominantly obtained at near-infrared wavelengths, whereas PDS~70 remains the only system with a confirmed millimeter detection of CPD emission. Using high-contrast L-band observations, \cite{jorquera2020} investigate  planetary emission in the 10 DSHARP disks and report no secure detections;  the point-source candidate in Elias 24 remains unconfirmed.} 

While spiral structures have been observed in numerous systems in millimeter-continuum emission, molecular emission, and scattered light \citep[e.g.,][]{grady1999,Fukagawa_2004,muto2012,grady2013,chris2014,huang_spiral_2018}, their origins remain uncertain. In some cases, the spirals could be produced by gravitational instabilities \citep{lee2020,veronesi2021,carreno2021,speedie2024}, by companion-driven interactions \citep[e.g.,][]{dong2016,kurtovic2018disk,monnier2019,xie2023}, or by close encounters from stellar flybys \citep[e.g.,][]{Dai2015,kurtovic2018disk,dong2022,curone2025}.

One possible explanation for the difficulty in detecting planetary evidence in millimeter observations \aizw{would} be the intrinsic weakness of planetary signals, which can be easily masked by the \aizw{background emission originating from the disk itself}. The discovered CPDs are identified primarily in \aizw{disks with large cavities in mm-wavelength, typically associated with relatively evolved systems (ages $>3$-$4$ Myr)}, where the planetary signals are more readily detectable and planetary formation is likely more advanced. Indeed, a small fraction of Class I disks exhibit annular structures \citep{ohashi2023}, possibly because planet formation may not sufficiently progress and/or the disk background emission remains significant. \aizw{Also, at such high optical depths, disk substructures and CPDs are intrinsically obscured and this bias could be strengthened for highly inclined systems. }

To detect weak planetary signals, it is essential  to disentangle them from the disk's background emission, treating the disk emission as ``noise" and the planetary signals as ``signal". Given that many Class II disks exhibit annular structures, we may assume that the disk's background emission is axisymmetric, in contrast to non-axisymmetric signatures of planets. Under this assumption, axisymmetric models of dusty disks are used to reproduce the bulk emission, allowing faint planetary signals to be detected in the residual images \citep[e.g.,][]{andrews2021, speedie_obs_2022}. 

In \citet{aizawa2024}, we present \aizwrev{an} analytical framework for deriving the axisymmetric structure of continuum emission from a disk under the assumption of a geometrically thin disk. This method not only recovers the radial profile with finer resolution than the CLEAN-based approach, as previously demonstrated by \citet{jennings2020}, but also determines both the disk geometry and the hyperparameters for the Gaussian process, with the former being essential for detecting faint planetary signals. 

In this paper, using the code, {\tt protomidpy} that implements the method proposed by \citet{aizawa2024}, we investigate the non-axisymmetric structures in the DSHARP disks \citep{andrews2018}, whose sensitivity and angular resolution are well suited for studying faint asymmetric signals. We note that \citet{andrews2021} also search for \aizw{circumplanetary} emission in nine disks in the DSHARP sample; however, our study expands the sample and utilizes our proposed automated methodology for detecting faint asymmetries. 

The paper is structured as follows. In Section~\ref{sec:model}, we describe our target selection, data, and the methods used to generate residual images. Section~\ref{sec:asym_dsharp} presents the axisymmetric modeling results for 16 DSHARP disks along with their residuals. In Section~\ref{sec:revisit_spiral}, we provide detailed analyses of systems exhibiting spiral features, while Section~\ref{sec:ind} examines the remaining disks. Section~\ref{sec:corr} explores possible correlations between stellar and disk properties and the disk substructures. In Section~\ref{sec:age_bright}, we investigate a link between \aizw{inner-disk} brightness temperatures and stellar ages. \aizw{Section \ref{sec:discussion} presents our discussion}, and Section~\ref{sec:summ} summarizes our conclusions.

\section{Axisymmetric Modeling to Produce Residual Images} \label{sec:model}
\subsection{Data}
We analyze data from the DSHARP observations \citep{andrews2018}, which include 20 individual targets. The DSHARP survey provides high-angular resolution ALMA Band 6 (1.25 mm) data with four spectral windows, three dedicated to continuum and one covering \aizwrev{$^{12}\mathrm{CO}\,(J=2\mbox{--}1)$}. In our analysis, we exclude four targets, AS 205, HT Lup, HD 143006, and HD 163296, from our axisymmetric modeling because they exhibit strong non-axisymmetric features that impede accurate modeling of the disk's underlying axisymmetric structure. \citet{kurtovic2018disk} conduct a detailed analysis of AS 205 and HT Lup, both binary systems, and identify spirals in the disks. Similarly, \citet{andrews2021} study HD 143006 and HD 163296 using a combination of axisymmetric modeling and ad-hoc subtraction of strong non-axisymmetric structures, revealing a faint spiral in HD 143006. Although these analyses with our method are intrinsically interesting, our study focuses on nearly axisymmetric systems to search for faint planetary signals, thus not analyzing these systems with our axisymmetric modeling. 

For our analysis, we download self-calibrated continuum measurement sets for 16 disks from the DSHARP observations\footnote{\url{https://bulk.cv.nrao.edu/almadata/lp/DSHARP}} \citep{andrews2018}. We average the visibilities for each spectral window into a single channel and apply a 30-second time averaging. We then visually inspect the visibilities to assess data quality. Notably, we find that the measurement sets for DoAr~25 are significantly contaminated by systematic noise, so we exclude those outliers from our analysis. Details are provided in Appendix \ref{sec:doar25_out}. 

\subsection{Modeling of axisymmetric geometrically-thin disk }

We model \aizwrev{an} axisymmetric geometrically-thin disk for each target. In interferometric observations, the measured visibilities are approximately the Fourier transform of the sky brightness distribution. In particular, if the brightness profile is axisymmetric, this Fourier transform reduces to a one-dimensional Hankel transform. 

\citet{jennings2020} develop an analytical method to reconstruct the brightness profile directly from the visibilities, demonstrating that this approach yields higher-resolution profiles than conventional imaging techniques such as CLEAN. However, their method requires fixing several additional parameters, namely, the disk center, inclination, position angle, and hyperparameters, which are critical for accurately characterizing faint asymmetries \citep{andrews2021}. To better explore these faint signals, \citet{aizawa2024} develop an analytical method for deriving the posterior distributions of these parameters, including the brightness profile. We therefore adopt the approach of \citet{aizawa2024} to investigate asymmetric features in DSHARP disks.

We use the open-source package {\tt protomidpy}\footnote{\url{https://github.com/2ndmk2/protomidpy}} that implements the algorithm developed in \cite{aizawa2024}. We assume a geometrically thin, axisymmetric disk characterized by a radial intensity profile $\bm{a}$ and geometric parameters $\bm{g}$. Specifically, the radial profile is represented as a vector
\begin{equation} 
\bm{a} = \{I(r_{k})\}, \quad k = 1, 2, \dots, N,
\end{equation}
where $I(r)$ is discretized at the collocation points $r_{k}$ for the Fourier-Bessel series, and $R_{\rm out}$ is assumed to be the outer boundary of the disk. We assume $N=200$ and $R_{\rm out}=2\arcsec$ for for most disks, except for Elias 27, where $N=300$ and $R_{\rm out}=3\arcsec$. 

The geometric parameters are defined as 
\begin{equation}
\bm{g} = \left(\Delta x_{\rm cen}, \Delta y_{\rm cen}, \cos i, \mathrm{PA}\right), 
\end{equation}
where $(\Delta x_{\rm cen}, \Delta y_{\rm cen})$ specify the disk's central position in the observational frame, measured in the RA and Dec coordinates. The inclination $i$ and the position angle $\mathrm{PA}$ determine the disk orientation: $\mathrm{PA}$ is defined as the angle of the major axis of the disk measured counterclockwise from north, and $i$ is the angle between the line of sight and the normal to the disk plane.

To prevent overfitting of the model to the data, we employ a  radial basis function kernel \citep{aizawa2024}
\begin{equation}
    K(r, r')= \alpha \exp \Bigl[-\cfrac{(r - r')^2}{2 \gamma^2}\Bigr]
\end{equation}
with hyperparameters
\begin{equation}
    \bm{\theta} = \left( \alpha, \gamma \right). 
\end{equation}
Here, $\alpha$ determines the overall amplitude, and $\gamma$ determines the length scale, which controls how rapidly correlations for brightness profile decay with radial separation. 

Assuming observational visibilities 
\begin{equation}
\bm{d} = \{ d_i (u_{i}, v_{i}) \} \quad \text{for } i = 1, 2, \dots, M,
\end{equation}
with normally distributed noise, the likelihood function for the data given the parameters $(\bm{g},\bm{\theta})$, after marginalizing over $\bm{a}$, becomes a multivariate Gaussian \citep{aizawa2024} (see also \cite{kawahara2020} in the context of global mapping of an exoplanet). Consequently, we can sample from the posterior distribution of $(\bm{g},\bm{\theta})$ using Bayes' theorem:
\begin{equation}
p(\bm{g}, \bm{\theta} | \bm{d}) \propto p(\bm{d} | \bm{g}, \bm{\theta}) \, p(\bm{g}, \bm{\theta}).
\end{equation}
After drawing samples $\{(\bm{g}, \bm{\theta})\}$ from $p(\bm{g}, \bm{\theta} | \bm{d})$, we obtain the posterior distribution for $\bm{a}$. Although it is possible to construct the full posterior distribution for the brightness profiles, the variances of the parameters $\{(\bm{g}, \bm{\theta}, \bm{a})\}$ are extremely small for the high signal-to-noise DSHARP data \citep{aizawa2024}. Therefore, we adopt the maximum a posteriori (MAP) estimates for $(\bm{g},\bm{\theta})$ as representative values, and subsequently, \aizwrev{use} a MAP estimate on $\bm{a}$ assuming the MAP estimate on $(\bm{g}, \bm{\theta})$. 

To reduce computational time, we downsample the data by log binning with $N_{\rm bin}=500$, following the approach in \citet{aizawa2024}. In the binning process, we adopt a maximum radial frequency for the logarithmic grid of $q_{\mathrm{max}} = 1.5\times10^{7}\,\lambda $ as defined in \citet{aizawa2024}. Because the recorded weights in the measurement sets are systematically underestimated, we manually reduce the weights by factors of 2-4 following Sec 4.1 in \cite{aizawa2024}. The applied factors are listed in Table \ref{table:disk_para} under the column of ``Weight factor".

In computing the posterior distribution for geometric and hyperparameters, we assume the following prior distributions following \cite{aizawa2024}. For the disk orientation parameters, we adopt uniform priors: $\cos i \sim \mathcal{U}(0, 1)$ and $\mathrm{PA} \sim \mathcal{U}(0, \pi)$. For the disk's central position, we assume a uniform prior $\mathcal{U}(-1\arcsec, 1\arcsec)$. For the Gaussian Process hyperparameter $\alpha$, we assume a log-uniform prior ranging from $10^{-4}$ to $10^{4}$. For the spatial scale hyperparameter $\gamma$, we adopt a uniform prior $\mathcal{U}(0.01\arcsec, 0.15\arcsec)$.

In the sampling, 16 walkers are prepared and then evolved for at least 10,000 steps, discarding the first 3,000 steps as burn-in. Convergence is assessed via the integrated autocorrelation time for the chain\footnote{\url{https://emcee.readthedocs.io/en/stable/tutorials/autocorr/}} following \cite{aizawa2024}. 

\subsection{Making deprojected residual images} \label{sec:make_res}
With a set of optimized parameters $\bm{a}$, $\bm{g}$, and $\bm{\theta}$, we subtract the axisymmetric model from the visibilities. The measurement sets are then shifted to align the image center with the disk center using the \texttt{fixvis} task in \texttt{CASA} \aizw{in the same manner as \cite{aizawa2024}}. For the modified measurement sets after subtracting the axisymmetric component, we create residual images using the \texttt{tclean} task in \texttt{CASA}. The images are produced with a pixel scale of 6 mas and a size of $1000 \times 1000$ pixels. \aizw{We apply the multiscale CLEAN algorithm using scale sizes of [0, 30, 120, 360, 720, 1440] mas.} We adopt 3.5$\sigma$ as the threshold for CLEAN and adopt Briggs weighting with a robust parameter of 0.5, without applying a $uv$-taper. \aizw{The same \texttt{tclean} settings are also used to produce the unsubtracted images shown in Figures~\ref{fig:gallery_images1}-\ref{fig:no_five}.}
 
To examine the residual images differently, we also generate a pseudo face-on, or deprojected, view of the image \citep{aizawa2024}. Specifically, we convert spatial frequencies by by  $(u, v) \rightarrow (\cos i ( \cos ({\rm PA}) u_{j} -\sin ({\rm PA}) v_{j}), \sin ({\rm PA}) u_{j} +\cos ({\rm PA}) v_{j})$, and make the residual images using the modified data. As discussed in \cite{aizawa2024}, the transformation $(u, v) \rightarrow (u', v')$ decreases the intensity in Jy sr$^{-1}$ by a factor of $\cos i$, while the values of intensities in Jy beam$^{-1}$ are unchanged before and after deprojection.

\section{Residual images for 16 DSHARP disks}\label{sec:asym_dsharp}
\subsection{Overview of residual maps and classification by morphology}
We list the estimated geometric parameters and hyperparameters for the 16 disks in Table \ref{table:disk_para}. Using the maximum a posteriori estimate, we generate residual images by subtracting the axisymmetric component from each disk image following the method described in Section \ref{sec:make_res}. 

\begin{figure*} 
\begin{center}
\includegraphics[width=0.82\linewidth]{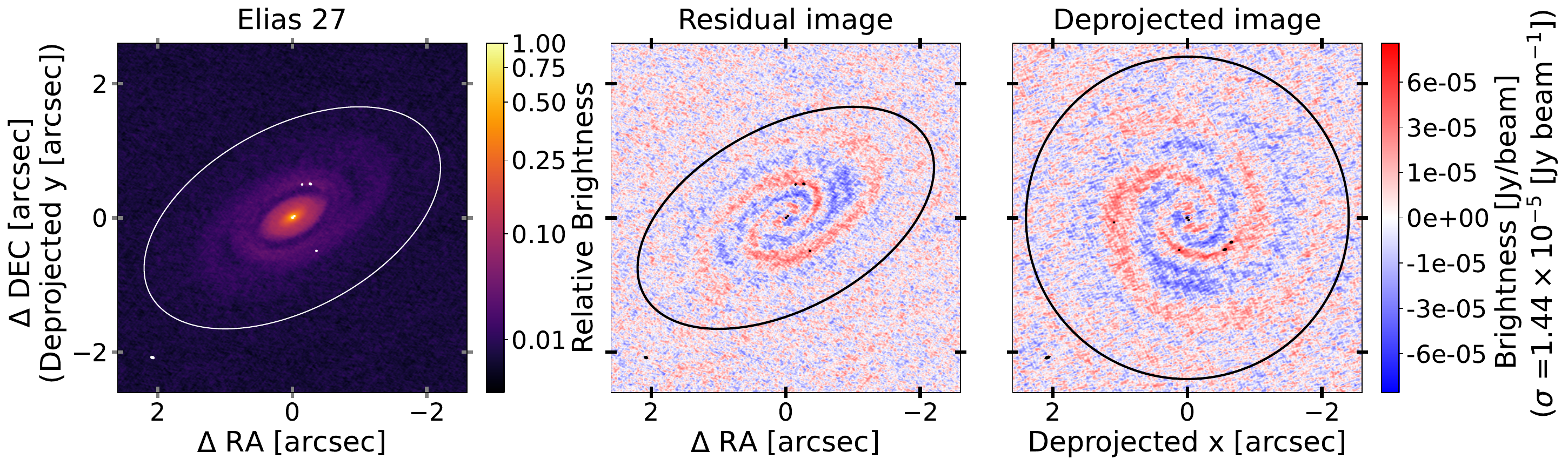}
\includegraphics[width=0.82\linewidth]{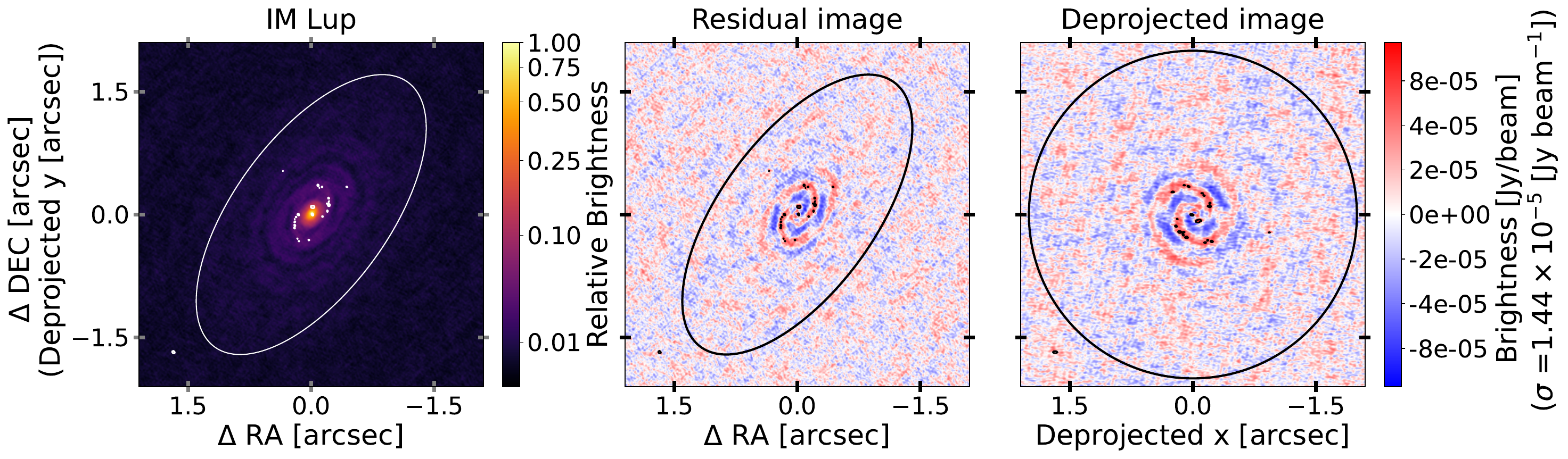}
\includegraphics[width=0.82\linewidth]{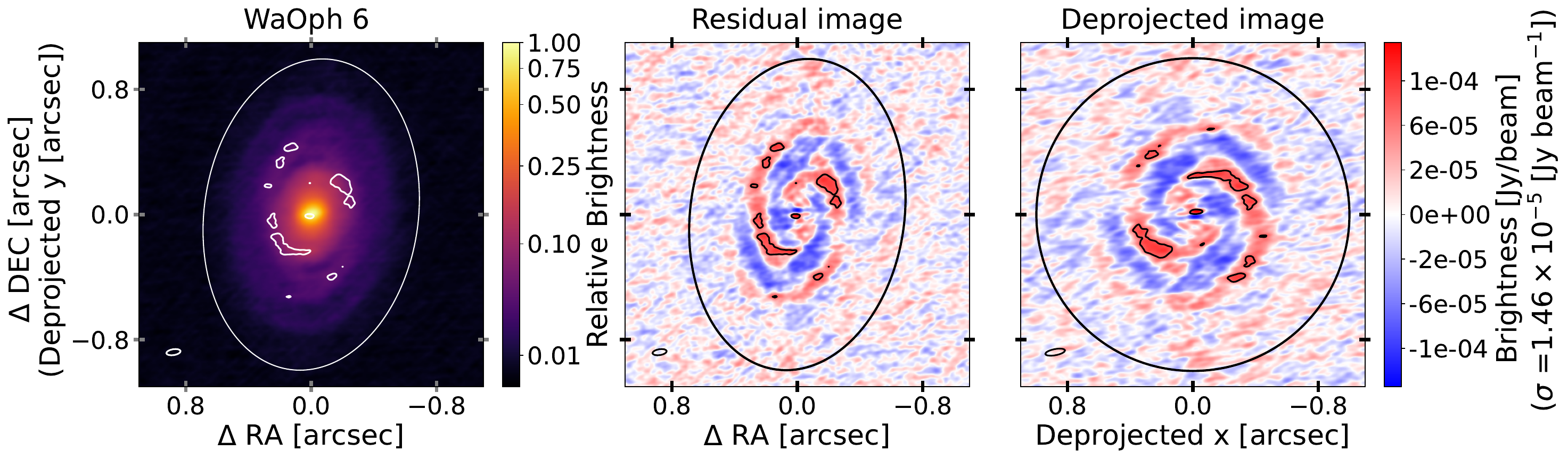}
\caption{Raw and residual images of three spiral disks with $5\sigma$ contours overplotted. The beam sizes are shown in the lower-left. The white/black ellipses show the approximate disk boundary, visually determined from the brightness profile. (Left) CLEANed images of the continuum emission before subtracting the axisymmetric models. (Center) Residual images obtained by subtracting the axisymmetric models generated with {\tt protomidpy}. (Right) The same residual images after deprojection. The dashed ellipse indicates the approximate extent of each disk. The brightness scale is shown with an asinh stretch to reduce dynamic range while avoiding saturation following \cite{andrews2018}.}
\label{fig:gallery_images1}
\end{center}
\end{figure*}

\begin{figure*} 
\begin{center}
\includegraphics[width=0.82\linewidth]{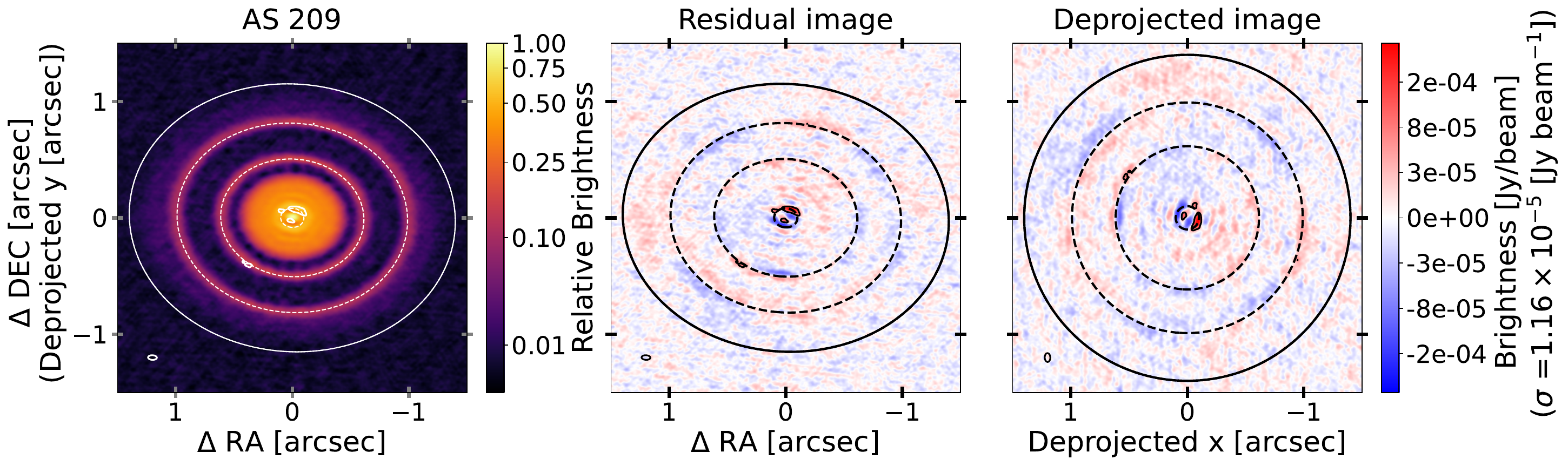}
\includegraphics[width=0.82\linewidth]{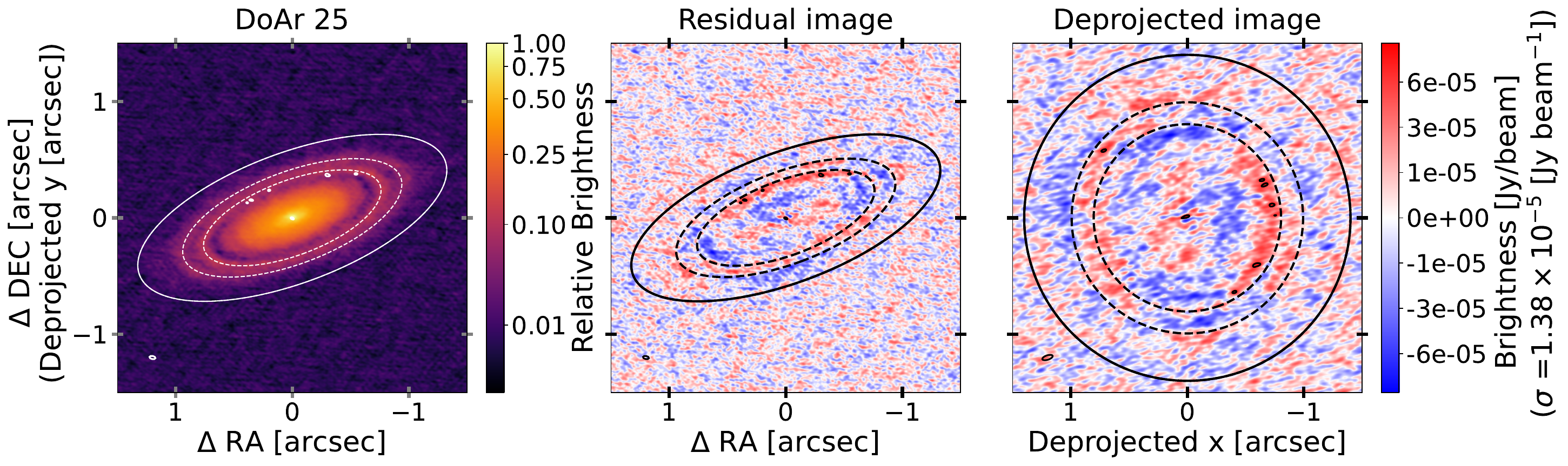}
\includegraphics[width=0.82\linewidth]{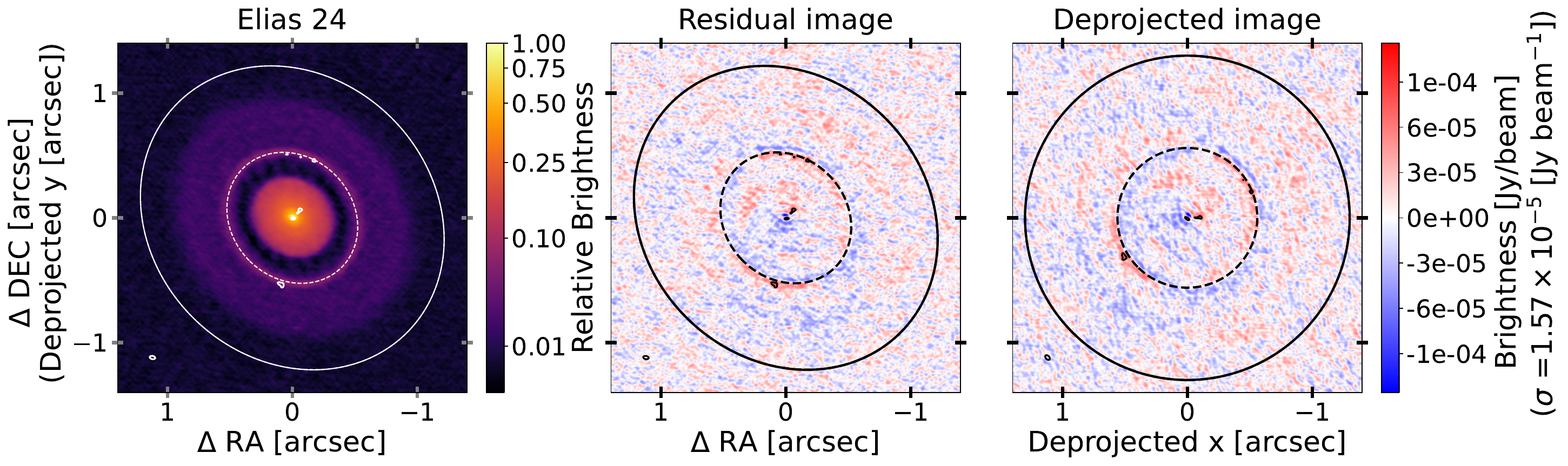}
\includegraphics[width=0.82\linewidth]{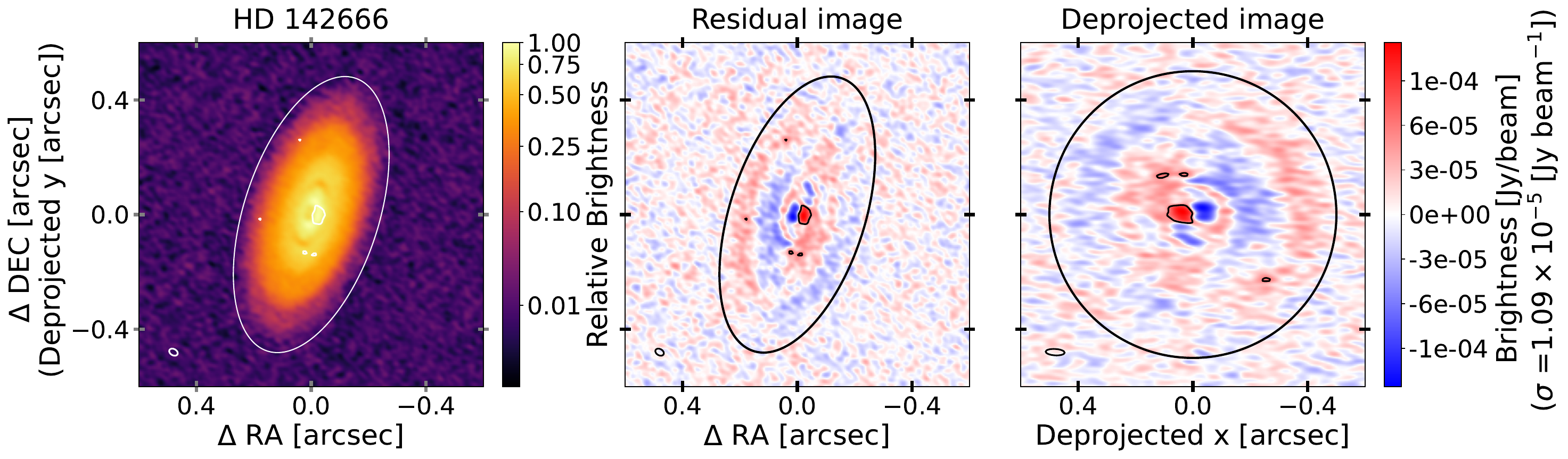}
\includegraphics[width=0.82\linewidth]{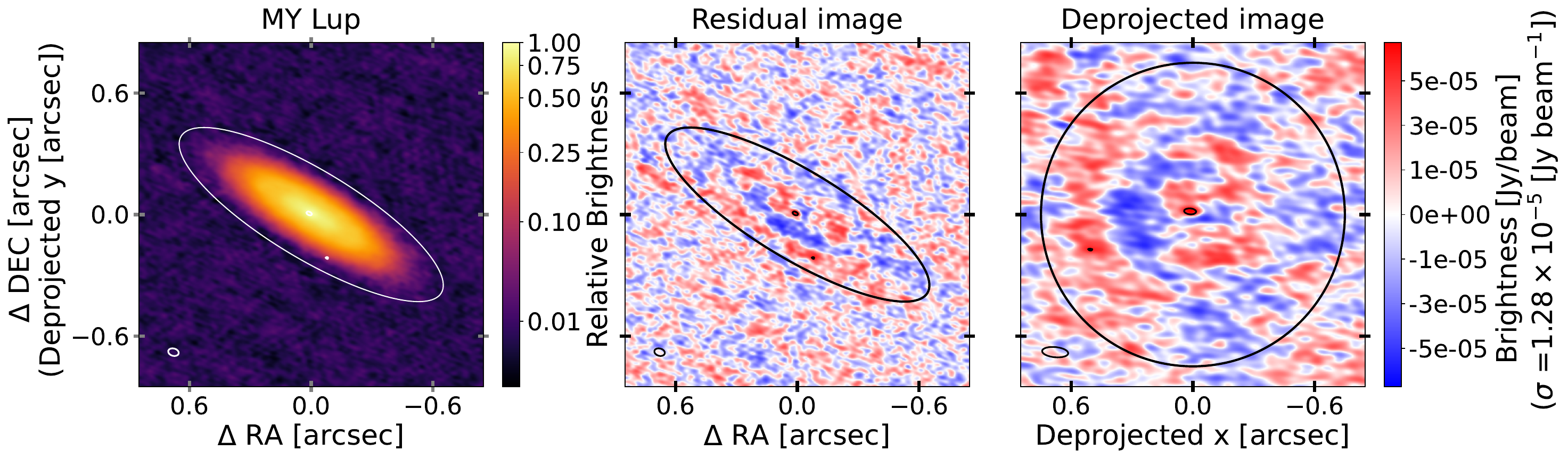}
\caption{Residual images for disks with notable vertical thickness in Section \ref{sec:geo_thick}. The dashed ellipses indicate the prominent rings. The format is the same as in Figure \ref{fig:gallery_images1}. }
\label{fig:gallery_images2}
\end{center}
\end{figure*}

\begin{figure*} 
\begin{center}
\includegraphics[width=0.82\linewidth]{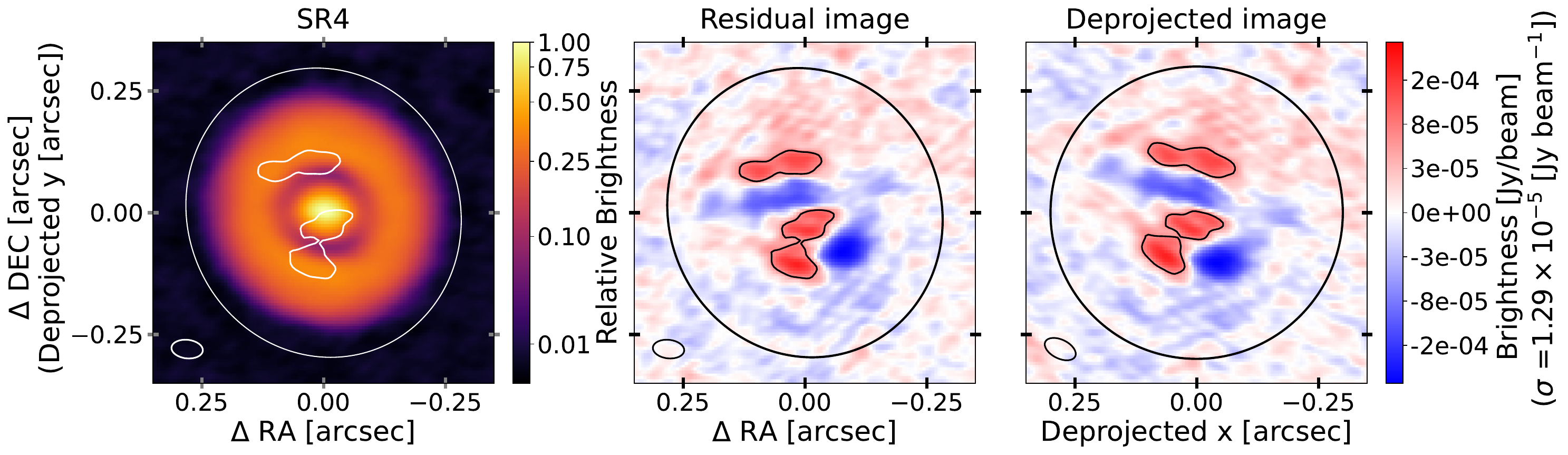}
\includegraphics[width=0.82\linewidth]{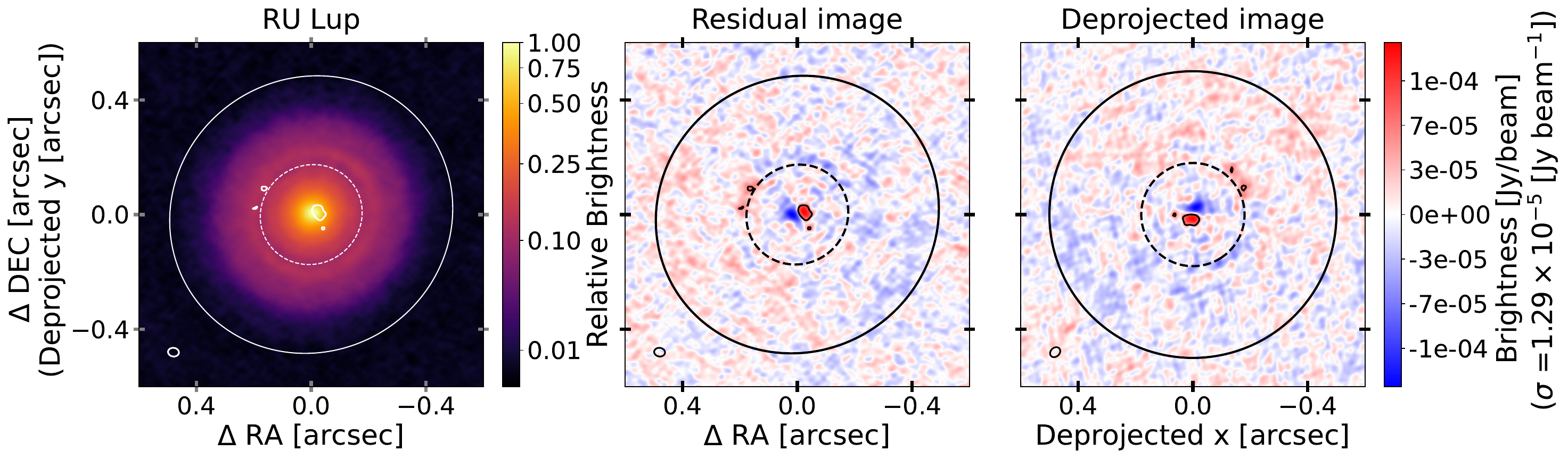}
\includegraphics[width=0.82\linewidth]{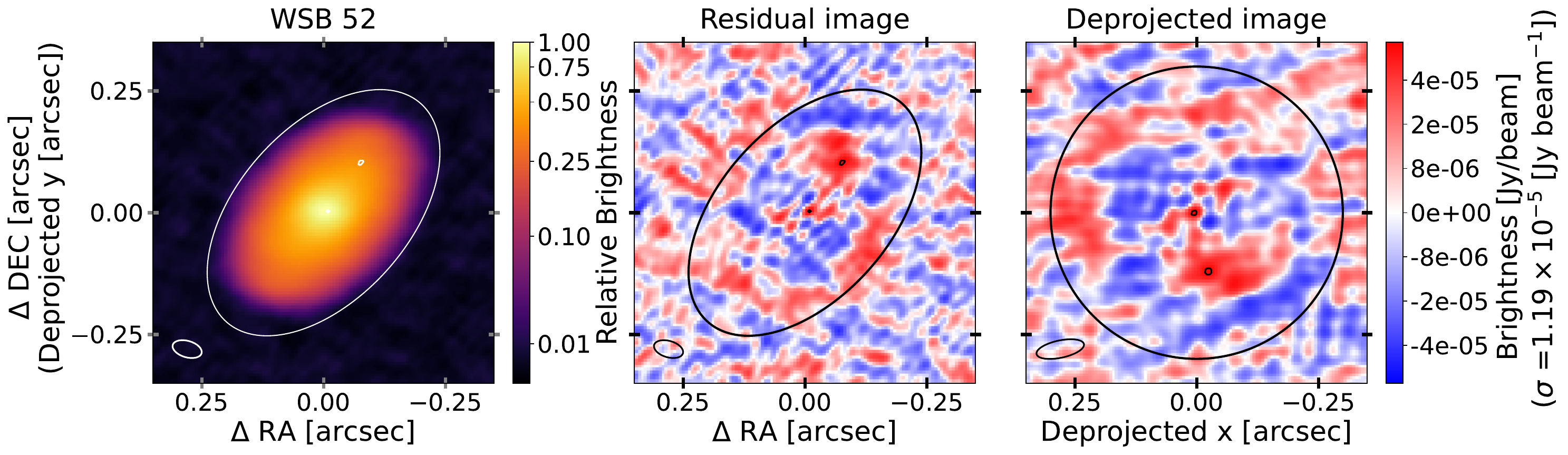}
\caption{Residual images for disks with $5\sigma$-residuals Section~\ref{sec:five_sigma}. The dashed ellipse indicates the prominent gap in RU Lup, with a pair of positive excesses located at the ring-gap boundary. The format is the same as in Figure \ref{fig:gallery_images1}. }
\label{fig:gallery_images3}
\end{center}
\end{figure*}

\begin{figure*} 
\begin{center}
\includegraphics[width=0.82\linewidth]{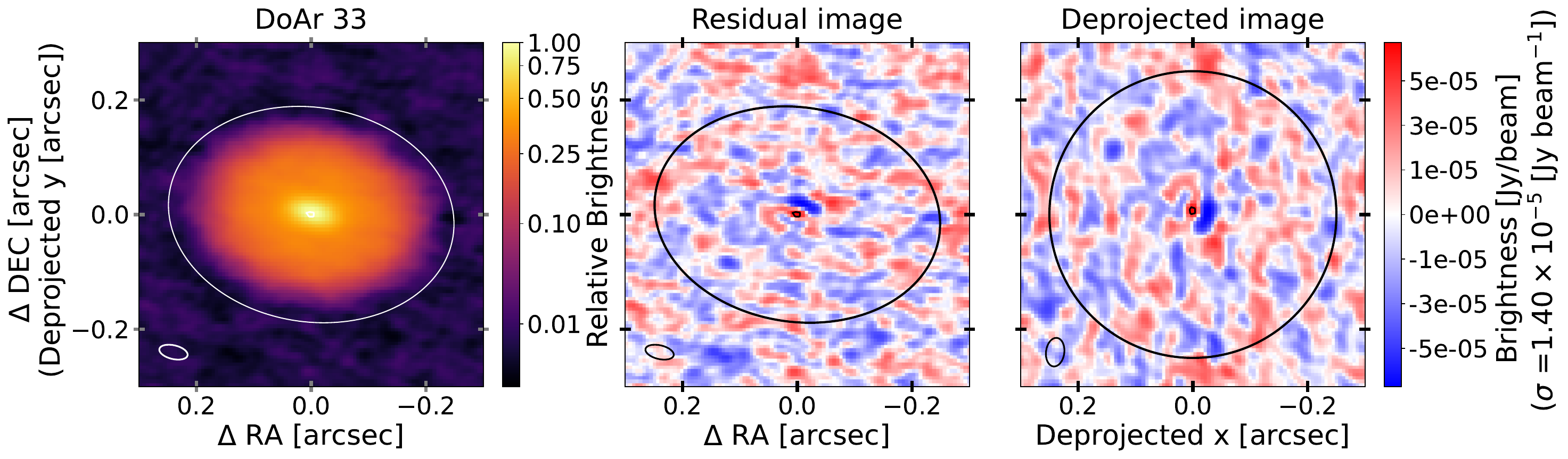}
\includegraphics[width=0.82\linewidth]{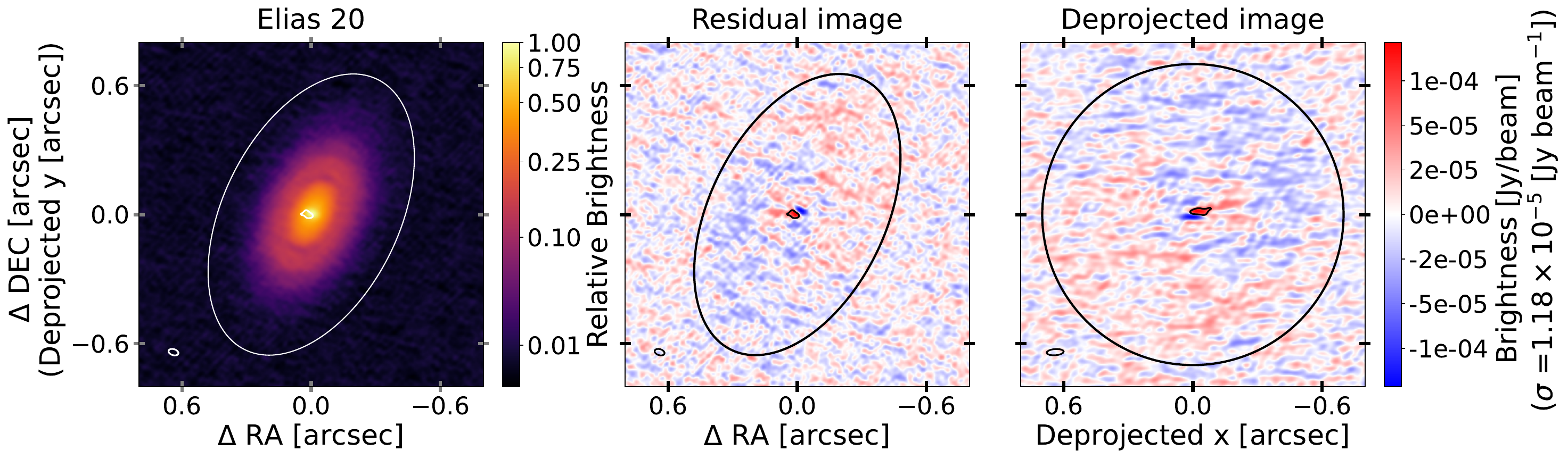}
\includegraphics[width=0.82\linewidth]{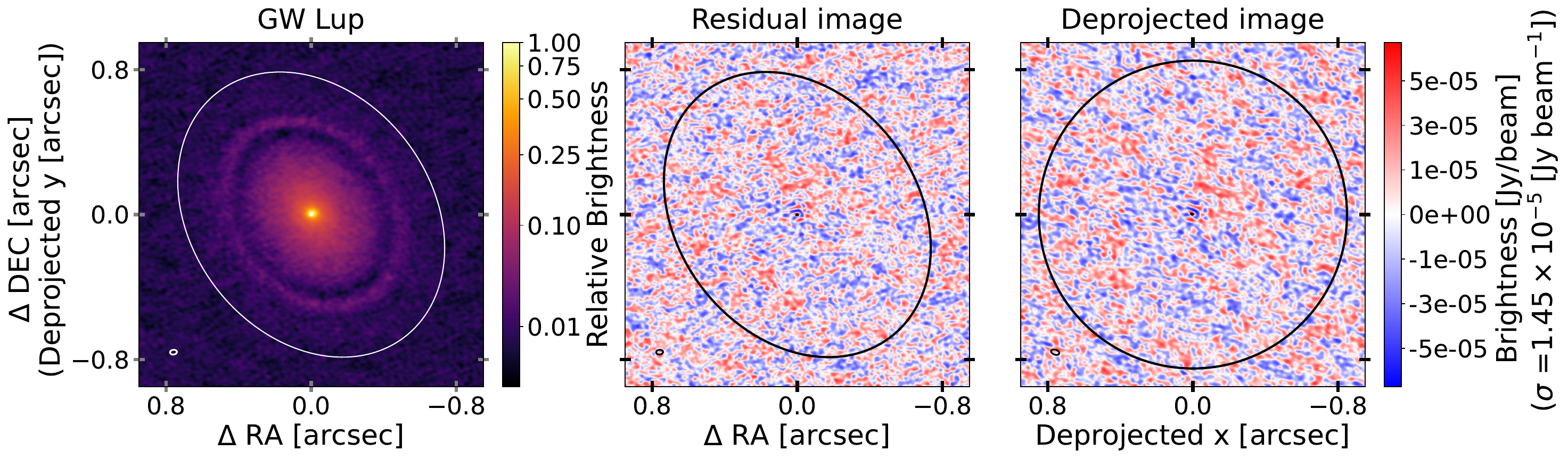}
\includegraphics[width=0.82\linewidth]{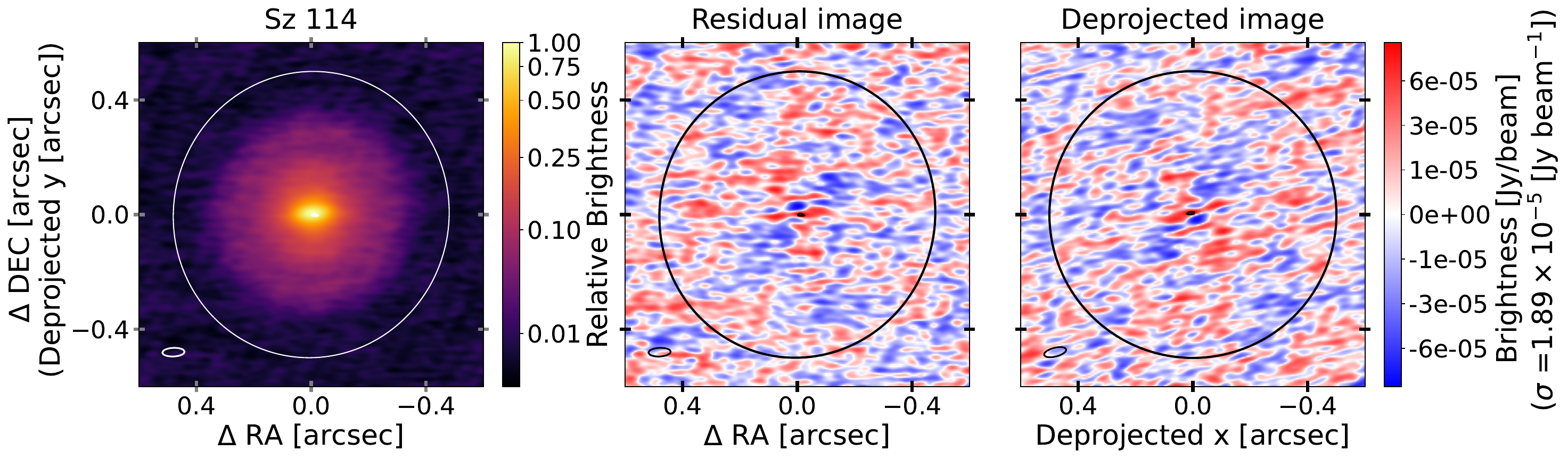}
\includegraphics[width=0.82\linewidth]{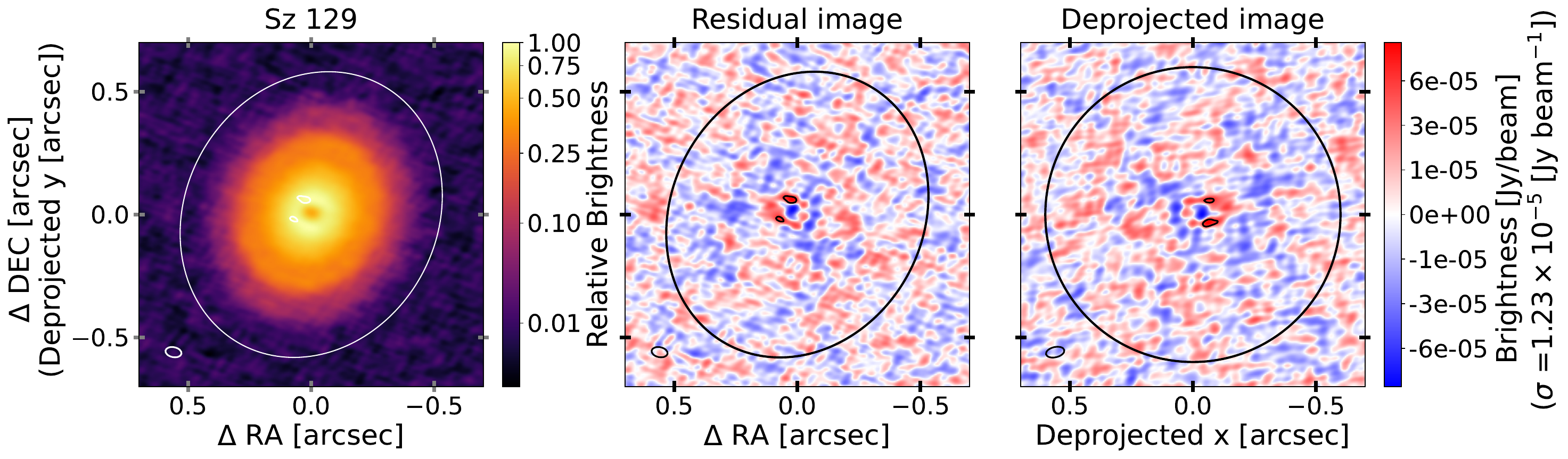}
\caption{Residual images for disks with weak signals in Section \ref{sec:weak_signal}. The format is the same as in Figure \ref{fig:gallery_images1}.  }
\label{fig:no_five}
\end{center}
\end{figure*}

Figures~\ref{fig:gallery_images1}-\ref{fig:no_five} show the residual images for each disk, grouped by their dominant features: spiral arms (Section~\ref{sec:revisit_spiral}), vertical thickness (Section~\ref{sec:geo_thick}), other systems with $5\sigma$ residuals beyond innermost regions (Section~\ref{sec:five_sigma}), and the remaining systems with weak asymmetries (Section~\ref{sec:weak_signal}). 

Elias 27, IM Lup, and WaOph 6 host spirals and are classified into the spiral group. Some systems show residuals attributed to \aizw{the} disk's finite vertical extent \cite[e.g.,][]{doi2021}, which cannot be modeled in our framework. In this case, the thickness produces the residuals symmetric with the disk's minor axis  (see Section~\ref{sec:geo_thick}). We identify \aizw{five} systems with significant vertical residuals as described in Section~\ref{sec:geo_thick}. The remaining systems are classified according to whether they exhibit $5\sigma$ residuals beyond the innermost regions near the stars, where nearly all systems, except GW~LUp, show such features. 

\aizw{We also compare the obtained residual maps with those produced using geometric parameters in \cite{huang_ring_2018}, as shown in Figure \ref{fig:comp_image}. Specifically, we recompute the visibility models using the optimized regularization parameters but assuming the geometric parameters in \cite{huang_ring_2018} when producing the residual images.} Overall, the morphologies of the residual maps agree well, although our analysis yields noticeably lower residual noise in several systems \aizw{because of differences in the geometric parameters}. In Appendix~\ref{sec:comparison_figure}, we also compare our recovered geometric parameters and brightness profiles with those from previous studies \citep{huang_ring_2018,andrews2021,jennings2022}. \aizw{The geometric parameters are largely consistent between this study and \cite{andrews2021} but show marginal discrepancies with \cite{huang_ring_2018}.} The radial and visibility profiles are also broadly consistent, with some differences. In the radial profiles, our estimated disk central intensities at the innermost radius, $r_1 = 0.0076 \arcsec$, are generally higher than those in \citet{jennings2022}.  We discuss the origins of discrepancies in Appendix \ref{sec:jeff_comp_tb}, confirming the consistencies between two studies. 

To further examine the residual structure in three disks, Elias 27, IM Lup, and WaOph 6, that exhibit pronounced even-symmetric spiral features characterized by $m=2$ modes \citep{perez2016, huang_spiral_2018, carreno2021, brown2021}, we construct odd- and even-symmetric residual maps by imaging the imaginary and real components of the visibilities separately (cf. \cite{aizawa2024}). In this approach, maps produced from the real visibilities isolate the even-symmetric residuals ($m=2,4 \dots$), whereas those produced from the imaginary visibilities reveal the odd-symmetric structures  $m=1,3 \dots$). Figure \ref{fig:odd-extraction} presents these odd- and even-symmetric residual maps alongside the original residual images for the three disks in the deprojected frame. 

\begin{deluxetable*}{lcccccccc}
\tablecaption{Disk parameters and weight factors. \label{table:disk_para}}
\tablewidth{0pt}
\tablehead{
  \colhead{Name} & 
  \colhead{$\gamma$ [\arcsec]} & 
  \colhead{$\log_{10}\alpha$} & 
  \colhead{$\Delta x_{\rm cen}$ [mas]} & 
  \colhead{$\Delta y_{\rm cen}$ [mas]} & 
  \colhead{$\cos i$} & 
  \colhead{($i$ [deg])} & 
  \colhead{PA [deg]}  & 
 \colhead{\begin{tabular}{c}
Weight \\
factor
\end{tabular}}}

\startdata
AS 209 & $0.0277^{+0.0009}_{-0.0009}$ & $-1.22^{+0.09}_{-0.08}$ & $0.76^{+0.07}_{-0.07}$ & $-1.22^{+0.05}_{-0.05}$ & $0.8210^{+0.0002}_{-0.0002}$ & $34.82^{+0.02}_{-0.02}$ & $85.77^{+0.04}_{-0.04}$ & 3.44 \\
DoAr 25 & $0.064^{+0.006}_{-0.004}$ & $0.0^{+0.3}_{-0.2}$ & $30.7^{+0.1}_{-0.2}$ & $-489.01^{+0.09}_{-0.09}$ & $0.3991^{+0.0003}_{-0.0003}$ & $66.48^{+0.02}_{-0.02}$ & $110.51^{+0.02}_{-0.02}$ & 3.53 \\
DoAr 33 & $0.089^{+0.005}_{-0.005}$ & $2.2^{+0.3}_{-0.3}$ & $1.8^{+0.1}_{-0.1}$ & $0.48^{+0.09}_{-0.10}$ & $0.748^{+0.002}_{-0.002}$ & $41.6^{+0.2}_{-0.2}$ & $81.3^{+0.3}_{-0.3}$ & 3.43 \\
Elias 20 & $0.039^{+0.002}_{-0.002}$ & $0.9^{+0.2}_{-0.2}$ & $-53.91^{+0.07}_{-0.06}$ & $-488.84^{+0.07}_{-0.07}$ & $0.5843^{+0.0007}_{-0.0007}$ & $54.24^{+0.05}_{-0.05}$ & $153.96^{+0.06}_{-0.06}$ & 3.50 \\
Elias 24 & $0.063^{+0.002}_{-0.002}$ & $3.1^{+0.2}_{-0.2}$ & $107.40^{+0.06}_{-0.06}$ & $-382.62^{+0.06}_{-0.06}$ & $0.8680^{+0.0004}_{-0.0004}$ & $29.78^{+0.05}_{-0.05}$ & $44.7^{+0.1}_{-0.1}$ & 2.82 \\
Elias 27 & $0.106^{+0.003}_{-0.003}$ & $3.2^{+0.2}_{-0.2}$ & $-2.5^{+0.1}_{-0.1}$ & $-1.53^{+0.08}_{-0.08}$ & $0.5547^{+0.0008}_{-0.0008}$ & $56.31^{+0.05}_{-0.06}$ & $119.19^{+0.05}_{-0.05}$ & 3.13 \\
GW Lup & $0.064^{+0.004}_{-0.005}$ & $2.8^{+0.3}_{-0.3}$ & $0.6^{+0.1}_{-0.1}$ & $0.3^{+0.1}_{-0.1}$ & $0.780^{+0.001}_{-0.001}$ & $38.8^{+0.1}_{-0.1}$ & $37.1^{+0.2}_{-0.2}$ & 3.44 \\
HD 142666 & $0.0205^{+0.0007}_{-0.0007}$ & $-1.02^{+0.08}_{-0.08}$ & $10.00^{+0.07}_{-0.07}$ & $15.5^{+0.1}_{-0.1}$ & $0.4714^{+0.0005}_{-0.0005}$ & $61.88^{+0.03}_{-0.03}$ & $162.18^{+0.03}_{-0.03}$ & 3.29 \\
IM Lup & $0.072^{+0.005}_{-0.004}$ & $2.7^{+0.3}_{-0.3}$ & $1.12^{+0.05}_{-0.05}$ & $-0.70^{+0.05}_{-0.05}$ & $0.6387^{+0.0006}_{-0.0006}$ & $50.31^{+0.04}_{-0.04}$ & $143.82^{+0.06}_{-0.06}$ & 2.72 \\
MY Lup & $0.085^{+0.008}_{-0.006}$ & $-0.7^{+0.3}_{-0.2}$ & $-77.3^{+0.2}_{-0.2}$ & $62.7^{+0.1}_{-0.1}$ & $0.2907^{+0.0004}_{-0.0004}$ & $73.10^{+0.02}_{-0.02}$ & $58.94^{+0.02}_{-0.03}$ & 3.54 \\
RU Lup & $0.0297^{+0.0007}_{-0.0007}$ & $-0.05^{+0.09}_{-0.09}$ & $-15.30^{+0.04}_{-0.04}$ & $85.55^{+0.04}_{-0.04}$ & $0.9535^{+0.0006}_{-0.0006}$ & $17.5^{+0.1}_{-0.1}$ & $123.7^{+0.3}_{-0.3}$ & 3.58 \\
SR 4 & $0.022^{+0.001}_{-0.001}$ & $0.2^{+0.1}_{-0.1}$ & $-59.50^{+0.06}_{-0.06}$ & $-508.55^{+0.05}_{-0.06}$ & $0.9307^{+0.0008}_{-0.0008}$ & $21.5^{+0.1}_{-0.1}$ & $23.0^{+0.3}_{-0.3}$ & 3.48 \\
Sz 114 & $0.094^{+0.005}_{-0.005}$ & $2.4^{+0.3}_{-0.3}$ & $-0.2^{+0.2}_{-0.2}$ & $0.8^{+0.1}_{-0.1}$ & $0.959^{+0.003}_{-0.003}$ & $16.5^{+0.6}_{-0.6}$ & $168^{+2}_{-2}$ & 3.40 \\
Sz 129 & $0.047^{+0.003}_{-0.003}$ & $-1.6^{+0.2}_{-0.1}$ & $3.1^{+0.1}_{-0.1}$ & $4.5^{+0.1}_{-0.1}$ & $0.8528^{+0.001}_{-0.001}$ & $31.5^{+0.1}_{-0.1}$ & $151.7^{+0.2}_{-0.2}$ & 3.54 \\
WaOph 6 & $0.080^{+0.003}_{-0.004}$ & $2.6^{+0.3}_{-0.3}$ & $-249.80^{+0.05}_{-0.06}$ & $-352.73^{+0.05}_{-0.05}$ & $0.6843^{+0.0007}_{-0.0007}$ & $46.82^{+0.05}_{-0.05}$ & $172.20^{+0.07}_{-0.07}$ & 3.66 \\
WSB 52 & $0.047^{+0.002}_{-0.002}$ & $0.3^{+0.2}_{-0.2}$ & $-119.32^{+0.06}_{-0.06}$ & $-433.04^{+0.05}_{-0.05}$ & $0.5854^{+0.0006}_{-0.0006}$ & $54.17^{+0.04}_{-0.04}$ & $138.34^{+0.06}_{-0.06}$ & 3.49 
\enddata
\end{deluxetable*}

\begin{figure*} \begin{center}
\includegraphics[width=0.48\linewidth]{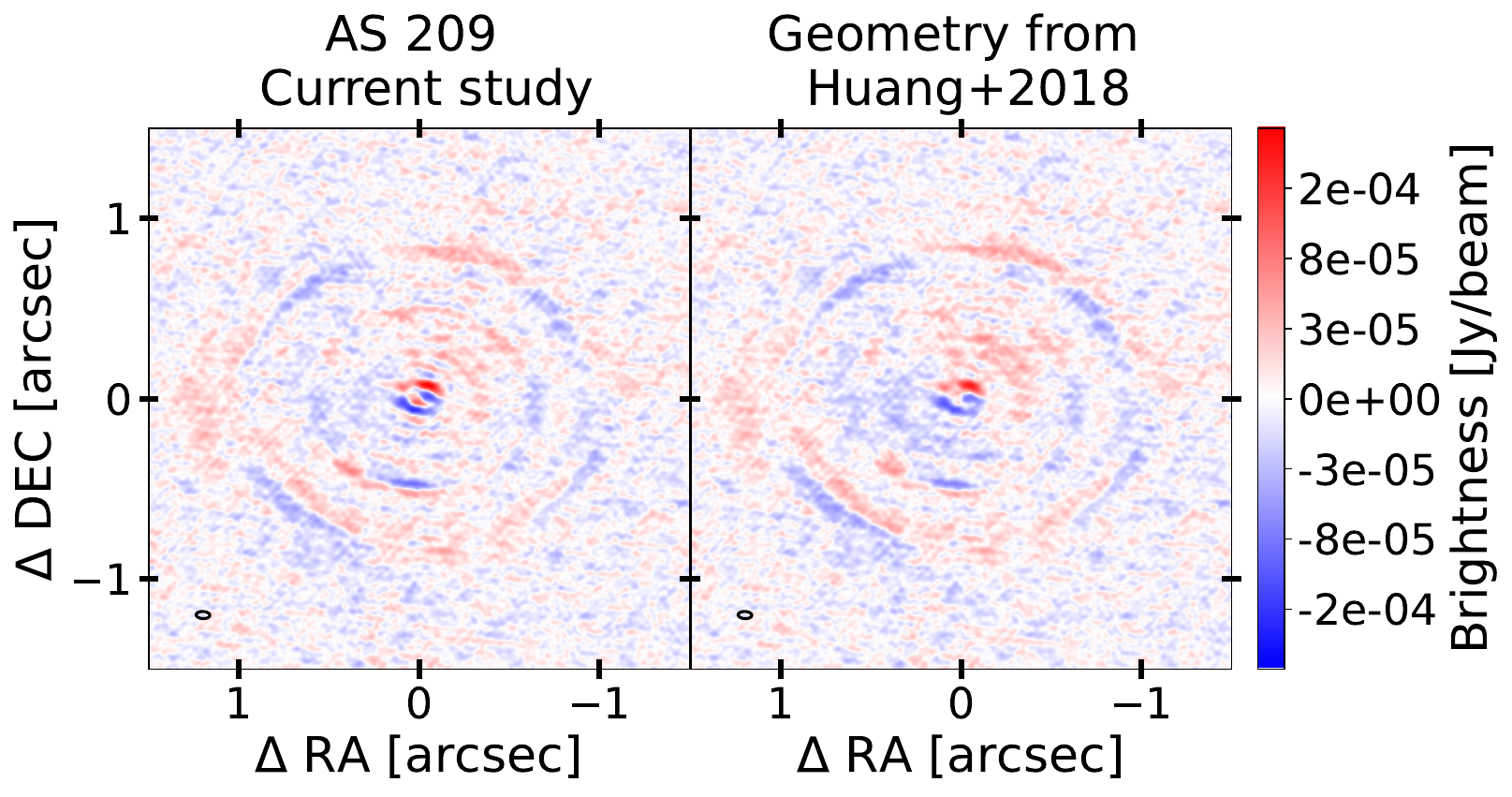}
\includegraphics[width=0.48\linewidth]{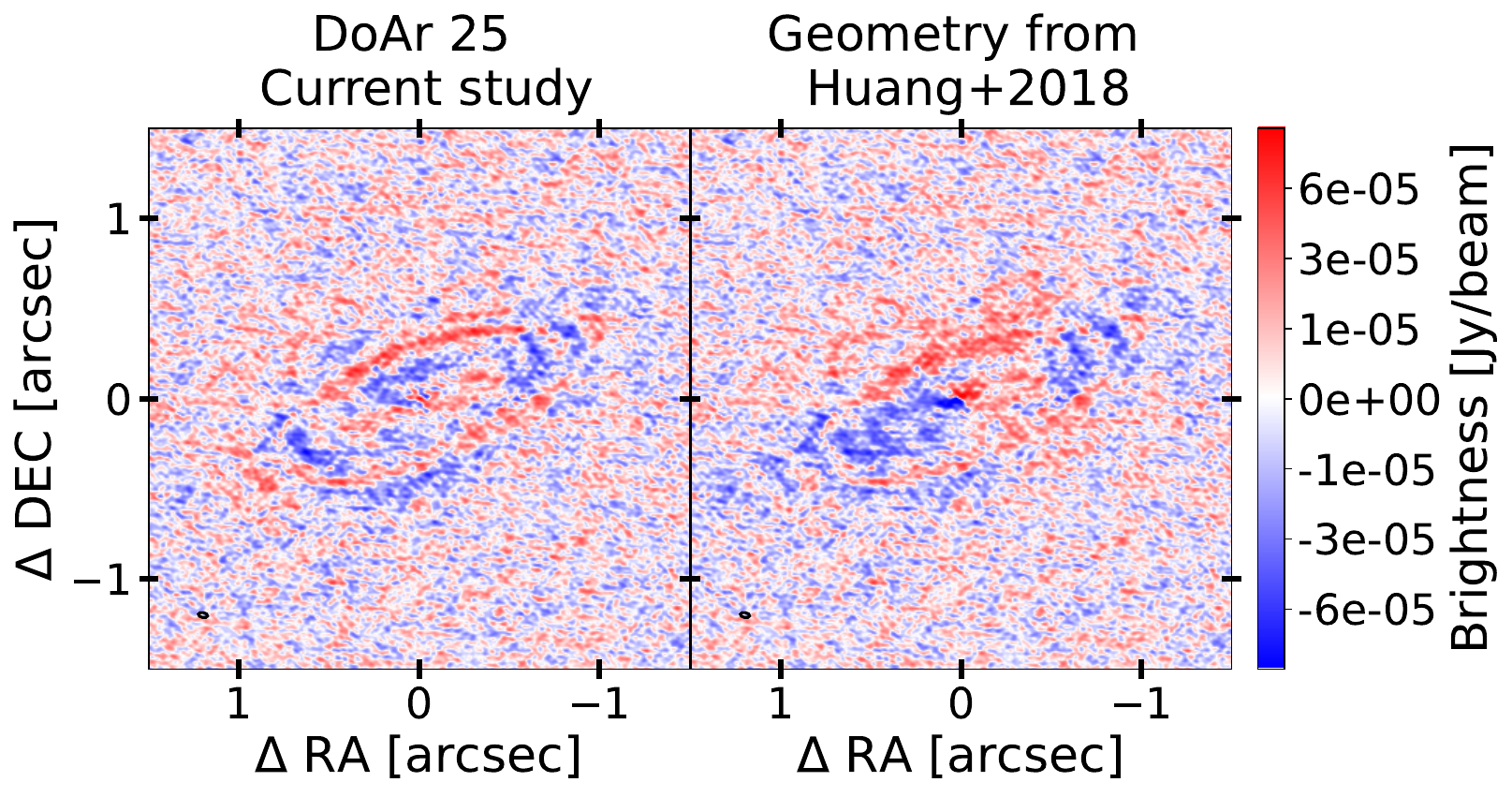}
\includegraphics[width=0.48\linewidth]{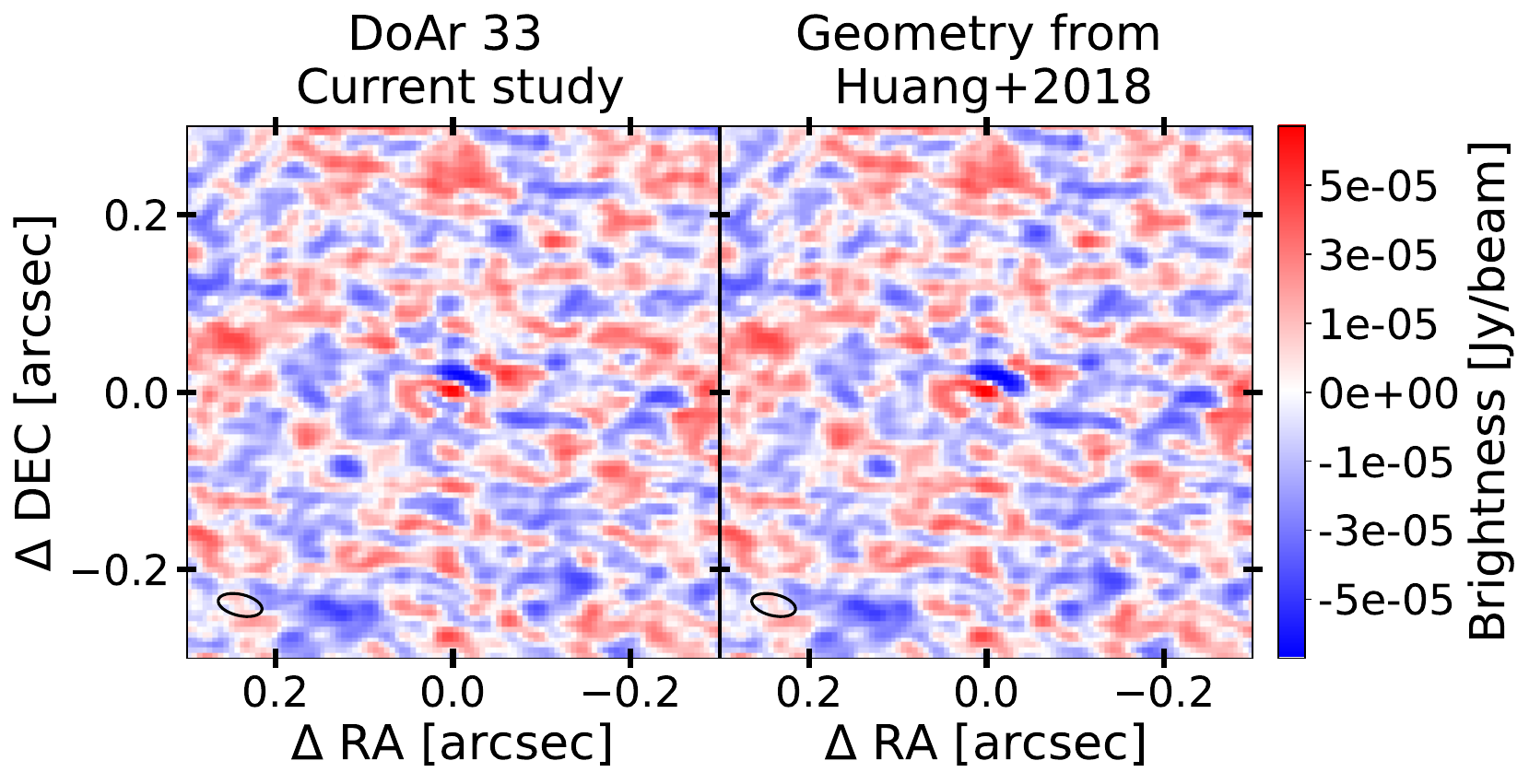}
\includegraphics[width=0.48\linewidth]{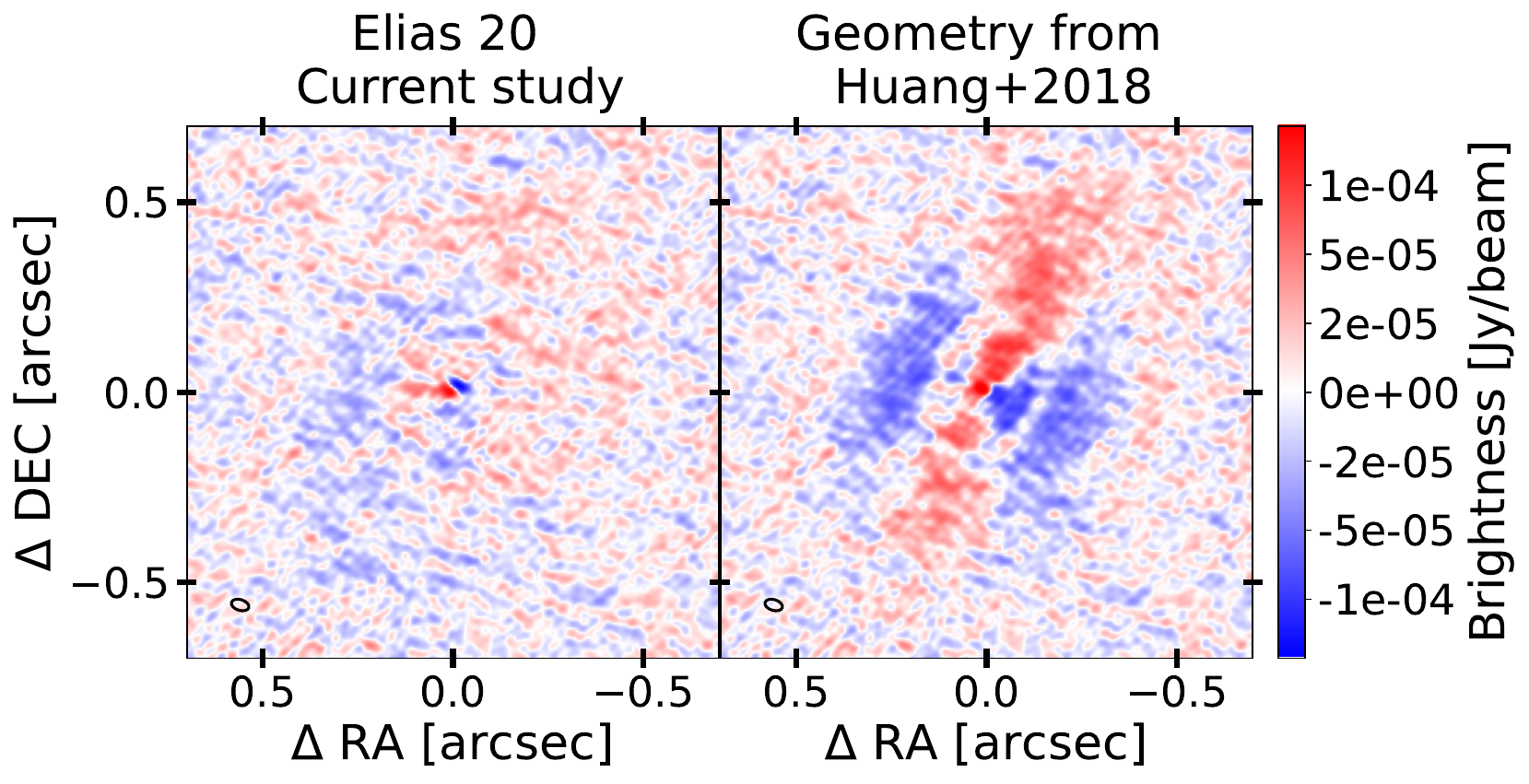}
\includegraphics[width=0.48\linewidth]{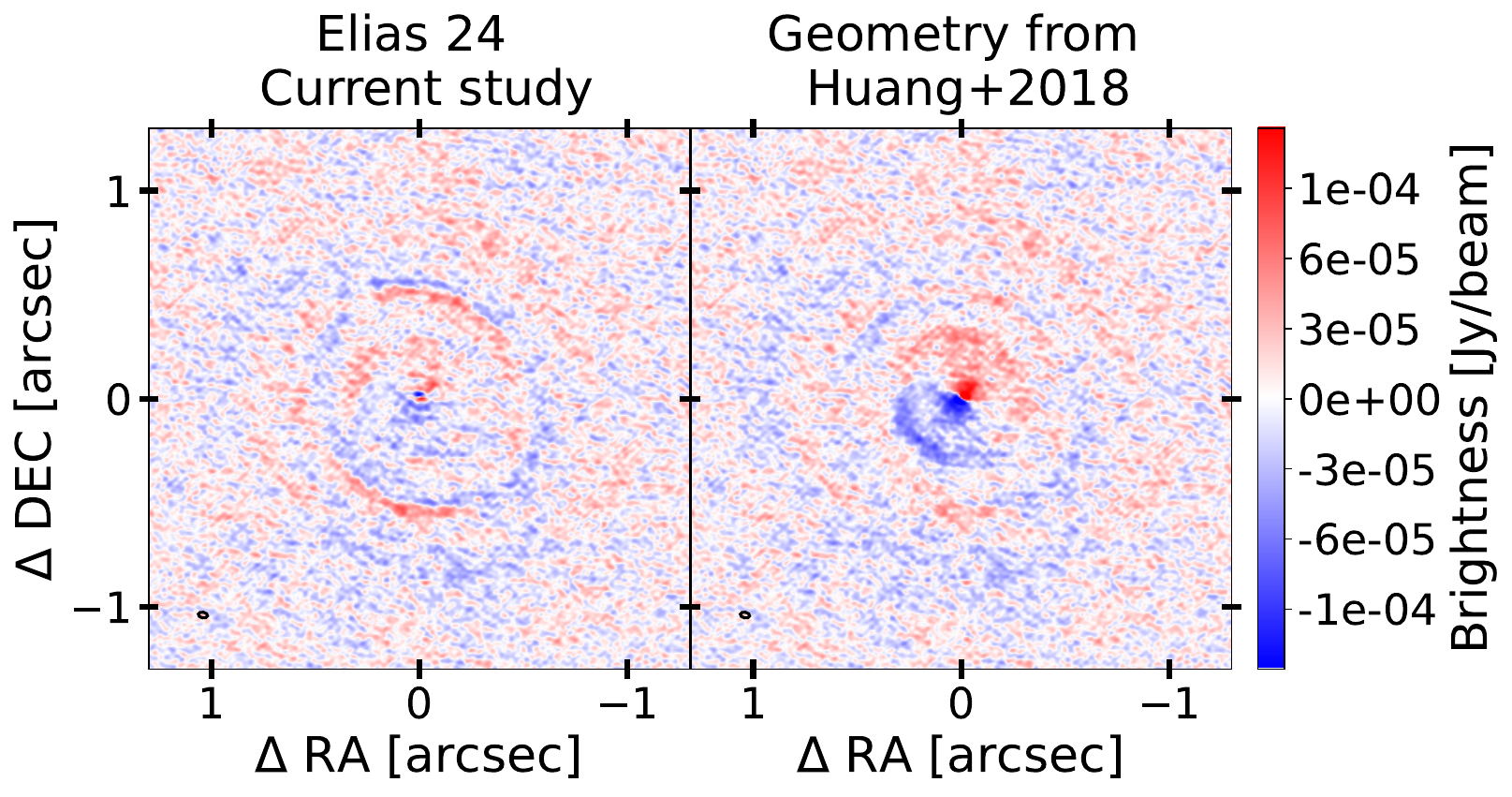}
\includegraphics[width=0.48\linewidth]{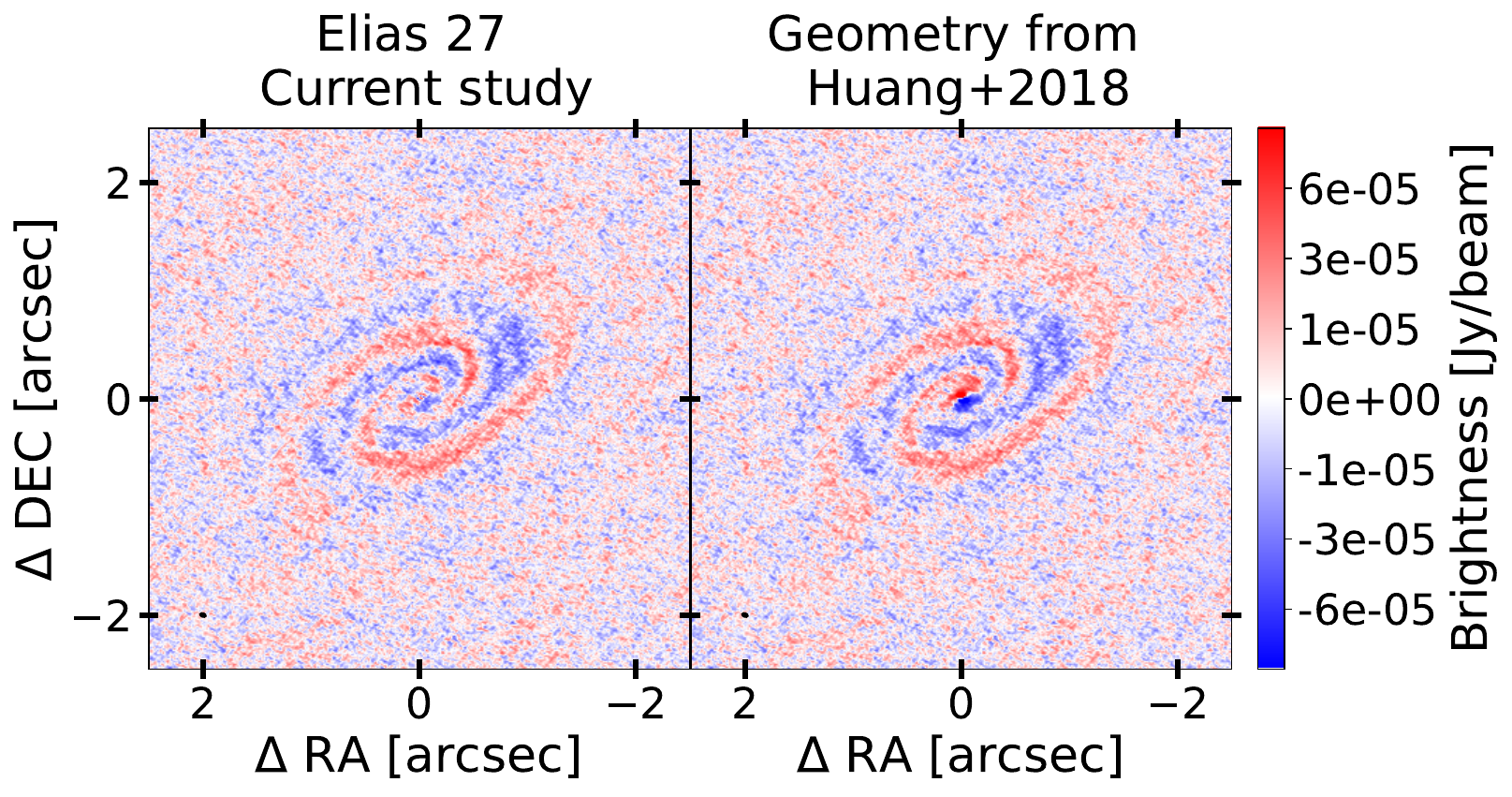}
\includegraphics[width=0.48\linewidth]{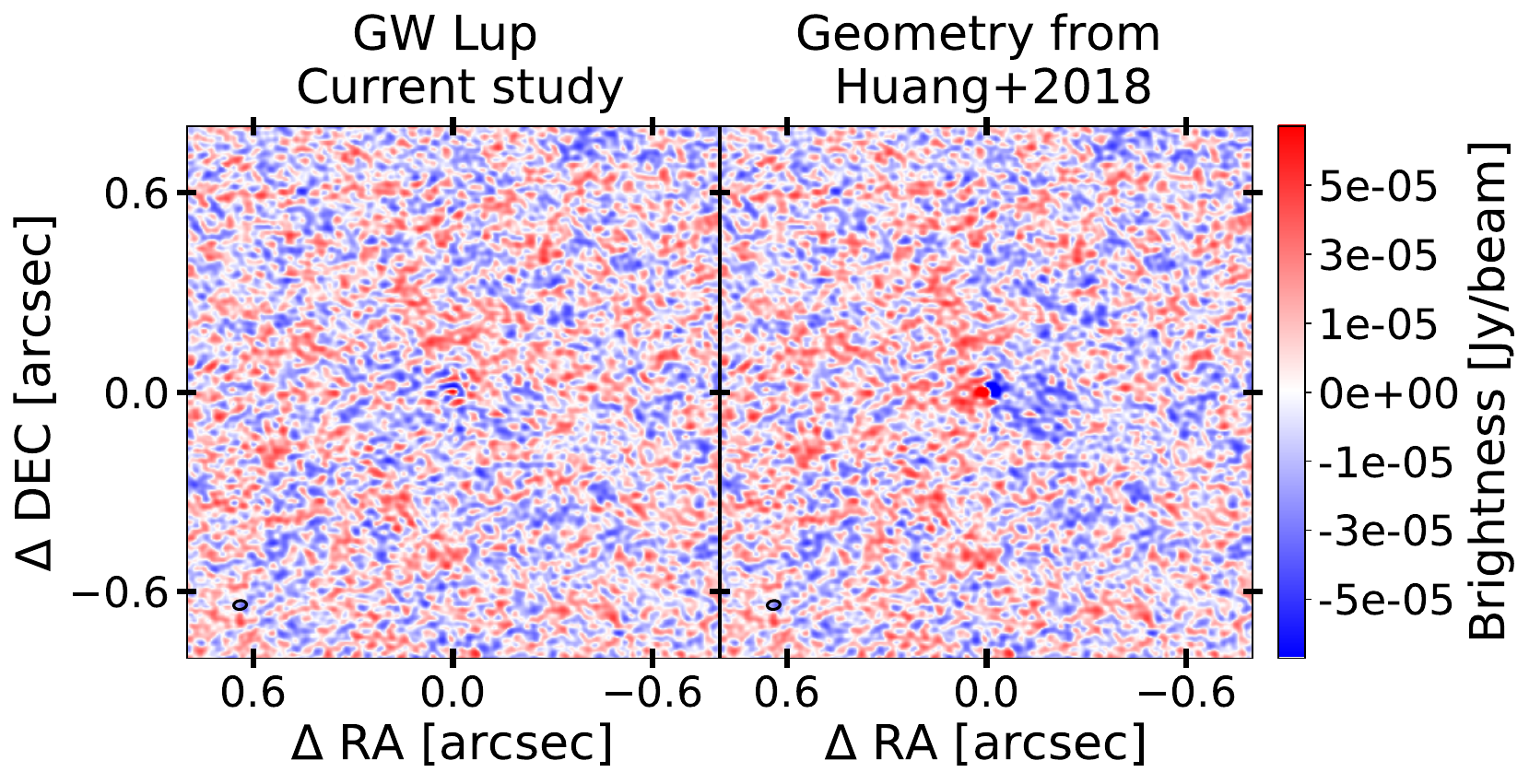}
\includegraphics[width=0.48\linewidth]{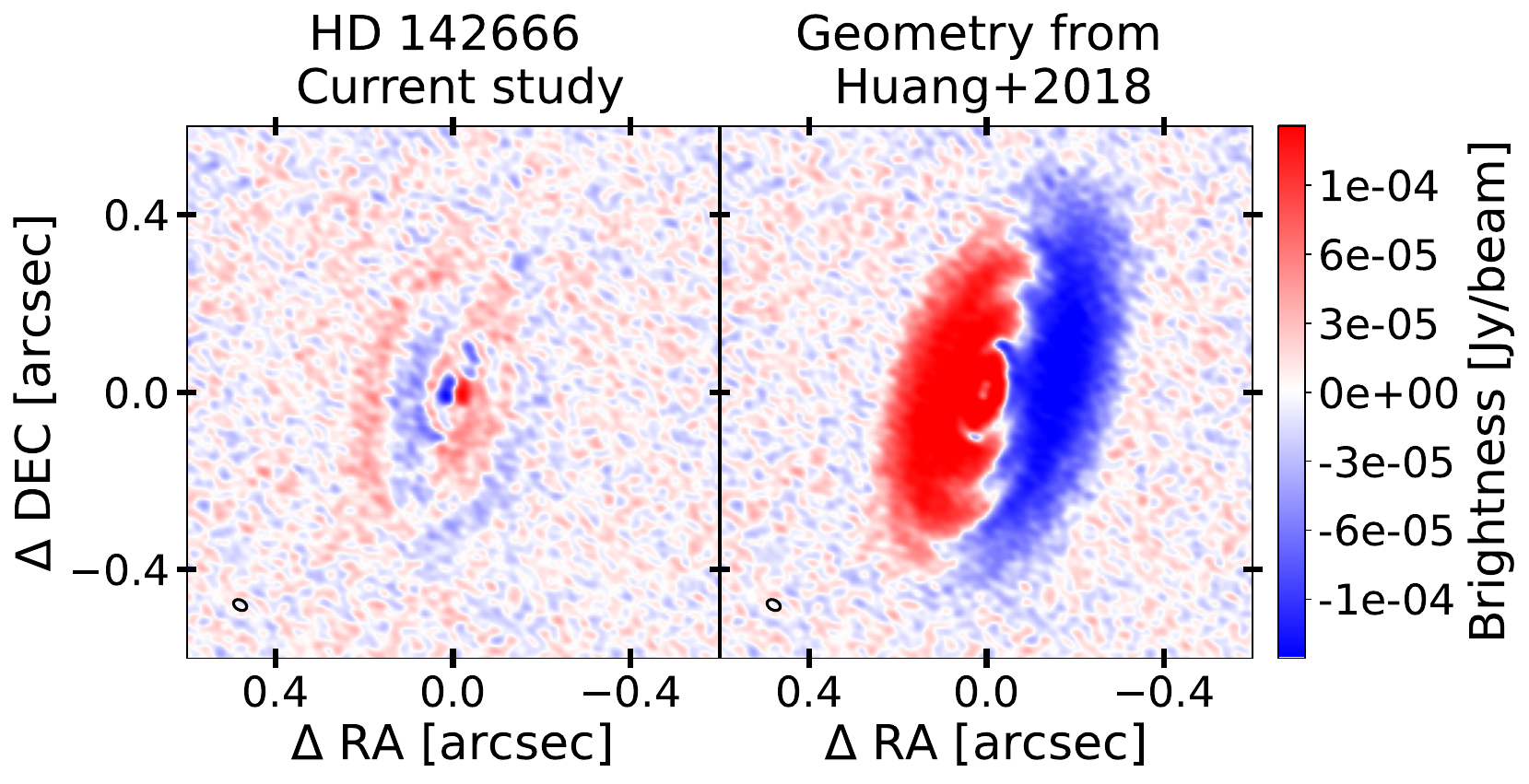}
\end{center}
\caption{Comparison of residual maps generated using the geometric parameters derived in this work and those from \citet{huang_ring_2018}. The left panels in each pair correspond to the central panels shown in Figures~\ref{fig:gallery_images1}-\ref{fig:no_five}, and the same color scale and formatting are adopted here. } 
\label{fig:comp_image} \end{figure*}
 \addtocounter{figure}{-1}

\begin{figure*} \begin{center}
\includegraphics[width=0.48\linewidth]{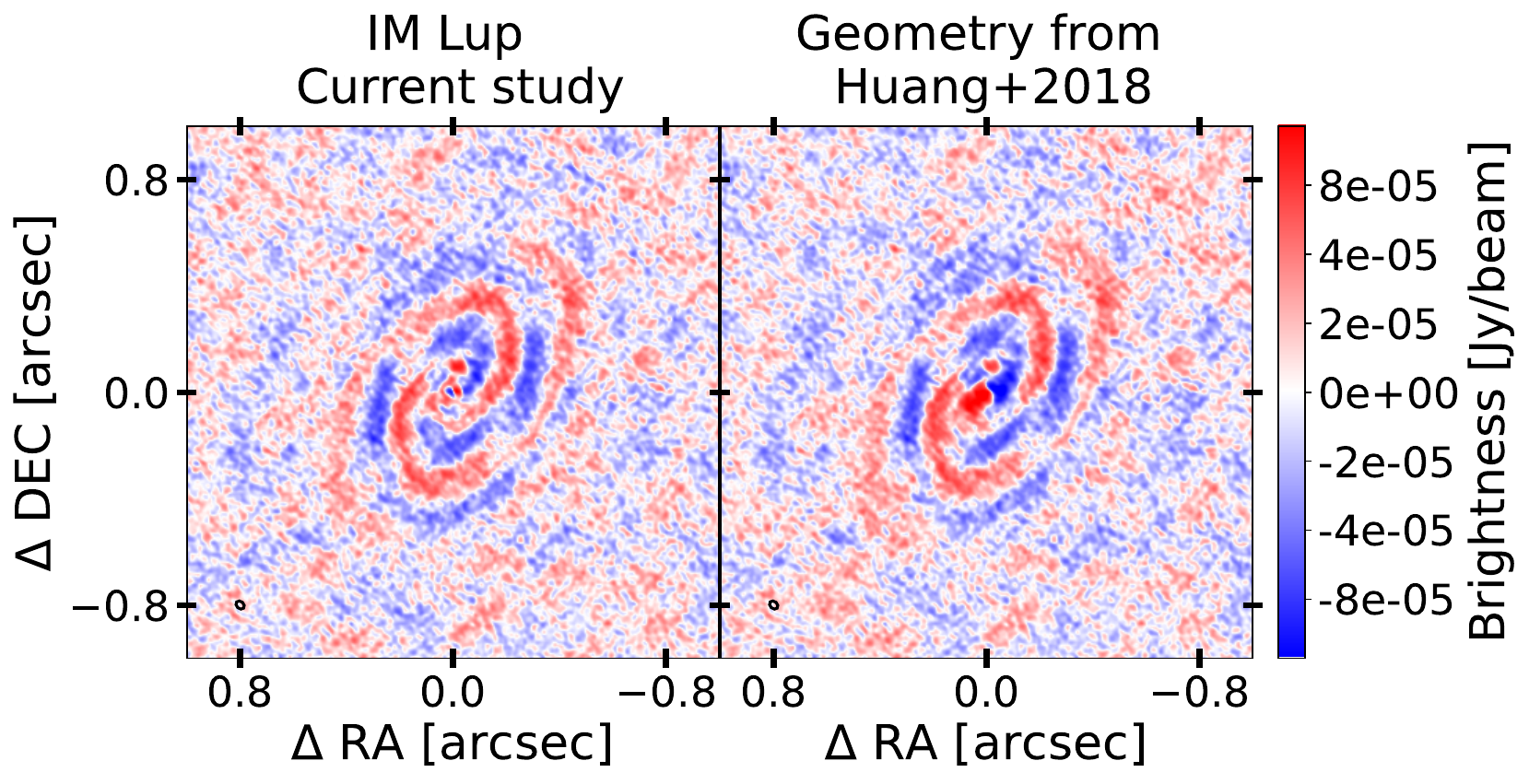}
\includegraphics[width=0.48\linewidth]{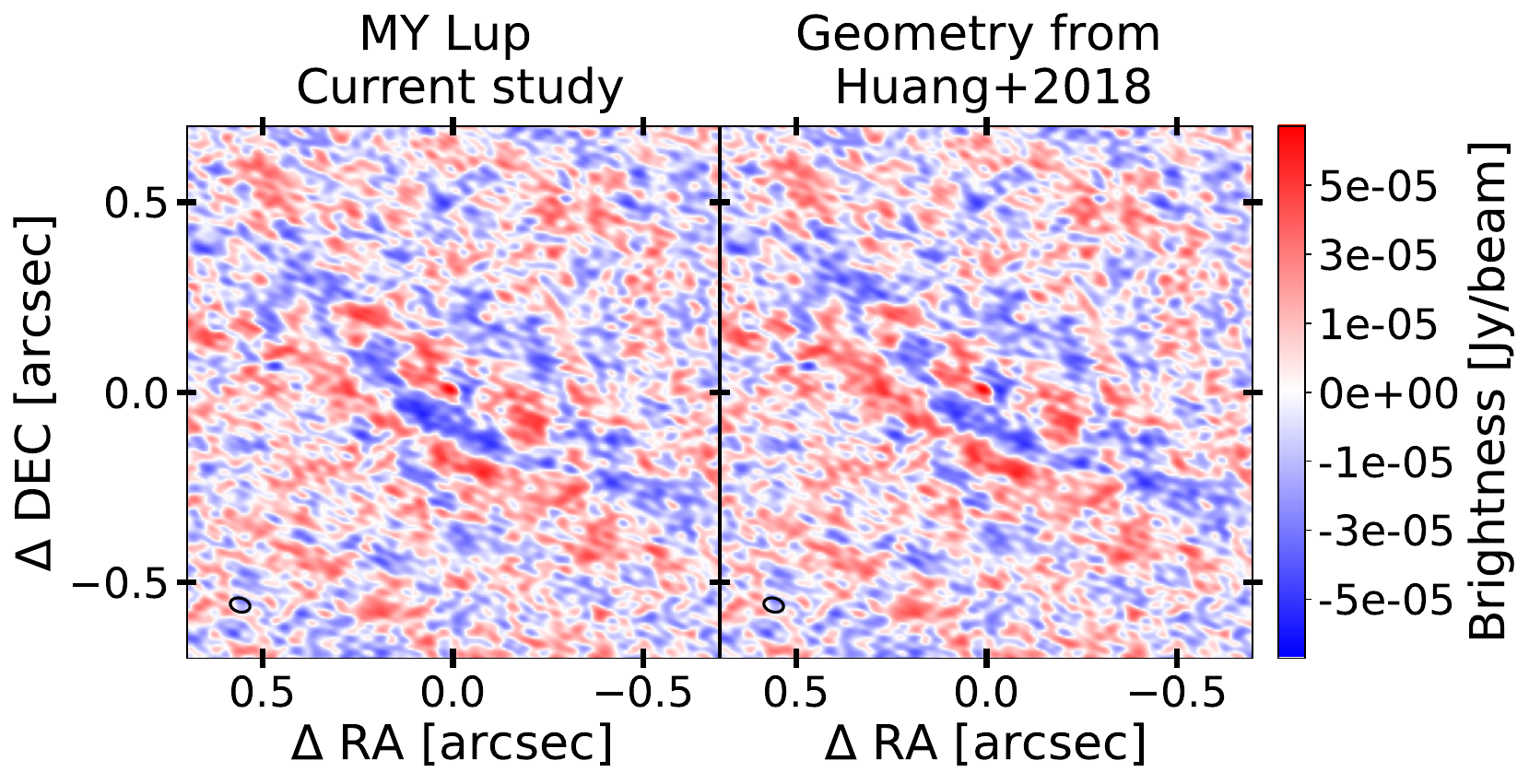}
\includegraphics[width=0.48\linewidth]{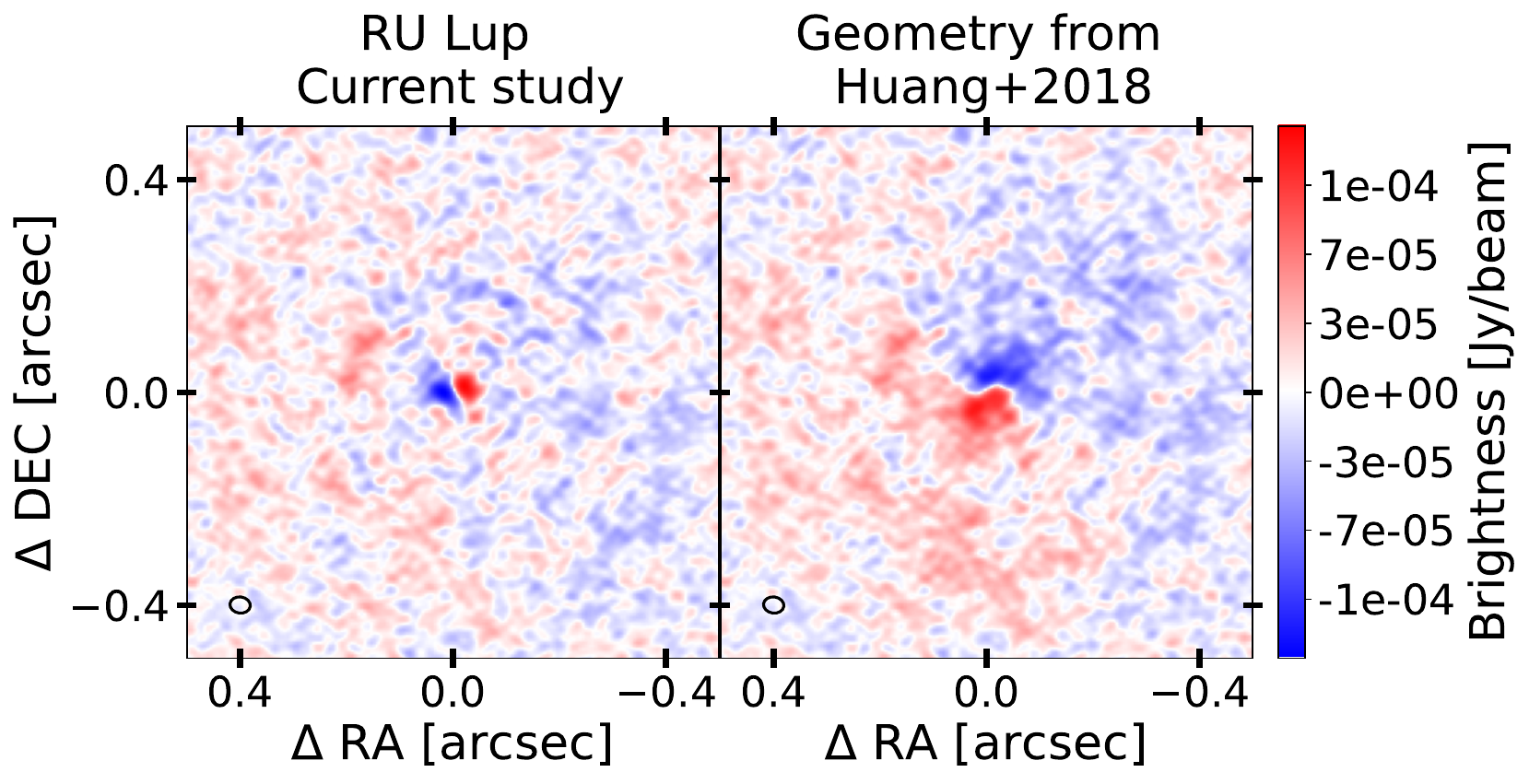}
\includegraphics[width=0.48\linewidth]{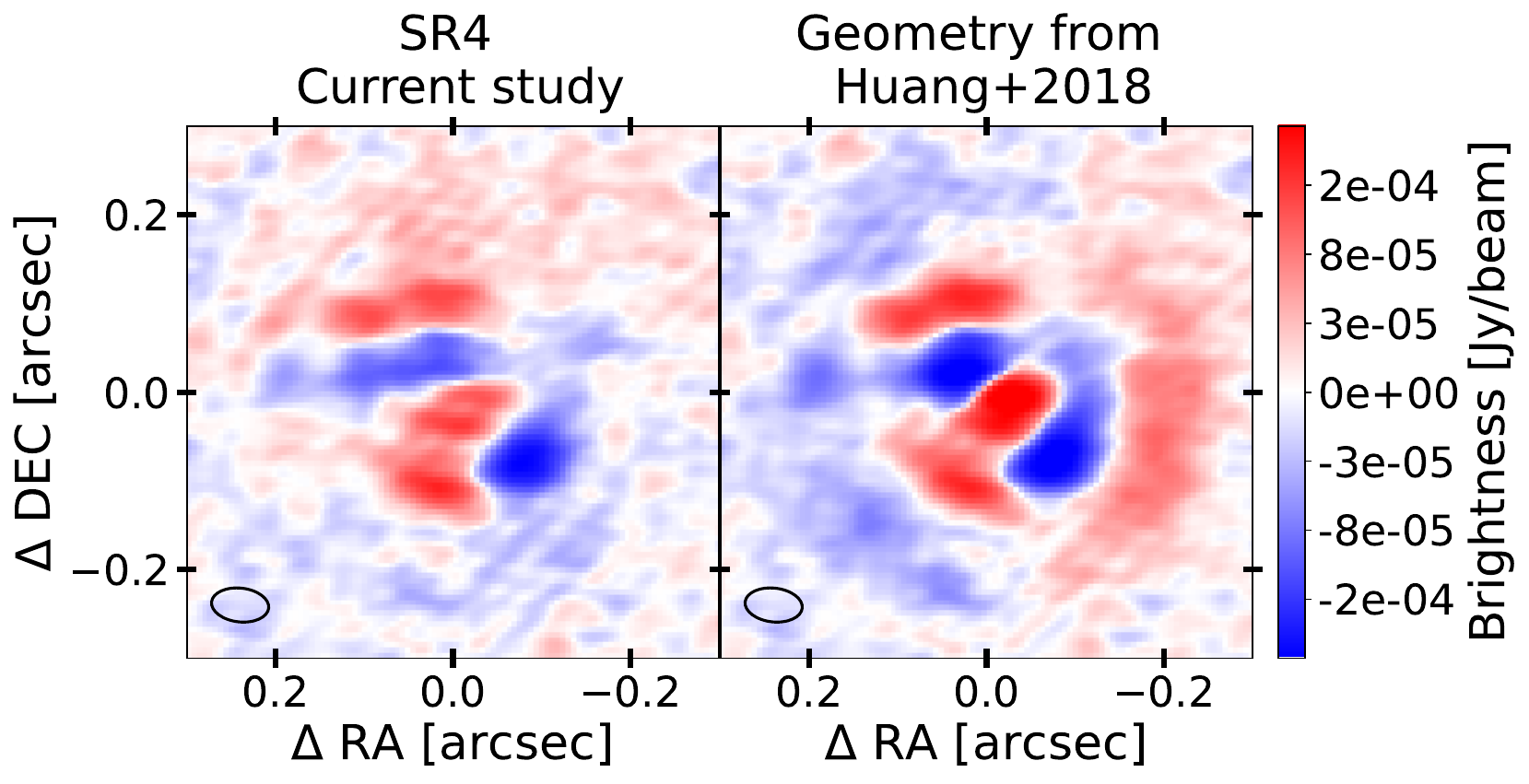}
\includegraphics[width=0.48\linewidth]{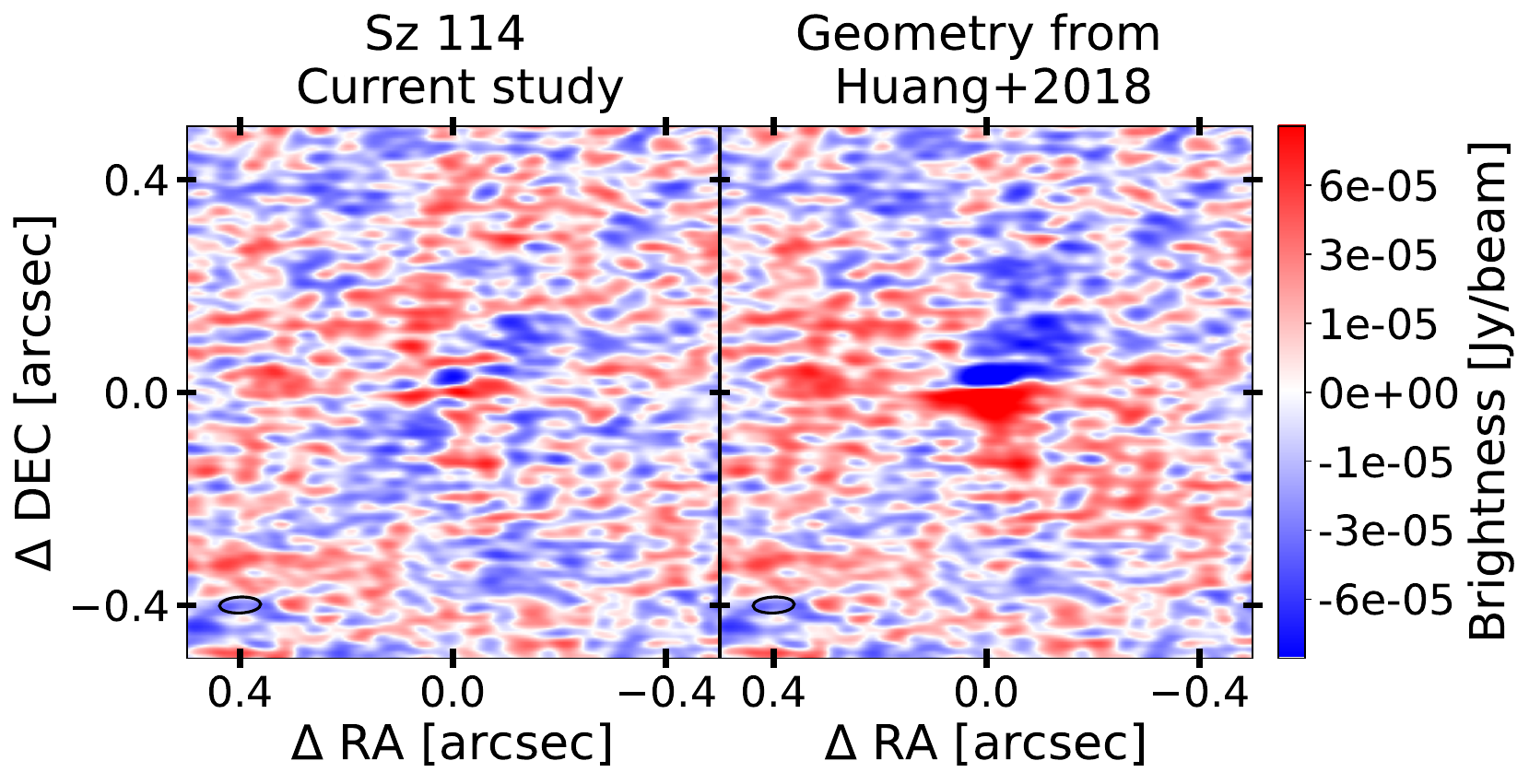}
\includegraphics[width=0.48\linewidth]{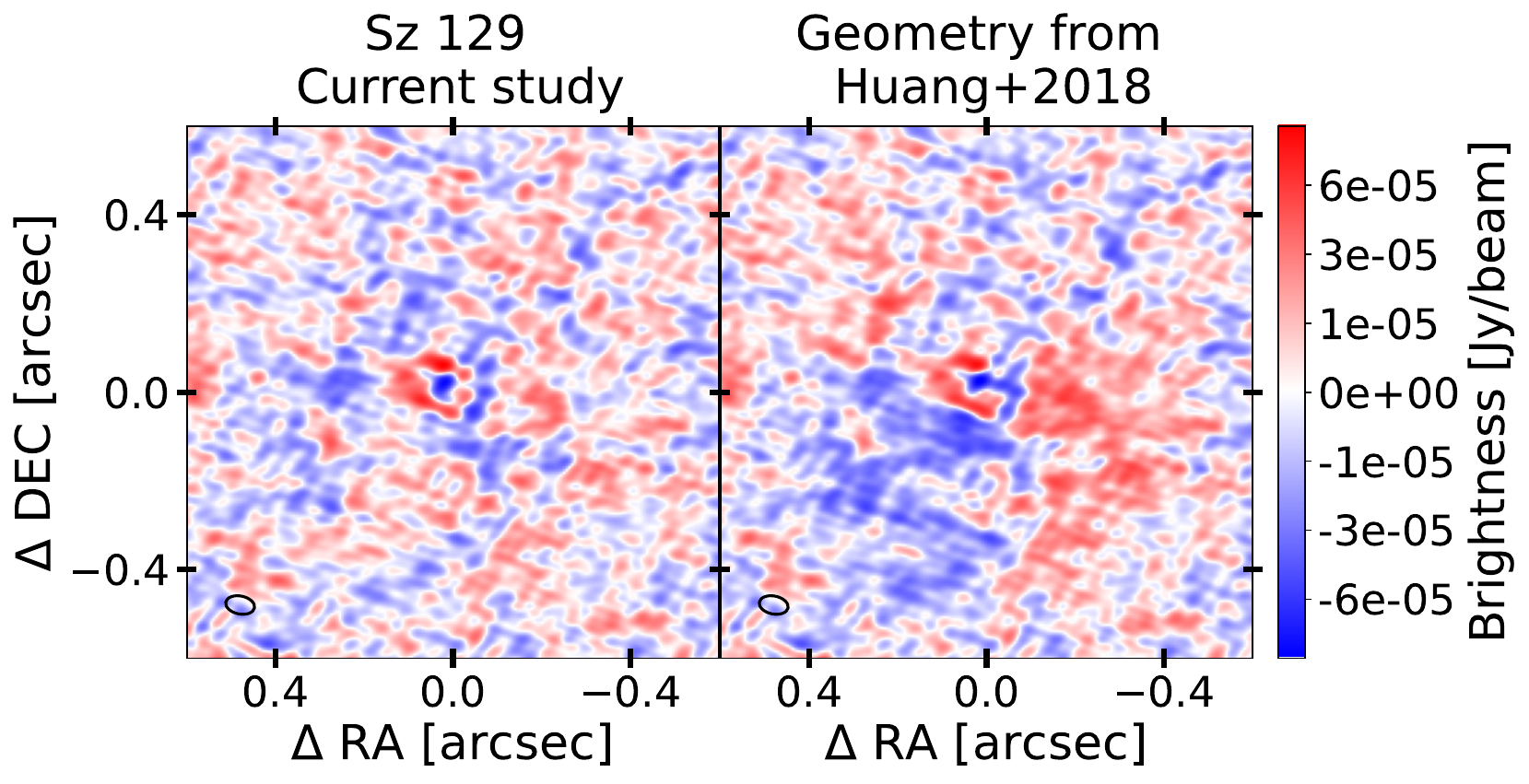}
\includegraphics[width=0.48\linewidth]{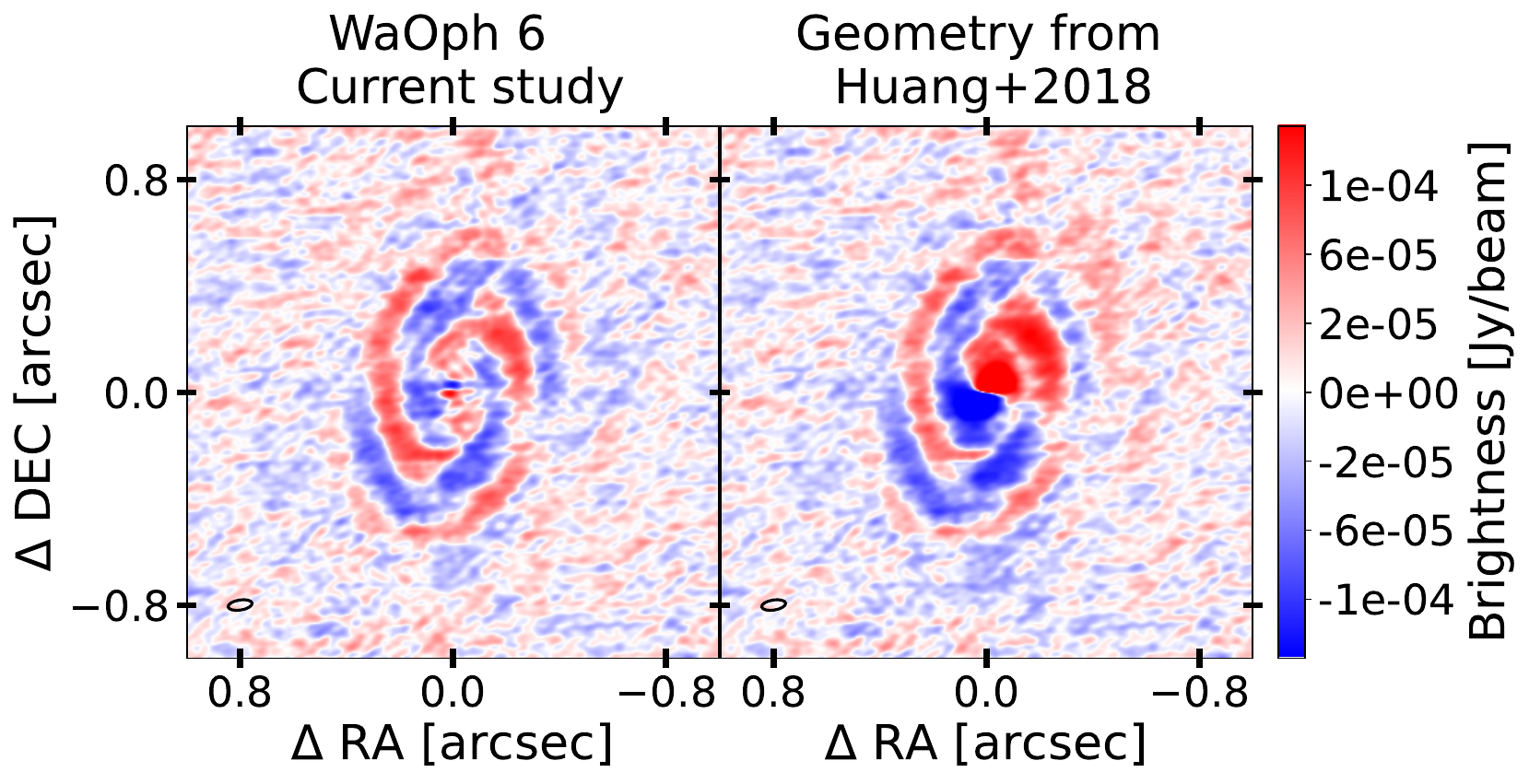}
\includegraphics[width=0.48\linewidth]{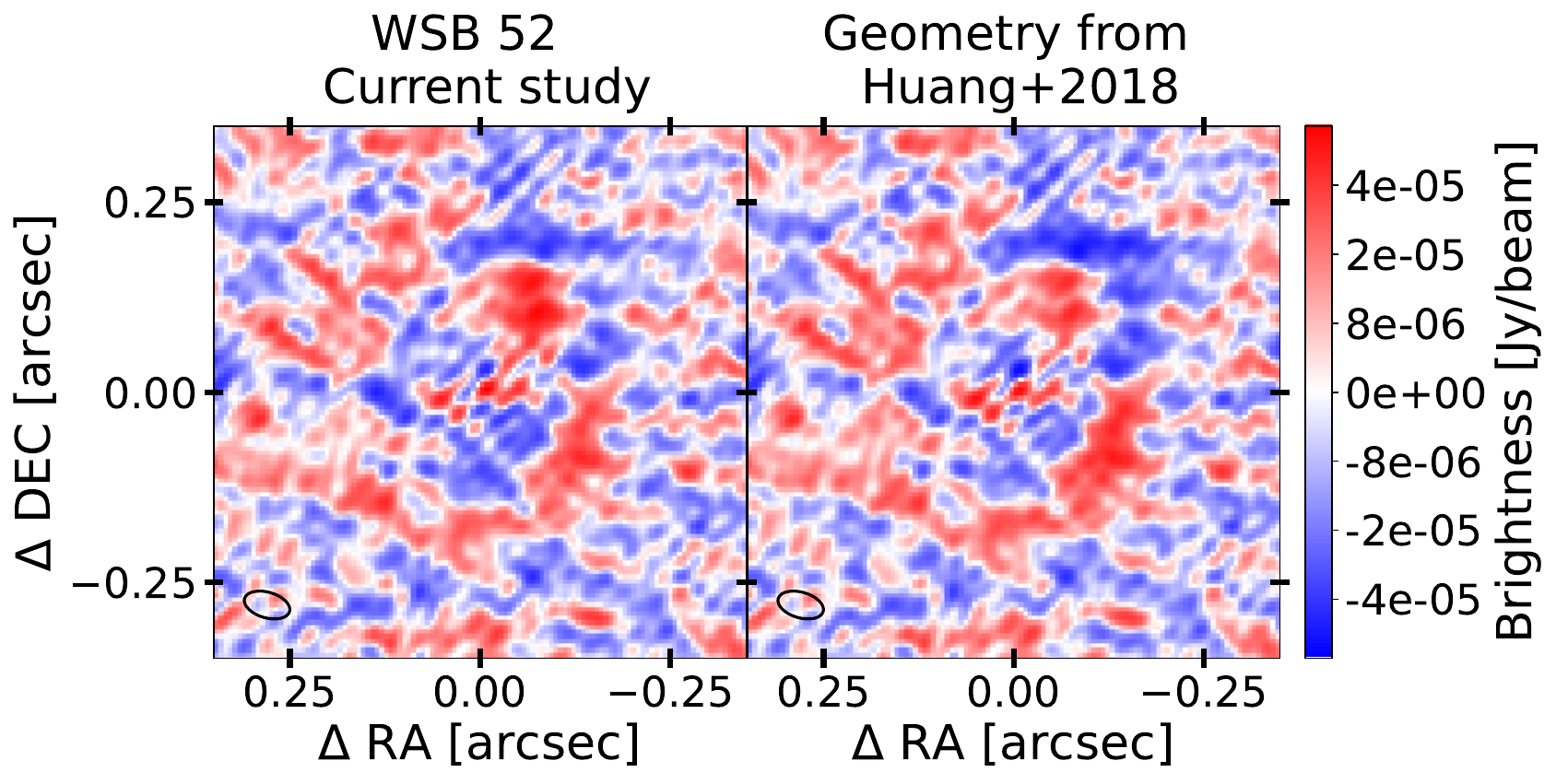}
 \end{center}
 \caption{(Continued). } 
 \end{figure*}

\subsection{No strong candidates for circumplanetary disk emission} 
\label{sec:no_cpd}

\begin{figure*} \begin{center}
 \includegraphics[width=0.9\linewidth]{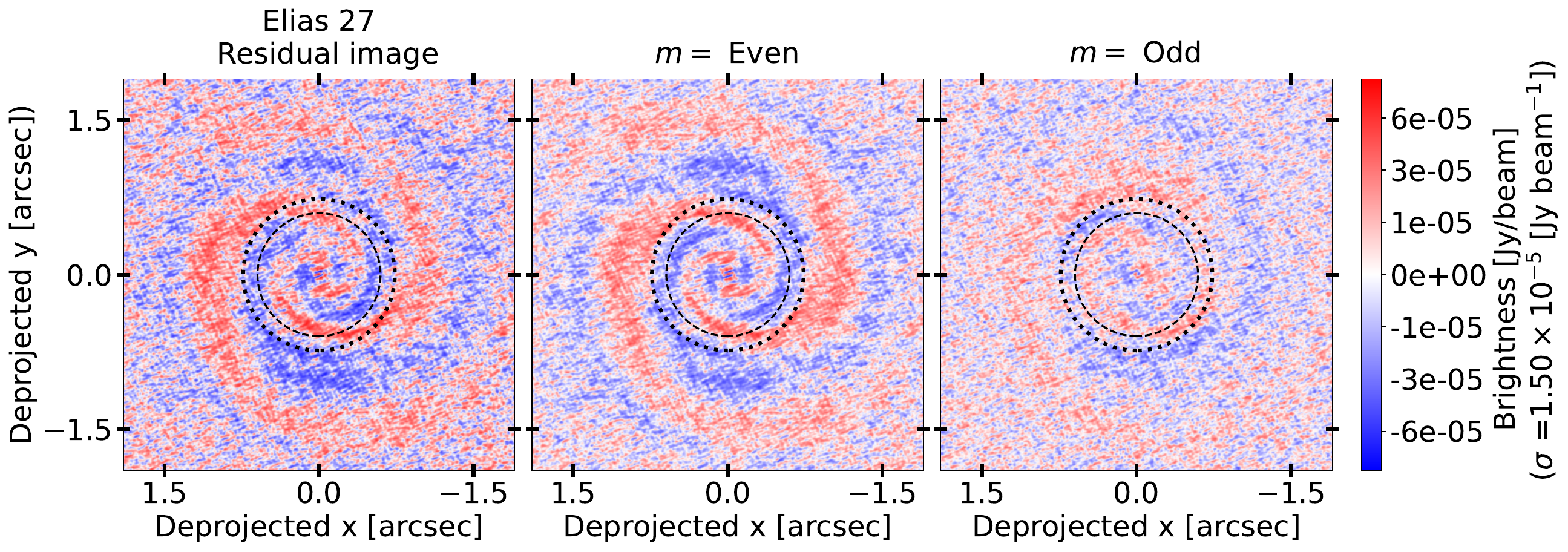}
 \includegraphics[width=0.9\linewidth]{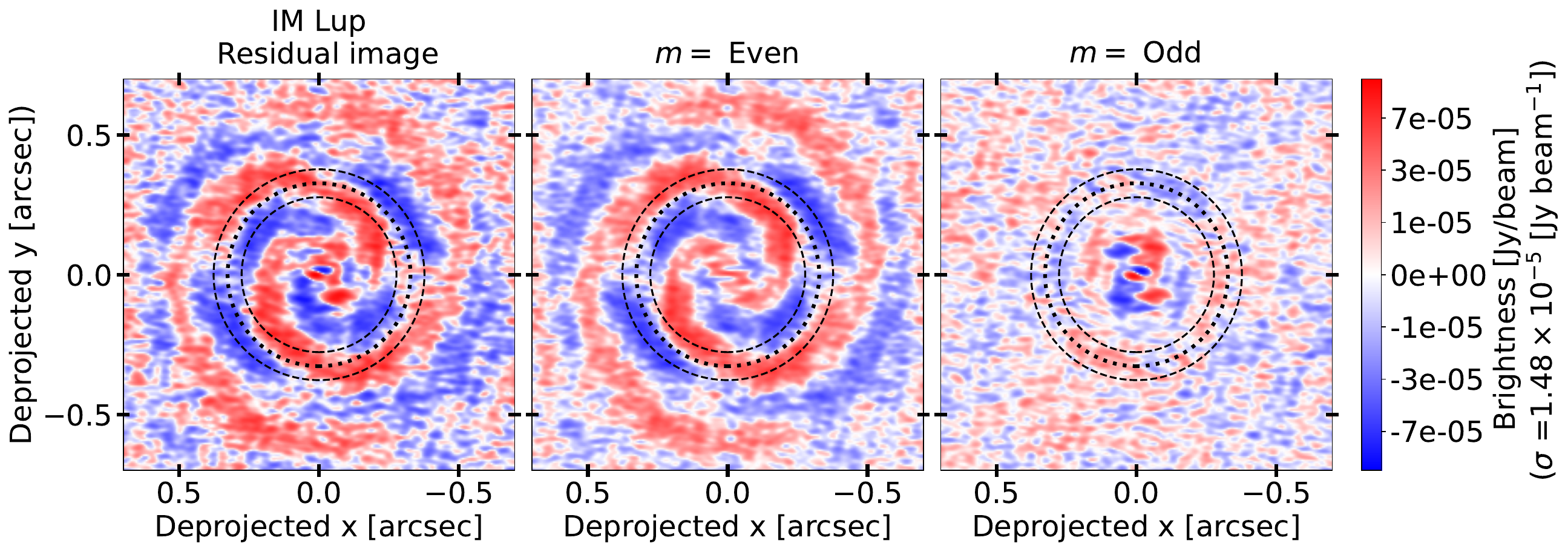}
\includegraphics[width=0.9\linewidth]{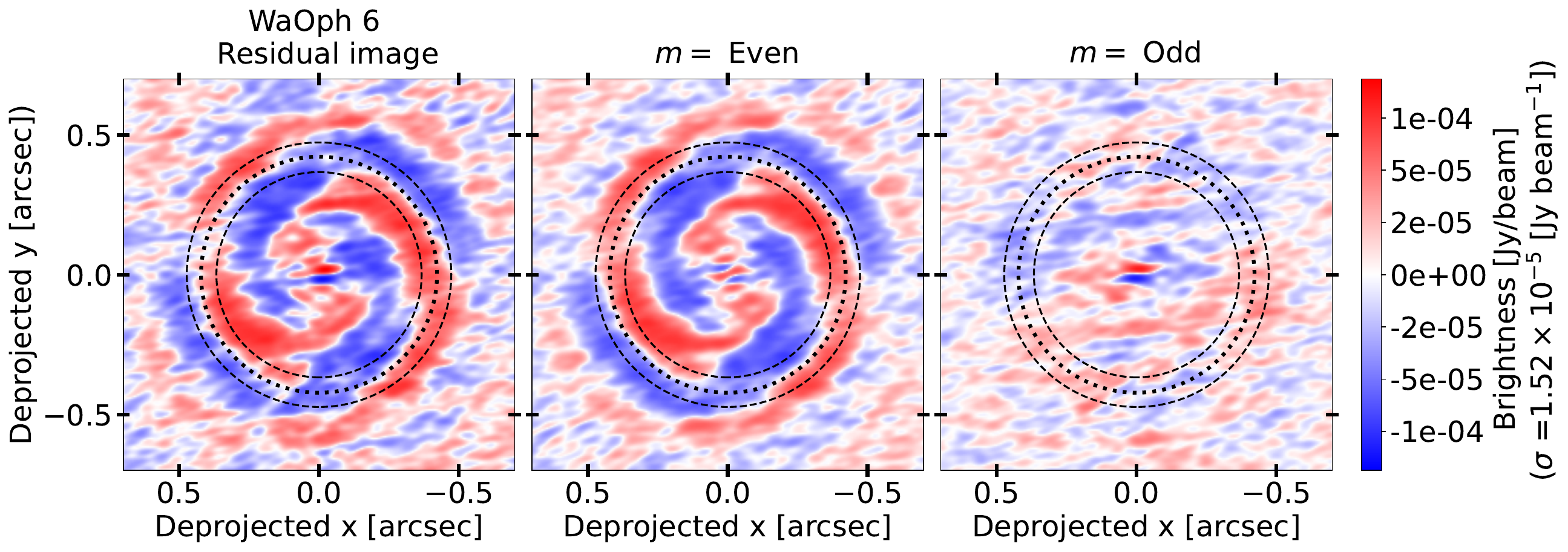}
\caption{Deprojected residual images of Elias 27, IM Lup, and WaOph 6.  (Left) The residual images after subtracting the best-fit axisymmetric models. (same as right panels in Figure \ref{fig:gallery_images1}). 
(Center) The even-symmetric (\textit{m} = Even) residual images produced with real part of residual visibilities.  
(Right) The odd-symmetric (\textit{m} = Odd) residual images produced with imaginary part of residual visibilities. The brightness scale is shown with an asinh stretch. \aizw{Dotted and dashed lines mark the rings and gaps, respectively, which are discussed in Section~\ref{sec:m2spiral} and Figure~\ref{fourier_ana} in the context of their correlations with the spiral amplitudes. } }\label{fig:odd-extraction}
 \end{center}
 \end{figure*}

We examine the residual images for point-source emission indicative of CPDs. In our analysis, we overlay 5$\sigma$ contours on all the residual images and visually inspect them for any flux excess. For Elias 27, IM Lup, and WaOph 6, we also search for any source in the odd-symmetric images in Figure \ref{fig:odd-extraction} as well. 
 
We detect $5\sigma$ excesses in all systems except GW Lup, but none of these features are consistent with compact, point-source emission. A majority of disks exhibit residuals above the 5$\sigma$ level in their innermost regions; however, we attribute these to complex, optically thick inner-disk structure, so we exclude them as CPD candidates. The others are associated with large-scale features, such as spiral arms and vertical disk thickness. Systems with $5\sigma$ residuals, RU Lup, SR 4, and WSB 52, show no evidence of point-source emission.  Taken together, these data provide no robust evidence for circumplanetary disks in our sample.

In a previous study, \citet{andrews2021} examine nine DSHARP disks, including HD 143006 and HD 163296, by analyzing residual maps for thirteen gap regions where protoplanets might reside. Although several candidate signals are identified, their low signal-to-noise ratios prevented definitive confirmation. Injection-recovery tests at the gap locations are used to quantitatively constrain CPD fluxes. In our sample, which includes all nine of their targets except HD 143006 and HD 163296, we likewise find no compelling CPD candidates. Adopting a $5\sigma$ threshold yields upper limits on CPD fluxes of 50-100~$\mu$Jy, comparable to those in \citet{andrews2021}.

\section{Detailed Analysis on Spiral Arms} \label{sec:revisit_spiral}
As presented in Figure \ref{fig:gallery_images1}, we confirm spiral patterns in Elias 27, IM Lup, and WaOph 6, in agreement with earlier studies \citep{perez2016, huang_spiral_2018, carreno2021, brown2021}. Here, we provide a detailed analysis of the morphology and Fourier decomposition.

\subsection{Formulation for Fourier Analysis} \label{sec:polar_fourier}
To characterize the spirals, we quantify the amplitudes and phases employing a polar Fourier transformation. We decompose a deprojected residual image, $I(x,y)$, into a Fourier series, analogous to the methods used in galaxy analyses \citep[e.g.,][]{binney2008}: 
\begin{equation} 
I(x,y) = \sum_{m=0}^\infty I_{m}(r) \cos \left[m\left(\phi - \phi_{m}(r)\right)\right], \label{eq:decomp}
\end{equation} 
where $I_{m}(r)$ represents the amplitude of the $m$-th mode, and $I_{0}(r)$ corresponds to the axisymmetric component. Note that the even-symmetric image, constructed using only the real part of the data, contains exclusively even-$m$ components, whereas the odd-symmetric image contains exclusively odd-$m$ components.

The amplitude and phase of the $m$-th component are computed as follows: 
\begin{align} I_{m}(r) &= 2 \sqrt{C_{m}^{2}(r) + S_{m}^{2}(r)}, \\ 
\phi_{m}(r) &=\frac{\mathrm{atan2}\left[S_{m}(r), C_{m}(r)\right]}{m}, 
\end{align} 
where $\mathrm{atan2}$ is the two-argument arctangent function. Note that the factor $1/m$ in the phase definition is missing in \citet{aizawa2024}. The coefficients $C_{m}(r)$ and $S_{m}(r)$ are defined by 
\begin{align} 
C_{m}(r) &=\frac{1}{2\pi} \int_{0}^{2\pi} I(r,\phi)  \cos(m\phi) d\phi, \\ 
S_{m}(r) &= \frac{1}{2\pi} \int_{0}^{2\pi} I(r,\phi) \sin(m\phi) d\phi,
\end{align} 
with $I(r,\phi)$ being the deprojected residual image in polar coordinates. We interpolate the deprojected residual images onto a regular polar grid and compute the integrals to obtain $C_{m}(r)$ and $S_{m}(r)$, from which $I_{m}(r)$ and $\phi_{m}(r)$ are derived.

\subsection{The most dominant mode with m=2} \label{sec:m2spiral}

\begin{figure*}
\begin{center}
\includegraphics[width=0.45 \linewidth]{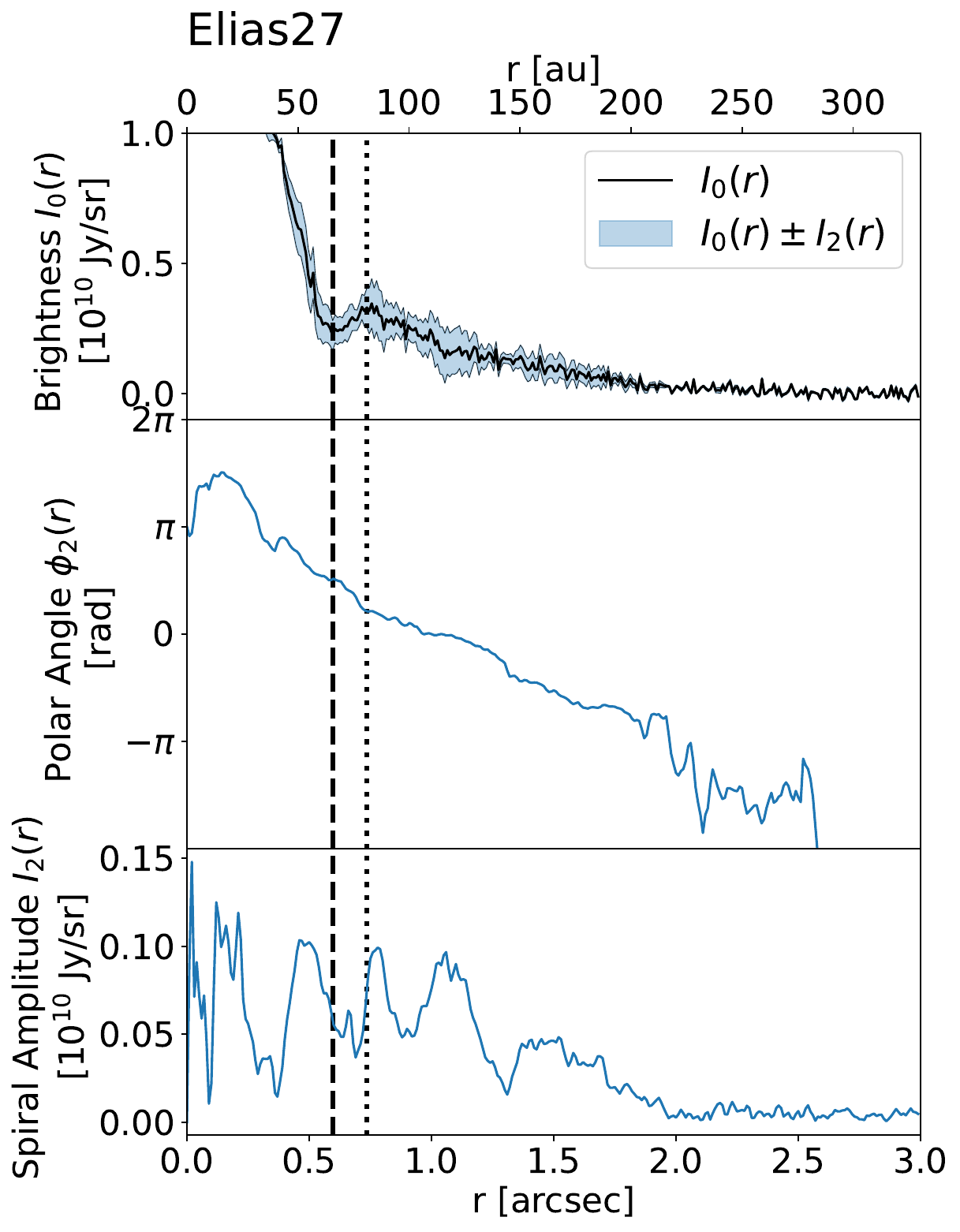}
\includegraphics[width=0.45 \linewidth]{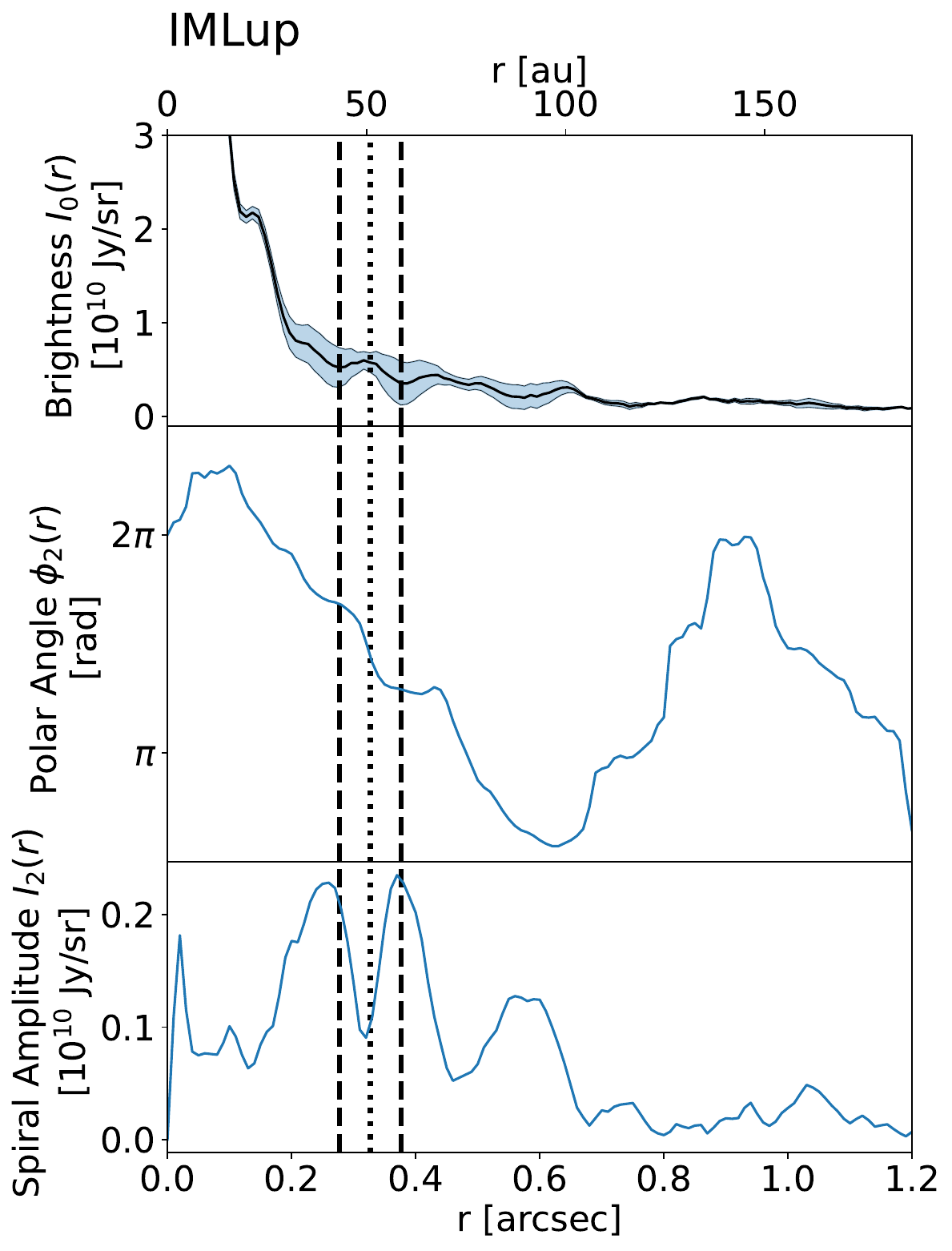}
\includegraphics[width=0.45 \linewidth]{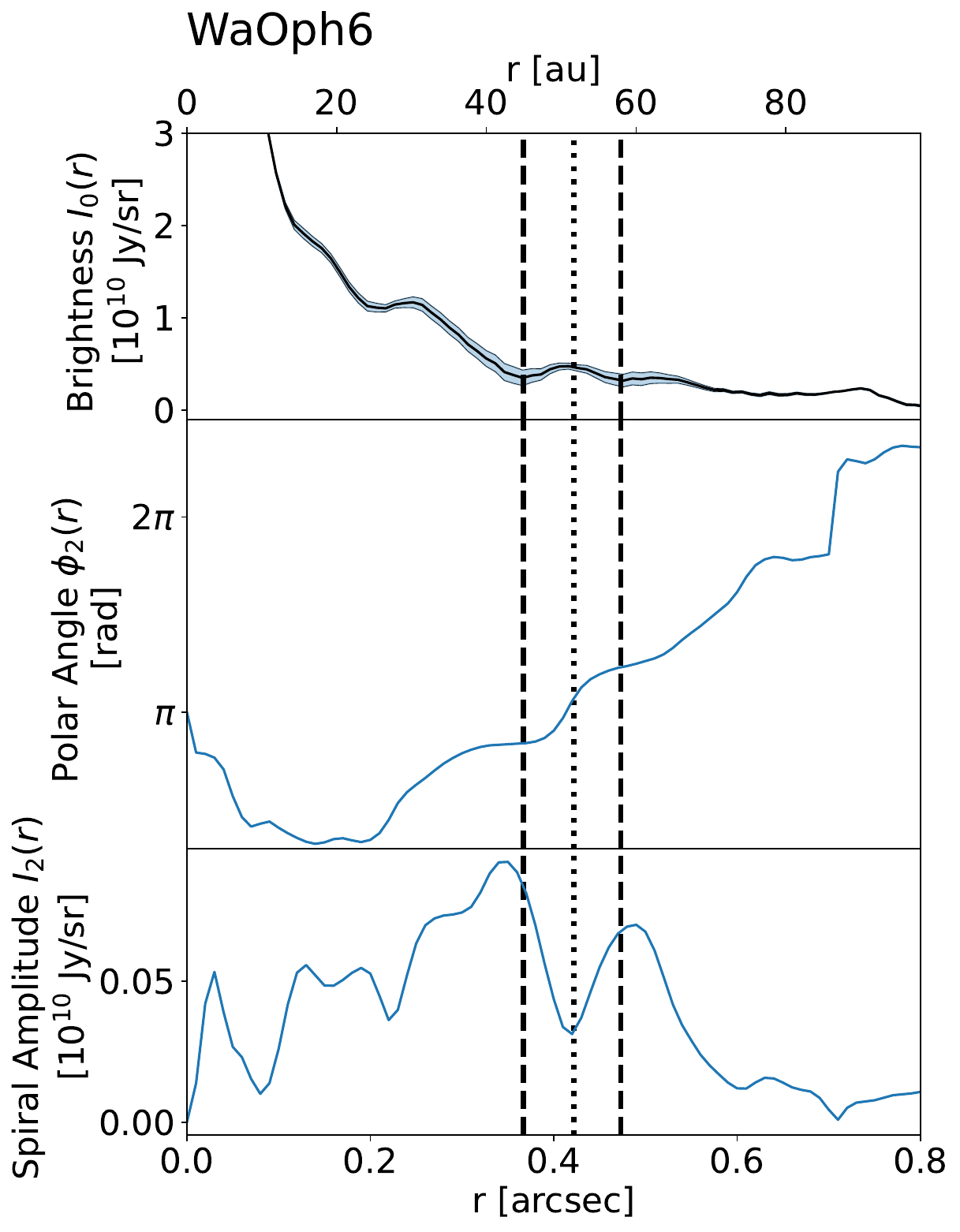}
\end{center}
\caption{Fourier analysis for $m=2$ spirals in Elias 27, IM Lup, and WaOph 6. Top panels show the brightness profiles  $I_0(r)$ calculated by {\tt protomidpy}, and the two panels show spiral phase $\phi_2(r)$ and amplitude $I_2(r)$. Dotted and dashed lines specify the locations of possible rings and gaps.  } 
\label{fourier_ana}
\end{figure*}

We investigate the properties of the most dominant mode, $m=2$. Figure \ref{fourier_ana} presents the brightness radial profiles $I_0(r)$ obtained with {\tt protomidpy},  spiral phases $\phi_2(r)$, and spiral amplitudes $I_2(r)$. Angular distances are converted into astronomical units (au) using distances from Gaia DR3 \citep{gaia2023}: $d = 155.8\pm0.5$ pc for Elias 27, $110.1\pm10.3$ pc for IM Lup, and $122.5\pm0.3$ pc for WaOph 6.


Our Fourier analyses roughly recover the spiral signals out to 2.5, 0.7, and 0.7$\arcsec$, corresponding to approximately 340, 80, and 85 au for Elias 27, IM Lup, and WaOph 6, respectively. All of the three disks exhibit clear radial variations in the spiral amplitude $I_2(r)$ with several local minima and maxima. For IM Lup and WaOph 6, the phases $\phi_2(r)$ exhibits a winding, staircase-like variation. The variations are particularly evident around 40-60~au in both cases. 

Notably, we identify possible correlations among gaps, rings, and spiral phases/amplitudes. In IM Lup, the spiral phases vary steeply at local minimum of the spiral amplitudes or at the ring at  $r=51$ au (dotted line in Figure \ref{fourier_ana}), but change more gradually at local maxima or at the gaps at $r=43$ and $59$ au (dashed line in Figure \ref{fourier_ana}). Similarly, in WaOph 6, the same trend is seen at the ring at $r=52$~au and at the gaps at $r=45$ and $58$ au. In contrast, Elias 27 does not exhibit similarly clear correlations \aizw{either at a prominent gap at $65$ au or at a ring at $81$ au}.

\begin{figure*}
\begin{center}
\includegraphics[width=0.9 \linewidth]{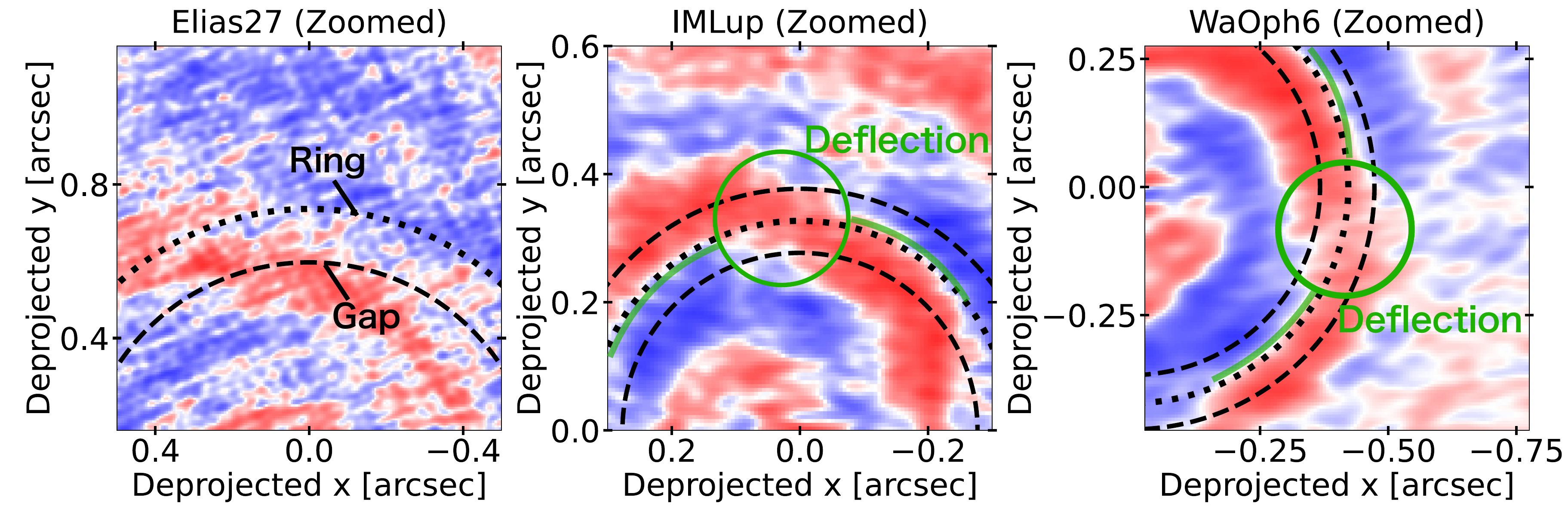}
\end{center}
\caption{ Zoomed views of even-symmetric images for  Elias 27, IM Lup, and WaOph 6. In each panel, the color scales are same as Figure \ref{fig:odd-extraction}. The horizontal and vertical axes are deprojected offsets from the disk center. Black dashed and dotted curves trace the continuum gaps and rings, respectively.  Green circles highlight the approximate position of spiral deflection. Green lines highlight alignment of rings with spiral arm edges. } 
\label{fig:zoomed_in_ring_spiral}
\end{figure*}

Figure~\ref{fig:zoomed_in_ring_spiral} shows zoomed views of even-symmetric images corresponding to the possible correlations for IM Lup and WaOph 6.  For Elias 27, where no such correlations are identified, we instead present a zoomed-in view of the prominent gap at $65$ au. In each panel, the black dotted curve marks the radial location of a prominent ring, while the black dashed curve traces the corresponding gaps immediately interior and/or exterior to that ring, as also highlighted in Figure \ref{fourier_ana}. 

In IM Lup and WaOph 6, we correspondingly identify \aizwrev{deflection features} \aizw{in the spirals} for the correlations near the gap radii, as highlighted by green circles. At the \aizwrev{deflections}, the spiral pitch angles exhibit steep changes. The structures apparent in the even-symmetric images are present on both sides of the original residual images. Notably, at \aizwrev{the deflection} radii, the rings coincide with the edges of the spiral arms, as indicated by green lines. Along the ring radii, the spiral amplitude are small because they follow the edges of spirals with amplitudes of $\sim0$. This naturally produces the correlations between spiral amplitudes and phases. However, the origin of the \aizwrev{deflections} themselves are uncertain.

\cite{huang_spiral_2018} report possible correlations between spiral amplitudes and disk substructures: for WaOph 6, they find that the spiral contrast at ∼51 au reaches a local minimum at a ring (or interarm) feature, where the ring appears to connect the two spiral arms. For IM Lup, they report that local minima in the spiral amplitudes at ∼48, 72, and 98 au correspond to regions of decreasing spiral pitch angle and the presence of rings or interarm features. Our analysis confirms the reported correlations, and we further argue that the correlation might arise from the \aizw{deflections} in the spirals.

\subsection{Possible one-armed spiral in Elias 27} \label{sec:odd_sym_spiral}

\begin{figure*}
\begin{center}
\includegraphics[width=0.45\linewidth]{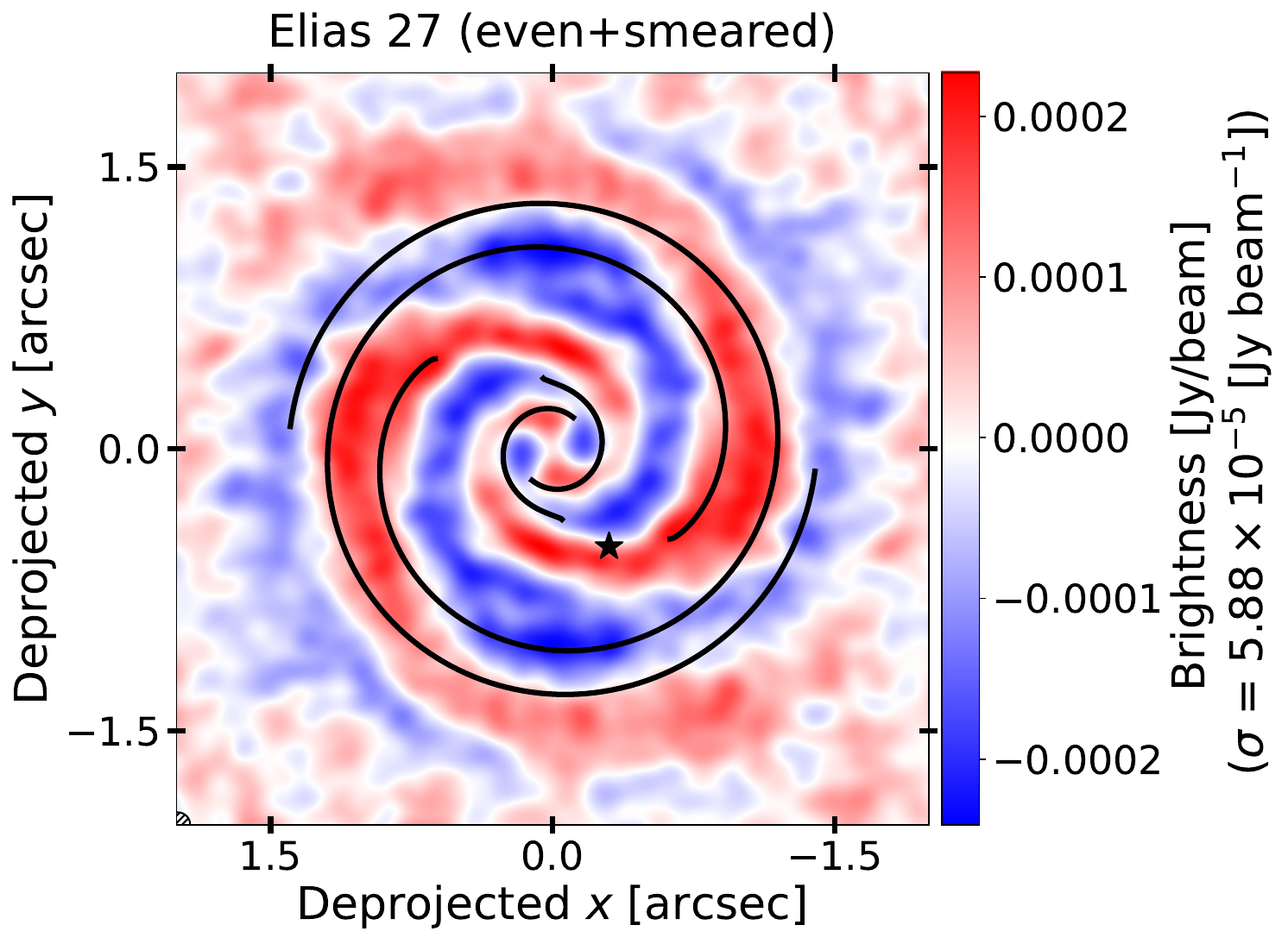}
\includegraphics[width=0.45\linewidth]{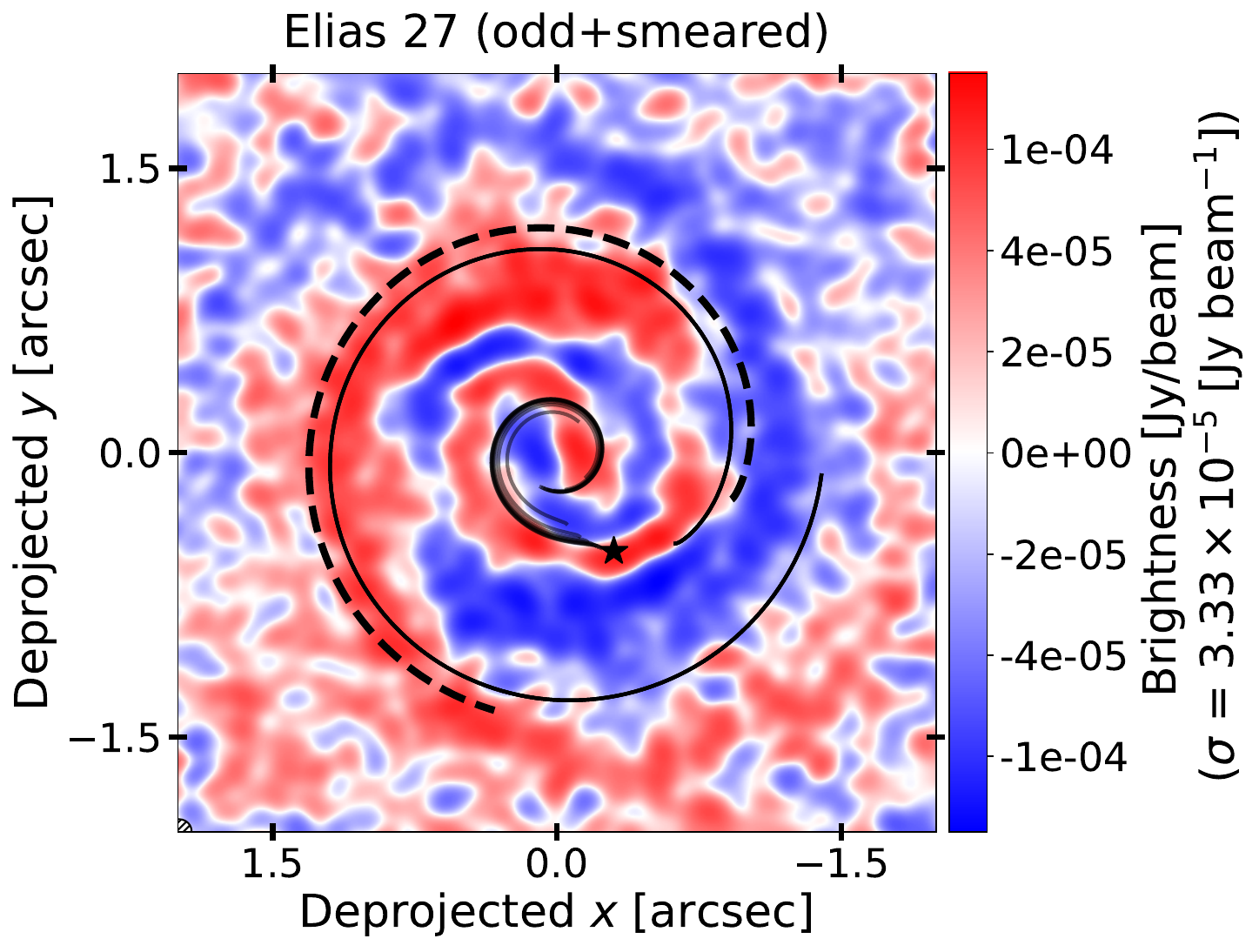}
\includegraphics[width=0.45\linewidth]{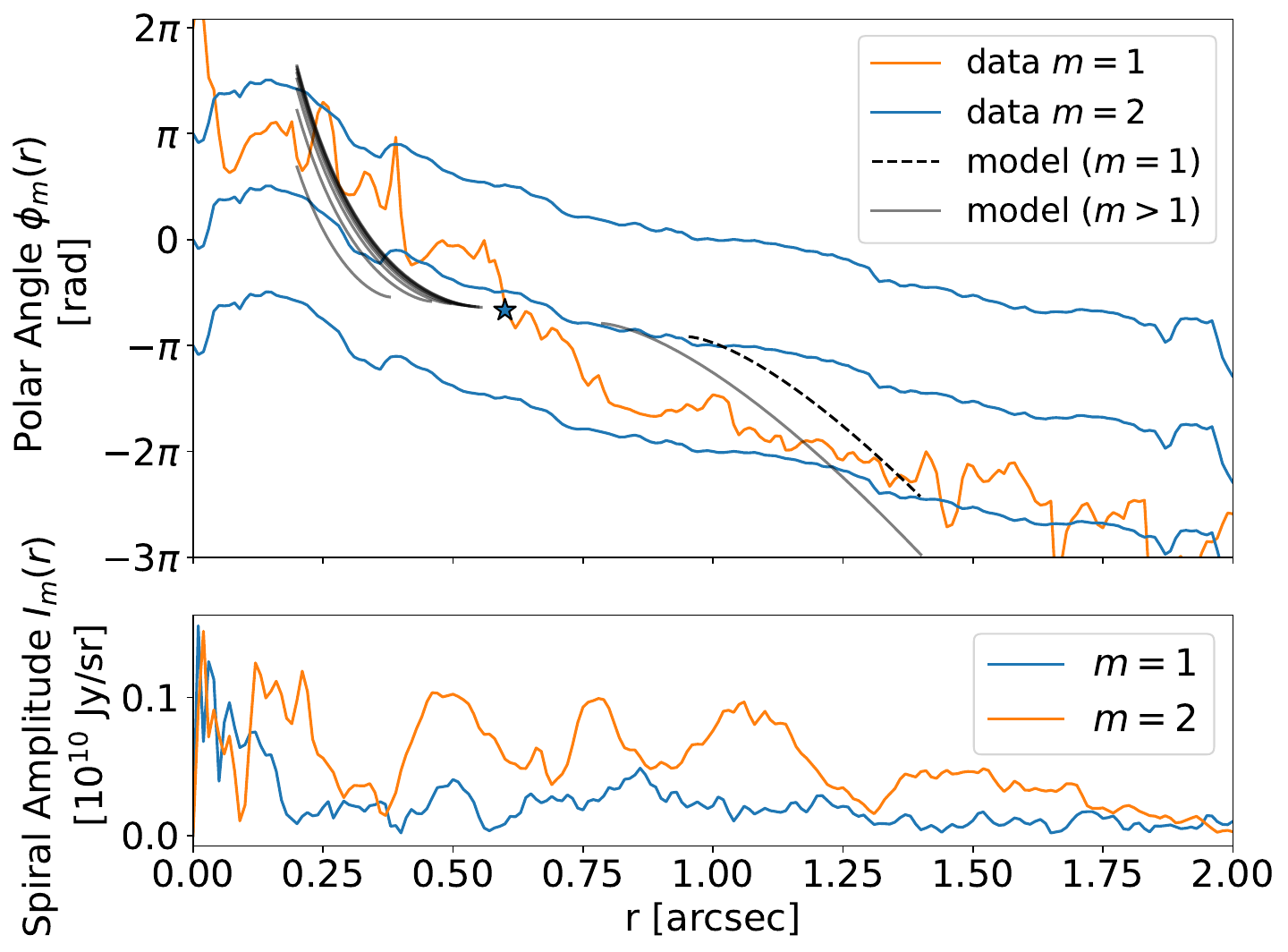}
\end{center}
\caption{(Upper panels) Smeared even-symmetric (upper left) and odd-symmetric (upper right) residual maps of Elias 27 and the overlaid planetary spiral models (black lines). In the upper left panel, only the $m=2$ spiral is shown; in the upper right panel, the $m=1$ spiral is indicated by a dashed line and the $m>2$ spirals by solid lines. In the current model, $m=1$ spiral appears only exterior to the planet's orbit. (Lower panel) Fourier analysis of Elias 27 for $m=1$ and $m=2$, with the corresponding planetary spiral models from the upper panels overlaid.
}
\label{fig:spiral_planet_fit_elias27}
\end{figure*}

In the odd-symmetric image of Elias 27, \aizw{where the $m=\text{even}$ modes are removed}, we identify a possible clockwise one-armed ($m=1$) spiral. To highlight this feature, we show a smoothed version of the original image in Figure \ref{fig:spiral_planet_fit_elias27} by convolving with a Gaussian kernel to achieve a circular beam of 0.13$\arcsec$. The lower panel of the figure shows the amplitude and phase profiles of the $m=1$ mode. The phase of the inner $m=1$ spiral aligns with one of the $m=2$ arms at small radii, whereas the phase of the outer $m=1$ spiral aligns with the other $m=2$ arm at larger radii. This is also seen in the image plane (see Figure \ref{fig:spiral_planet_fit_elias27}). On the other hand, the $m=1$ amplitudes are overall weaker than those of the $m=2$ mode, and there is no clear correlation between the amplitudes of the two modes. 

A planet candidate in the Elias 27 disk is reported based on its kinematic signature \citep{pinte2020}. To compare the spiral predicted for this candidate with our observations, we overplot an analytical spiral model in Figure \ref{fourier_ana}, adopting a simple passively-irradiated disk \aizw{whose  temperature profile is set by the radial distance, disk flaring, and stellar luminosity} \citep[e.g.,][]{ogilvie2002,bae_2018,speedie_obs_2022}. For the model, we visually place the planet in the gap so that it aligns with the spiral morphology; 
its radial separation from the center is $0.6\arcsec$ and its phase angle is $-120^{\circ}$ in the deprojected frame. This location differs from that in \citet{pinte2020}, whose estimated location is offset from the gap. Details of the model and disk assumptions are given in Appendix \ref{sec:spiral_model}.

In Figure \ref{fig:spiral_planet_fit_elias27}, the black lines show the spiral model overlaid on the observed images. Multiple lines correspond to different $m$ modes, and the $m=1$ model exists only in the outer spiral. In the even-symmetric image, the model spiral is more tightly wound than observed. In the odd-symmetric image, the modeled $m=1$ spiral is marginally consistent with the observed one-armed structure for $r < 1.3\arcsec$, but beyond that radius, the model spiral becomes increasingly more tightly wound than observed. \aizw{The comparison of the model and the data may imply that the $m=1$ spiral is driven by the planetary gravity, whereas the $m=2$ spiral could have a different physical origin.}

Similar one-armed spirals have been identified in systems in earlier phases: a Class I protostar, TMC1 A \citep{aso2021,xu2023} and a Class 0 triple system, L1448 IRS3B \citep{tobin2016}, possibly attributed to the gravitational instability. The $m=2$ spiral in Elias 27 can be attributed either to gravitational instability \citep[e.g.,][]{tomida2017,meru2017,forgan2018,veronesi2021,carreno2021} or to planetary perturbations \citep[e.g.,][]{meru2017,forgan2018}. \cite{forgan2018} argue that gravitational instability produces an (even-)symmetric spiral, whereas planet-induced spirals tend to be asymmetric. If a planet were responsible for the two-armed spiral in Elias 27, the phase of any one-armed feature should align with that of the stronger arm; yet observations reveal that the single-arm phase coincides with one arm at small radii and with the opposite arm at large radii, resulting in inconsistency. Thus, the origins of one-armed and two-armed spirals in Elias 27 might be different.   

Among the spiral disks, only Elias 27 exhibits the one-armed spiral, and the notable difference is that Elias 27 shows single deep dust gap, possibly curved by a protoplanet. \aizw{These observations, combined with the comparison to the analytical spiral model, might imply that the $m=2$ spiral could be excited by gravitational instability, whereas $m=1$ spiral could be driven by planetary gravity. } 
\aizw{Therefore, it would be interesting to} explore whether a (infall-driven) gravitationally unstable disk, naturally producing two-armed patterns, can coexist with an embedded planet that drives another spiral. For instance, \cite{pohl2015} show that the presence of a planet tends to reduce the opening angle of its induced spirals for the marginally unstable disks. Further theoretical and observational efforts are needed to uncover the true origin of the spiral structures in Elias 27.

\subsection{Possible bar structures at disk centers} \label{sec:bars}
\begin{figure*}
\begin{center}
\includegraphics[width=0.32 \linewidth]{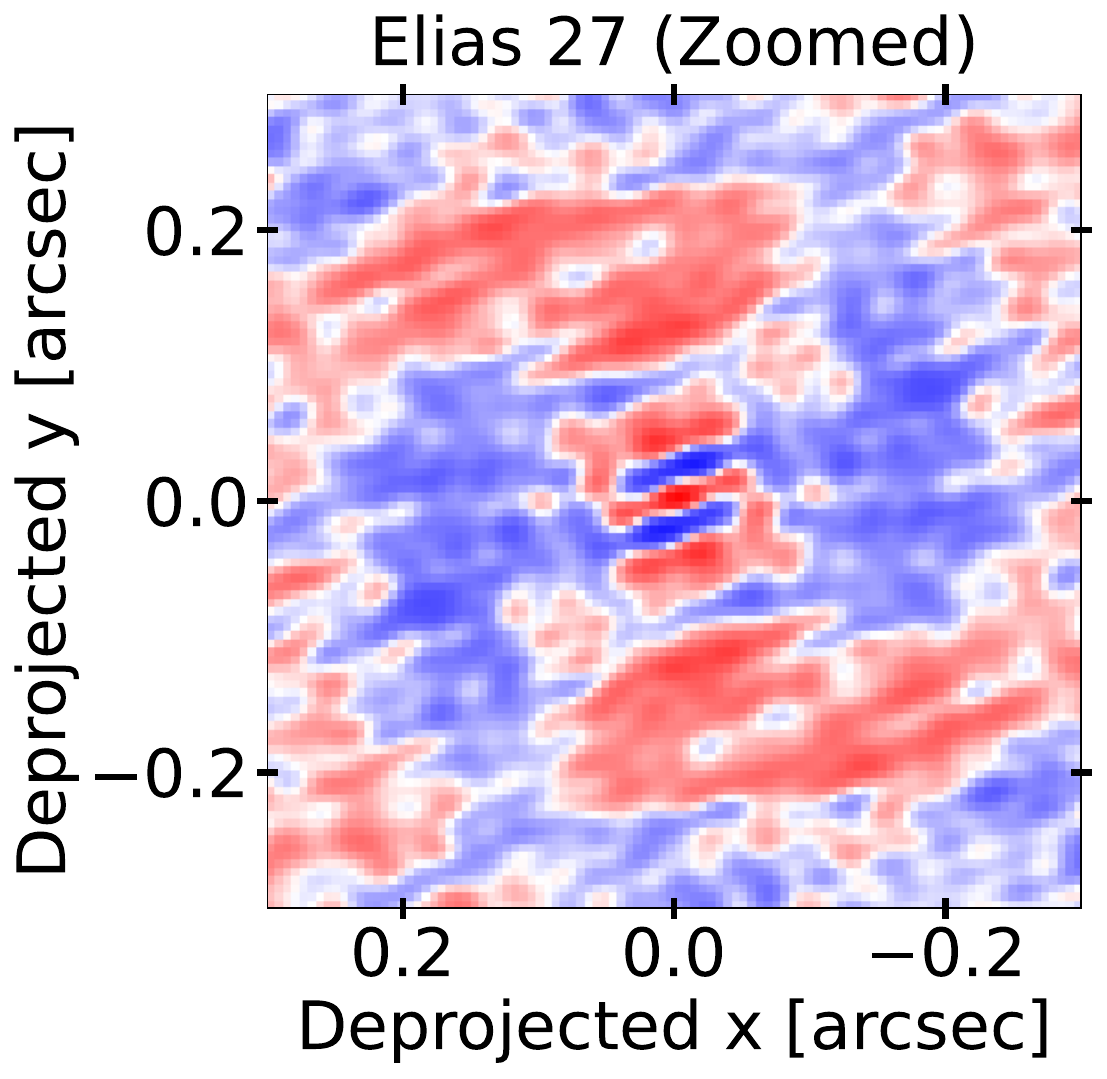}
\includegraphics[width=0.32 \linewidth]{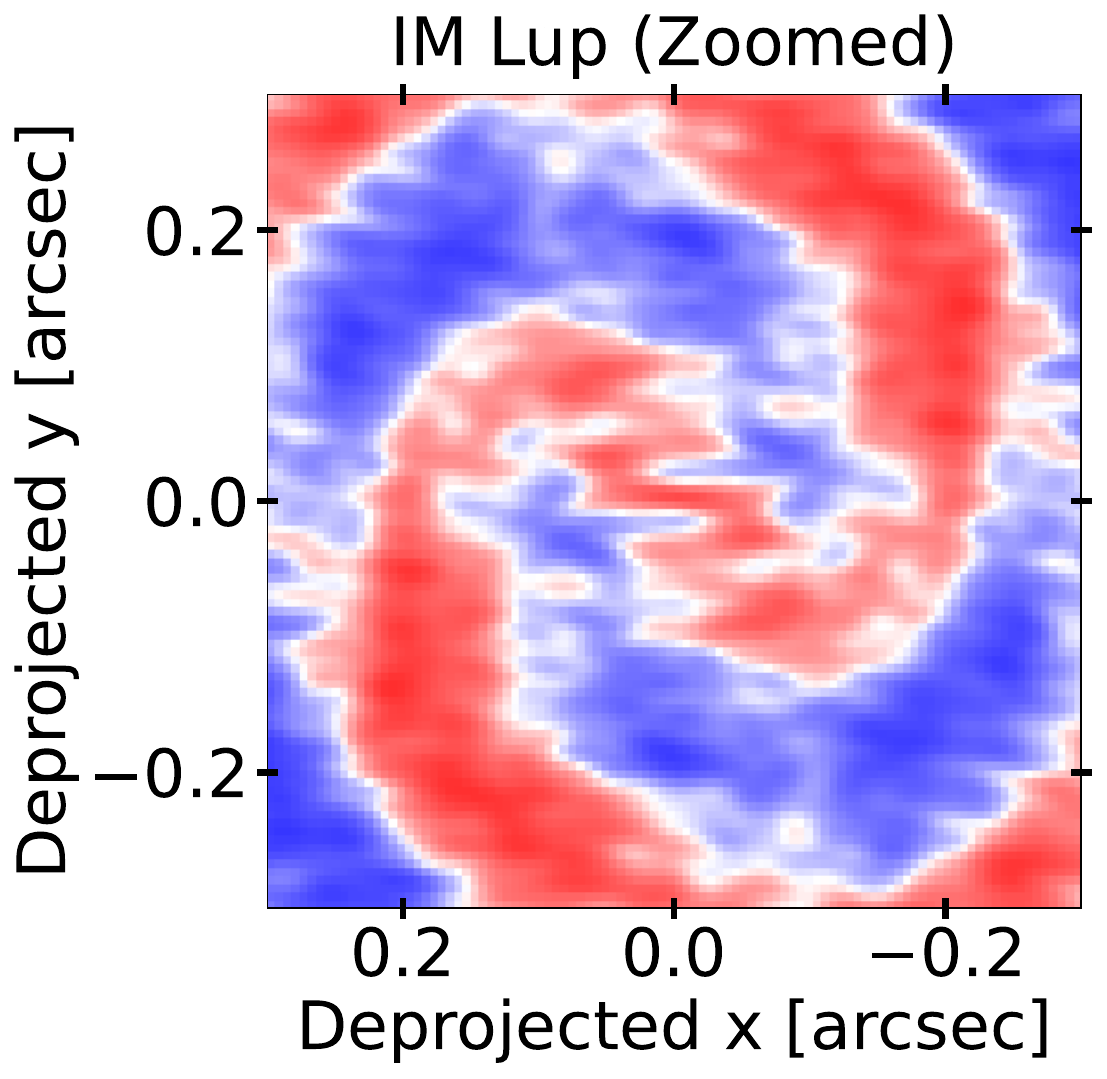}
\includegraphics[width=0.32 \linewidth]{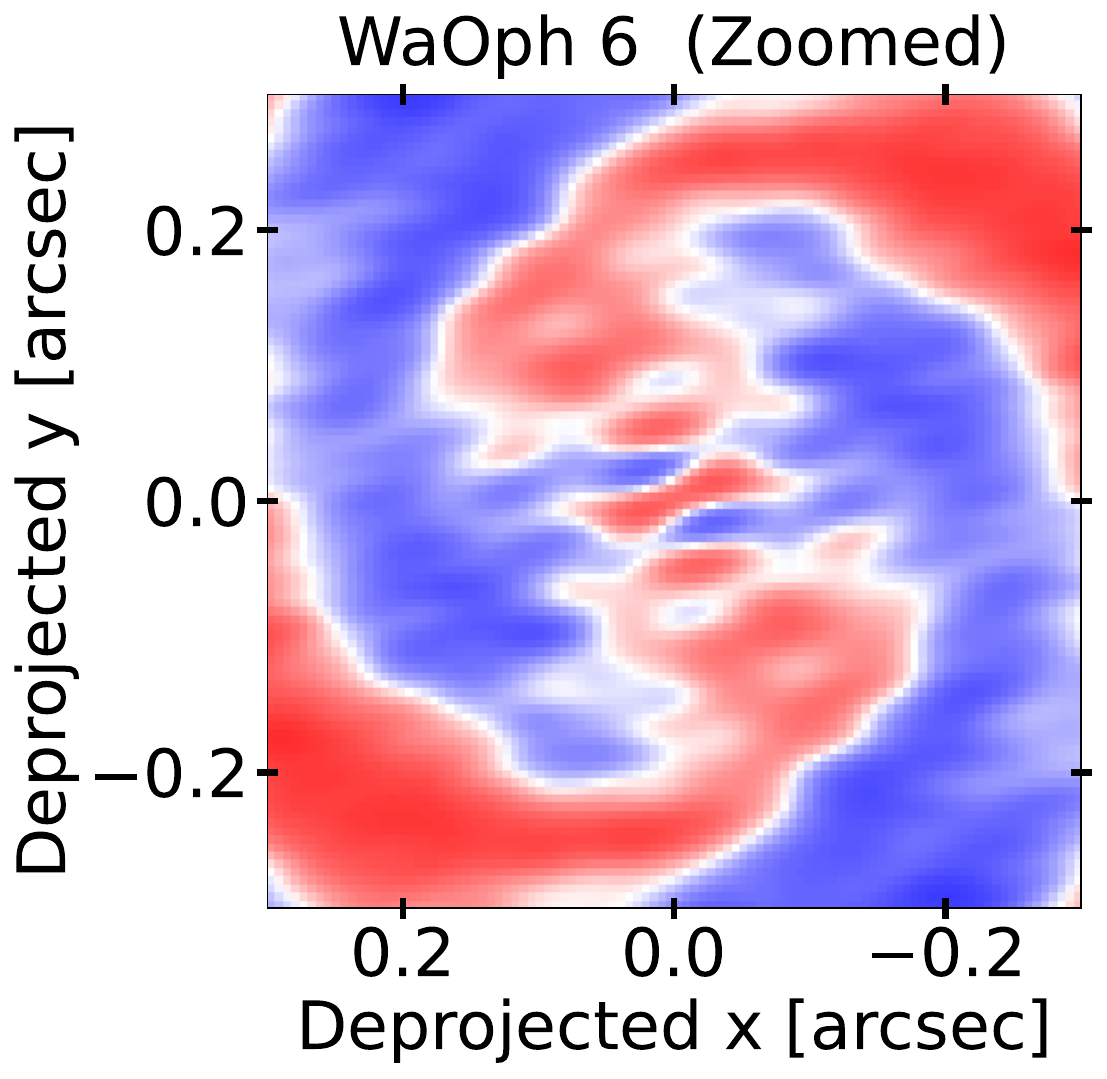}
\end{center}
\caption{ Zoomed-in views of possible bar-like structures in even-symmetric images. } 
\label{fig:bar_zoomed_in}
\end{figure*}

In the spiral structures,  we identify possible discontinuities in the arms in the innermost regions of the disks shown in Figure \ref{fig:odd-extraction}. Figure \ref{fig:bar_zoomed_in} presents zoomed-in views of bar-like features in the even-symmetric images. Consistent with these features, the spiral phases in Figure \ref{fourier_ana} appear to deviate from the smooth trends seen in the outer regions, instead converging to nearly constant values, as might be anticipated in the presence of bars. We manually determine the boundaries of the bar and spiral components using the residual images and phase plots. For Elias 27, the bar extends from 0 to 0.2$\arcsec$ and the spiral from 0.2 to 2.5$\arcsec$. For IM Lup, the bar is found between 0 and 0.15$\arcsec$, with the spiral spanning 0.15 to 0.7$\arcsec$. Finally, for WaOph 6, the bar ranges from 0 to 0.2$\arcsec$ and the spiral from 0.18 to 0.7$\arcsec$. 

Spiral breaks or bar-like features are reported in WaOph 6 \citep{brown2021} and Haro 6-11 \citep{huang2025}. Our results further support the possibility that bar structures may be prevalent in protoplanetary disks. Nevertheless, distinguishing genuine spiral arms from  spiral-bar transitions remains challenging. \cite{aizawa2024} perform injection-recovery simulations for $m=2$ spirals with an amplitude comparable to observed one, using an observational setup similar to that employed for WaOph~6, and find that the recovered phases are unreliable for $r<0.1$$\arcsec$. This suggests that constraining the phases in the inner radii is generally difficult, complicating the distinction between spiral arms and bars. Therefore, we caution that the observed breaks or bar-like features may be artifacts of insufficient signal-to-noise ratios. 

\section{Residuals for the Remaining Systems} \label{sec:ind}

\subsection{Evidence for vertical thickness:  AS~209, DoAr~25, Elias~24,  HD~142666, and MY~Lup} \label{sec:geo_thick}
Some of disks clearly exhibit signals of vertical thickness for dust disks. We systematically pick up such systems exploiting symmetry expected for geometrically thick disks. Specifically, for a disk viewed in its deprojected frame, geometric signals for an axisymmetric disk should be symmetric with respect to the minor axis, i.e.,
\begin{equation}
    I(x, y) = I(x, -y),  \label{eq:geo_sym}
\end{equation}
where $I(x,y)$ denotes the intensity at position $(x,y)$, and with the $x$-axis aligned along the minor axis and the $y$-aix aligned along the major axis. We can decompose $I(x, y)$ into symmetric and antisymmetric components, 
\begin{align}
    I(x, y) &=   I_{\rm sym}(x, y)  + I_{\rm asym}(x, y), \\
    I_{\rm sym}(x, y) &= \frac{I(x, y) + I(x, -y)}{2}, \\
    I_{\rm asym}(x, y) &= \frac{I(x, y) - I(x, -y)}{2}. 
\end{align}
which satisfy $I_{\rm sym}(x, y) = I_{\rm sym}(x, -y)$ and $I_{\rm asym}(x, y) = -I_{\rm asym}(x, -y)$. 

To select the disks with possible vertical thickness, we quantify the areas where the absolute values of the residuals exceed $2\sigma$, with $\sigma$ defined as the standard deviation of the original residual image. \aizw{In the folded images, the noise level is reduced by a factor of $\sqrt{2}$ compared to the original residual images, so the $2\sigma$ threshold is roughly equivalent to $2\sqrt{2}\simeq 3\sigma$ threshold in the original images.} We select disks that satisfy the following criteria \aizw{for all $n=2$, $2.5$, and $3$}: (a) the area of residuals for $I_{\rm sym}(x, y)$ beyond \aizw{$n\sigma$} levels in the symmetric image is at least twice that in the anti-symmetric image $I_{\rm asym}(x, y)$, and (b) the area beyond $n\sigma$ for for $I_{\rm sym}(x, y)$ is at least three times larger than the beam area.

\aizw{Five} disks around AS~209, DoAr~25, Elias~24, HD~142666, and MY~Lup in Figure \ref{fig:gallery_images2} meet these criteria. Previous studies report possible vertical structures in  HD~142666 \citep{huang_ring_2018} and in Elias 24 \citep{jennings2022}, while \cite{villenave2025} infer vertical dust scale heights for DoAr 25 and HD 142666. \aizw{\cite{huang_ring_2018} also report possible vertical structures in Sz 129. Although this system meets the criteria under a $2.5\sigma$ threshold is adopted, we exclude it from the present analysis for consistency. } 

\begin{figure*} \begin{center}
\includegraphics[width=0.48\linewidth]{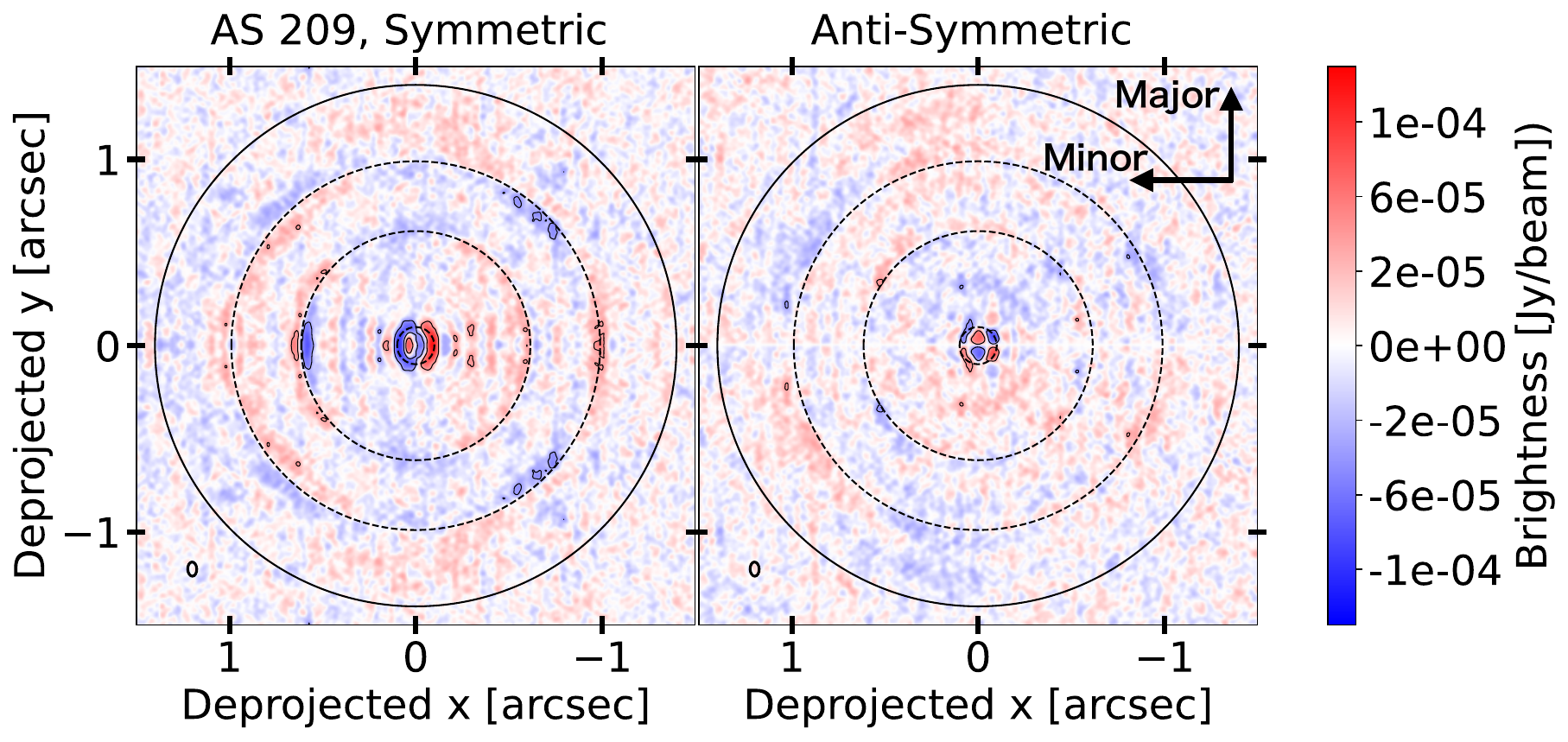}
\includegraphics[width=0.48\linewidth]{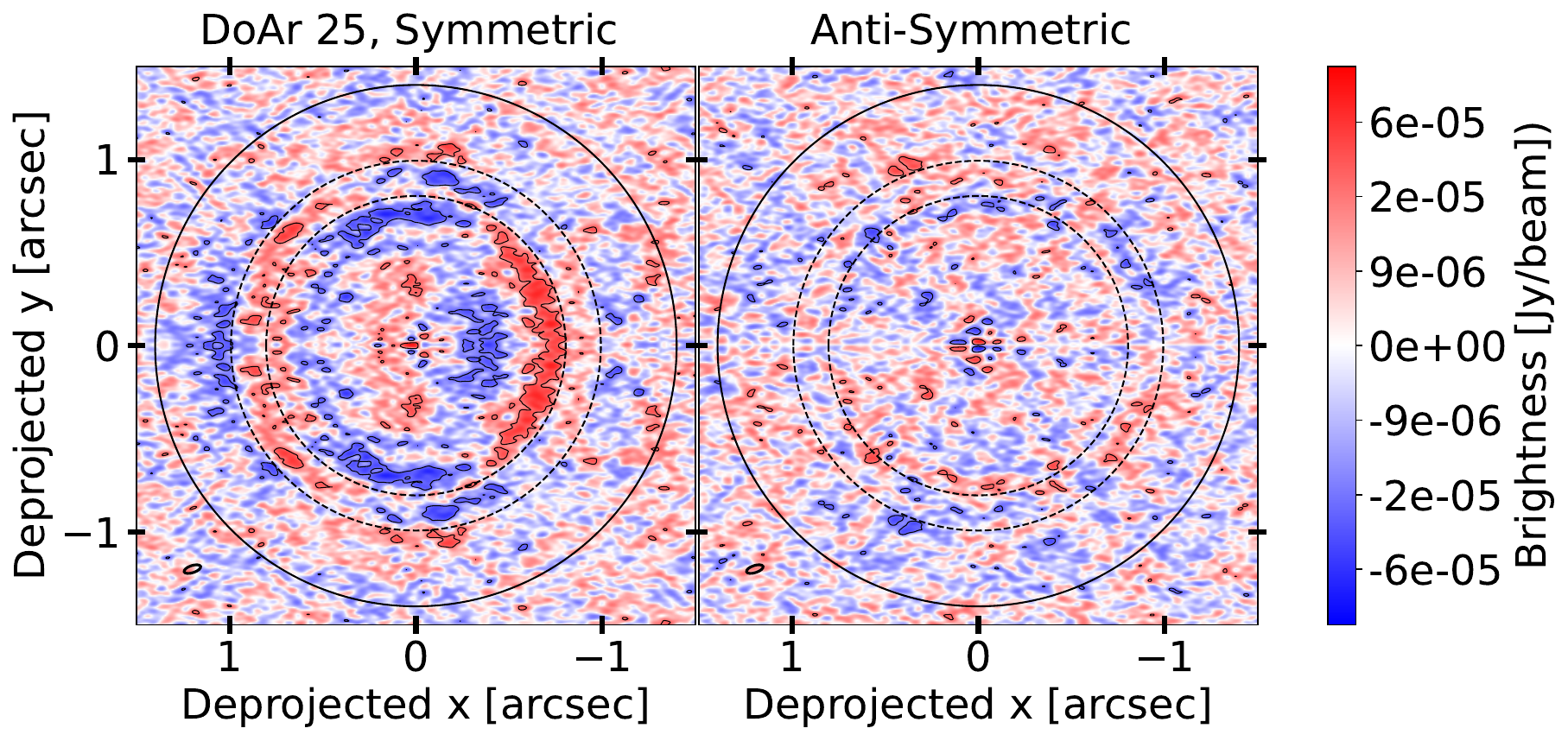}
\includegraphics[width=0.48\linewidth]{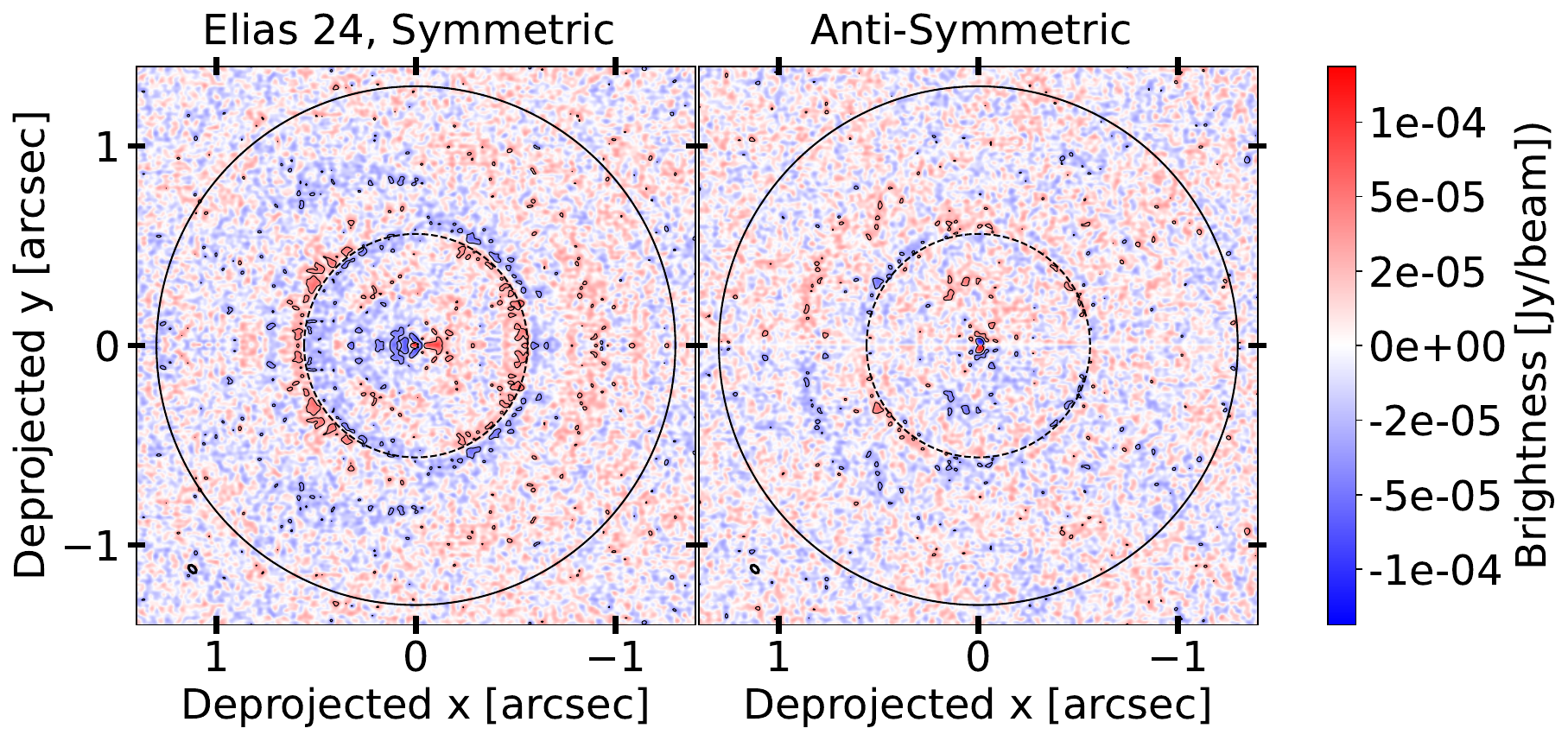}
\includegraphics[width=0.48\linewidth]{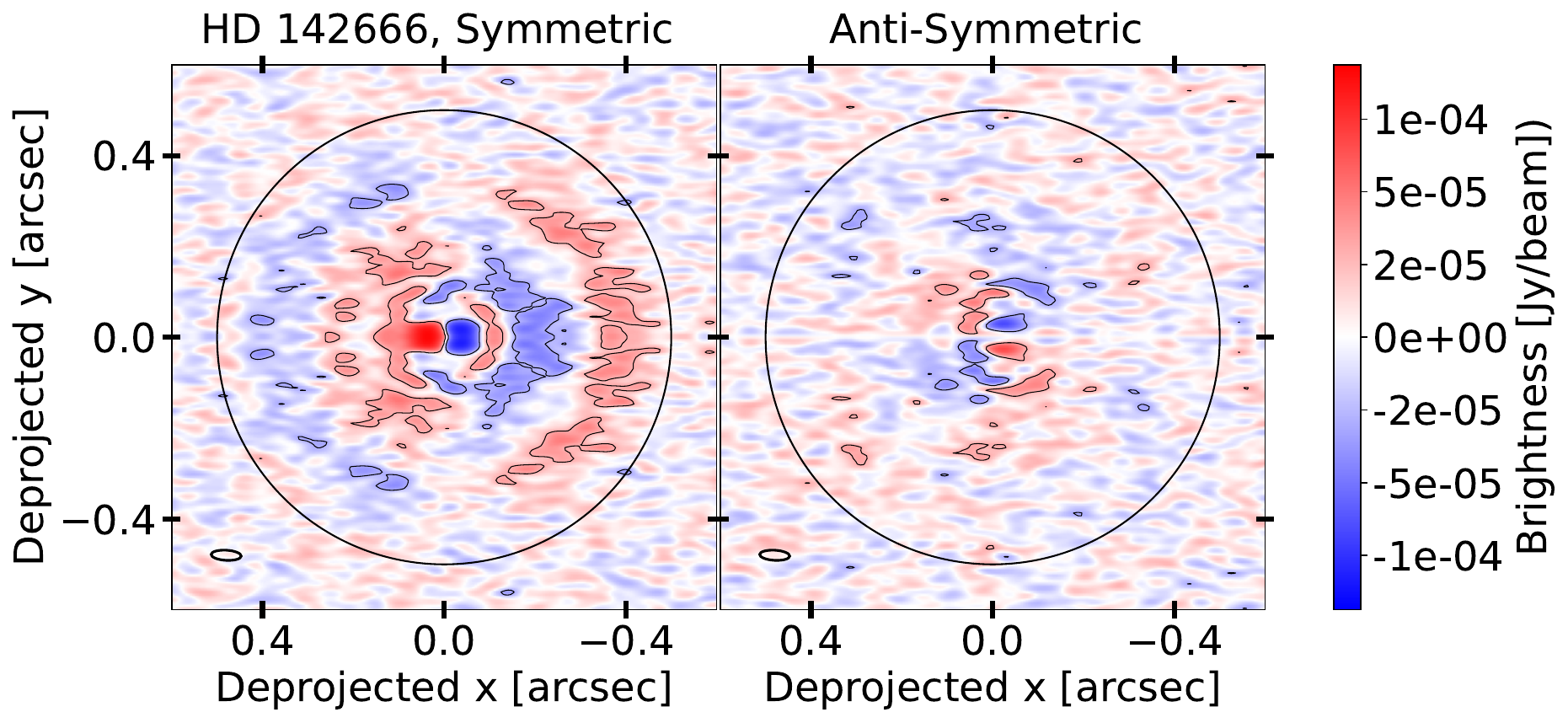}
\includegraphics[width=0.48\linewidth]{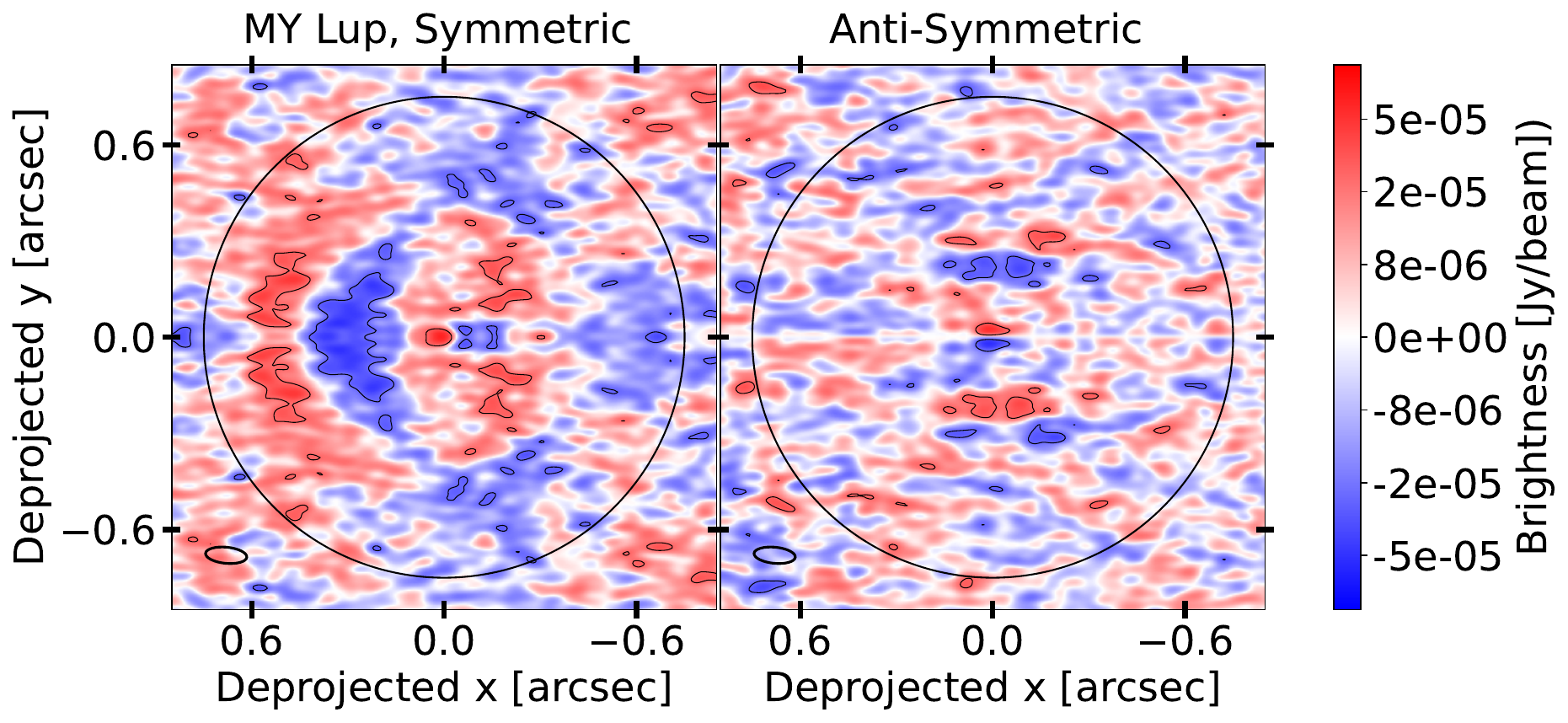}
 \end{center}
 \caption{Residual images for \aizw{five} disks (AS 209, DoAr~25, Elias~24, HD~142666, and MY~Lup), decomposed into symmetric and antisymmetric components. Residuals beyond $2\sigma$ are highlighted as contours. \aizw{Here, $\sigma$ is taken from the original residual images shown in Figure \ref{fig:gallery_images1}.} (Left) Minor-axis symmetric components, $I_{\rm sym}$, likely reflecting the geometrical thickness of the disks. (Right) Minor-axis anti-symmetric components, $I_{\rm asym}$, which are not expected to reflect the geometrical thickness.}\label{fig:geo_res}
 \end{figure*}

Figure~\ref{fig:geo_res} shows the deprojected views of $I_{\rm sym}(x,y)$ and $I_{\rm asym}(x,y)$ for the \aizw{five} disks. The residuals predominantly appear around gaps and rings in AS~209, DoAr~25, and Elias~24, while in HD 142666 and MY Lup the signals extend over broader regions. \aizw{For AS~209, DoAr~25, and Elias~24, the residuals are also  stronger in the inner disks than in the outer disks. Nevertheless, this does not necessarily indicate a greater vertical extent in the inner disks compared to the outer disks, because the vertical signals, obtained by subtracting the models, should increase with unsubtracted intensities, which are typically higher toward the inner regions. }

Among the systems, AS 209 exhibits particularly complex residual patterns (see also \citet{guzman2018}, \citet{jennings2020}, \citet{aizawa2024}). Nevertheless, the areas of $I_{\rm sym}$ residuals beyond the $2\sigma$ level is 3 times larger than that in $I_{\rm asym}$, suggesting that the vertical thickness can be the primary origin of the residuals. We note that additional features are also present in the asymmetric image, which may arise from radially varying geometric parameters, similar to the residual pattern reported in HL Tau \citep{2015ApJ...808L...3A}, which likewise shows multiple ring and gaps.  

Using three-dimensional dust-gas hydrodynamical simulations, \citet{bi2021} demonstrate that sufficiently massive, gap-opening planets drive meridional gas flows that can \aizwrev{lift} sub-millimeter grains to heights of up to $\sim$70 \% of the local gas scale height at the gap edges. This suggests that the rings might be indeed lifted up in AS 209, DoAr 25, and Elias 24, where the signals are identified near the ring and gaps.

Vertical dust scale heights for the disks can be constrained through comparison with radiative transfer modeling \citep{pinte2016,ohashi2019,doi2021,villenave2022,pizzati2023,Guerra-Alvarado2024,Ribas2024, villenave2025}. Within our sample, \cite{villenave2025} derive finite vertical dust scale heights for DoAr 25 and HD 142666. Our analysis suggests that, for the remaining \aizwrev{three} disks, AS 209, Elias 24, and MY Lup, the vertical dust scale heights \aizw{above the midplanes are significant enough to be measurable, making these systems ideal targets for further investigation. }

\subsection{Systems with $5\sigma$ residuals: RU~Lup, SR~4 and WSB 52}  \label{sec:five_sigma}

Figure \ref{fig:gallery_images3} presents the residual images for disks with $5\sigma$ residuals found outside the innermost regions. The residuals for RU Lup and SR~4 are found at the gap-ring boundaries.  Especially, residuals for RU Lup are identified as a pair of two positive excess. Gap-ring boundaries have large pressure gradients, which can trigger vortex formation via the Rossby-Wave instability (RWI) \citep[e.g.,][]{Lovelace1999}. \cite{liu2023} demonstrate that a pair of vortices can be formed by performing nonlinear two-fluid simulations for the so-called Type II dusty RWI, where the pair emerges a traveling mode that occurs in the presence of dust in relatively mild pressure bumps. 


In WSB~52, we identify a positive excess slightly above the $5\sigma$ threshold. The area for the (geometric) residuals beyond \aizw{$2.5\sigma$} in $I_{\rm sym}$ in Section \ref{sec:geo_thick} is four times larger than those in  $I_{\rm asym}$, but the area itself is only twice the beam size; therefore we do not classify this system into the vertical-thickness category in Section \ref{sec:geo_thick}. In this system, \citet{aizawa2025} report a jet-bubble-disk interaction, in which a jet-induced bubble expansion deforms the gas disk. The slight residual in the symmetric image may indicate vertical thickness, possibly due to the disk deformation, but the significance is not conclusive. 

\subsection{Insignificant asymmetries: DoAr 33, Elias 20, GW Lup, Sz 114, and \aizwrev{Sz 129} } \label{sec:weak_signal}
Figure \ref{fig:no_five} presents the residual images for systems with no significant asymmetries beyond $5 \sigma$ outside the innermost regions. Exceptionally in the residual image of Elias~20, we observe a gradual change in intensity from the lower left to the upper right, but it is below the $5\sigma$ level, so we categorize Elias~20 as disks into this category. For the other systems, DoAr~33, GW~Lup, Sz~114, we do not find notable features. Especially, GW~Lup does not exhibit $5 \sigma$ excess on the disk plane, implying the highest degree of axisymmetry among the DSHARP disks.

\section{Correlation analysis} \label{sec:corr}

\aizw{We present a correlation analysis} to investigate which parameters of the systems determine the condition for substructures. \aizw{In such an analysis, it is important to consider how the sample is drawn or determined from the full population and to what extent it is biased. In this respect, the DSHARP sample is biased toward bright disks, as it was selected by imposing a peak brightness threshold of 20 mJy per $0.3\arcsec$ beam (4.8 K) \citep{andrews2018}. The host stars span a wide range of properties, with $0.2-2 M_{\odot}$ and luminosities covering two orders of magnitude. Thus, while the sample encompasses diverse stellar and disk properties, the interpretation should be made  with the caution that the discussion is limited to disks that exceed the brightness threshold.}
\subsection{Formulation}

To study the correlations, we simply adopt the Pearson correlation coefficient for class $i$ and parameter $j$ for the sample size of $n$, 
\begin{equation}
r_{i,j}
= \frac{\sum_{k=1}^n \left(D_{i,k} - \bar{D}_i\right)\left(P_{j,k} - \bar{P}_j\right)}
       {\sqrt{\sum_{k=1}^n \left(D_{i,k} - \bar{D}_i\right)^2}
        \,\sqrt{\sum_{k=1}^n \left(P_{j,k} - \bar{P}_j\right)^2}}, \label{eq:r_ij}
\end{equation}
where $D_{i,k} \in \{0,1\}$ is the dummy variable for class $i$; $D_{i,k}=1$ if the $k$-th target is in the $i$-th class, and $D_{i,k}=0$ otherwise. $P_{j,k}$ is the $j$th physical parameter for the $k$-th target. $\bar{D}_i$ and $\bar{P}_j$ denote the averages of $D_{i,k}$ and $P_{j,k}$ over all targets $k$, respectively. 

For $D_{i,k}$, we consider three classes following the previous sections: ``spiral" in Section \ref{sec:revisit_spiral}, ``thick disk" in Section \ref{sec:geo_thick}, and ``weak asymmetry" in \ref{sec:weak_signal}. In addition, we consider another class, ``arc", for HD 163296 and HD 143006, which are not analyzed by our method but important sample for discussing asymmetric substructures in a systematic manner. By defining $D_{i,k}$ to be binary variable, we can naturally represent both the presence and the absence of objects in a class. We remove binary systems, HT Tau and AS 205 because binaries are important but complicating factors to interpret. The sample size is thus $n=20-2=18$. Among the targets, RU Lup, SR 4, and WSB 52 in Section \ref{sec:five_sigma} do not belong to any of classes defined above, thus satisfying $D_{i,k}=0$. 

For $P_{i,k}$, we consider nine parameters: stellar age $\log t_{\star}$, stellar mass $\log M_{\star}$, stellar luminosity $\log L_{\star}$, stellar surface temperature $\log T_{\star}$, accretion rate $\log \dot{M}$, total flux, dust disk size, disk inclination $i$, and \aizw{inner-disk} brightness temperature $T_{\rm B, inner}$, defined as the brightness temperature at the innermost radius $r_1$ of the recovered brightness profiles in our model. For HD 143006 and HD 163296, we adopt the brightness temperature computed from brightness profiles in the literature \citep{jennings2022} because we do not model them in the current paper. Except for $T_{\rm B, inner}$, we adopt values from the literature: dust disk sizes and inclinations from \cite{huang_ring_2018} and all other parameters from \cite{andrews2018}. For Elias 20, only an upper limit of $\log t_{\star}<5.9$ Myr  is available, and we adopt $\log t_{\star}=5.4$ Myr for the correlation analysis. For MY Lup and HD 142666, which \aizw{have only upper limits} $\log \dot{M}/[M_{\odot}/{\rm yr}]<-9.6$ and $\log \dot{M}/[M_{\odot}/{\rm yr}]<-8.4$, we instead substitute $\log \dot{M}/[M_{\odot}/{\rm yr}]=-10.1$ and $\log \dot{M}/[M_{\odot}/{\rm yr}]=-8.9$, respectively. \aizw{Nevertheless, excluding these systems does not largely} affect the results. 

We evaluate statistical significance for the correlation coefficient $r$ for the $n$ sample is assessed via the usual $t$-statistic
\begin{equation}
t = r \,\sqrt{\frac{n-2}{1 - r^2}} \label{eq:t_stat}
\end{equation}
with $n-2$ degrees of freedom \citep[e.g.,][]{hogg2013introduction}.  

\begin{figure*} 
\begin{center}
\includegraphics[width=0.7\linewidth]{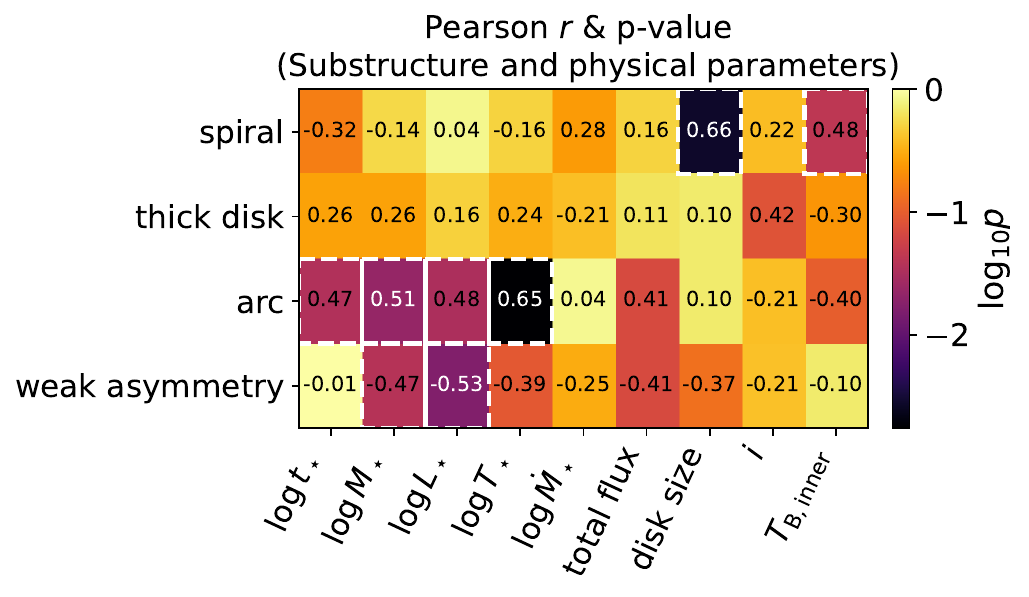}
\caption{Pearson correlation coefficients ($r_{i,j}$, shown as numbers in each cell) between the presence of different substructure types (spiral arm, thick disk, arc, and weak asymmetry) and various stellar and disk parameters. The color scale represents the logarithm of the p-value ($\log_{10} p$), with darker colors indicating higher statistical significance. White dashed boxes highlight correlations with $p<0.05$. }
\label{fig:corr1}
\end{center}
\end{figure*}

Figure \ref{fig:corr1} shows $r_{ij}$ and the $p$-values for system parameters and substructure classes. We highlight possible correlations $(r_{i,j}\neq 0)$ with $p<0.05$ by dashed white squares in the figure. The implications for the disk substructures from the correlation analysis are discussed in the next subsection. 

\begin{figure*} 
\begin{center}
\includegraphics[width=0.63\linewidth]{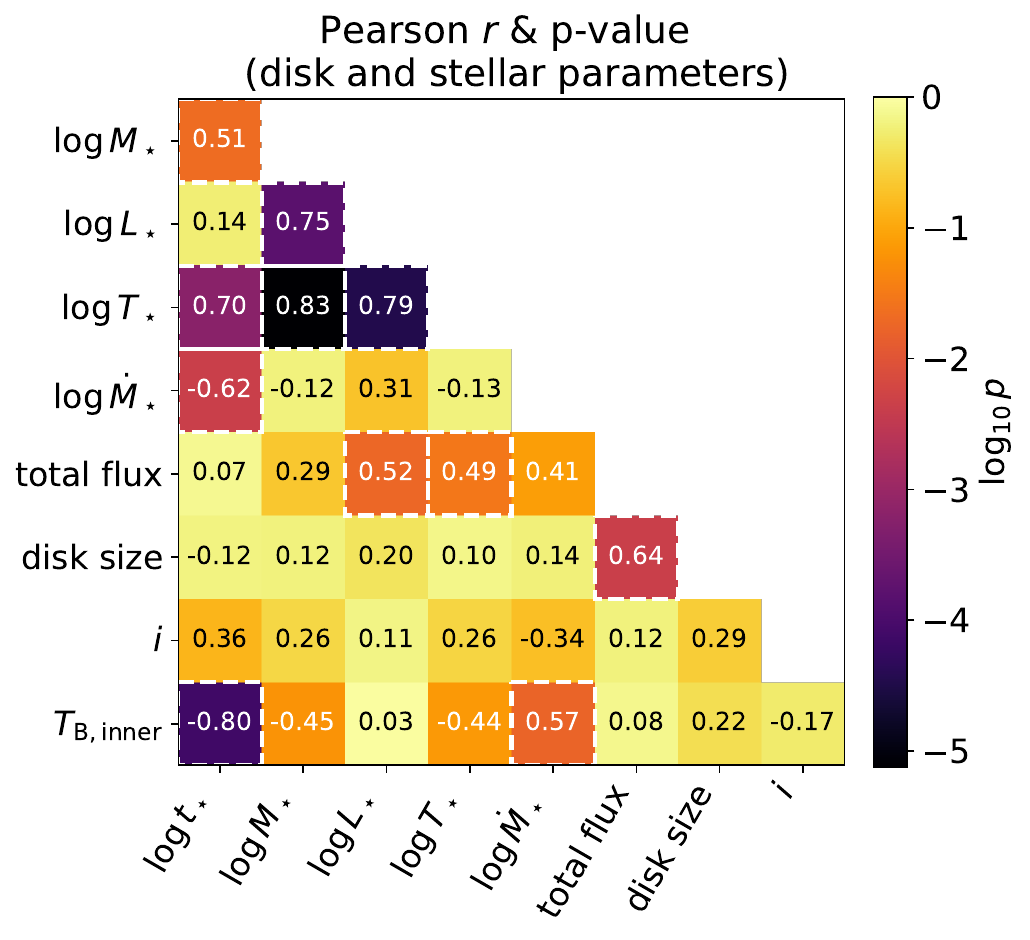}
\caption{ Pearson correlation coefficients ($r_{i,j}$, number in each cell) among stellar and disk physical parameters. The color scale represents the logarithm of the $p$-value ($\log_{10} p$), with darker colors indicating higher statistical significance. White dashed boxes highlight correlations with $p<0.05$. }
\label{fig:corr2}
\end{center}
\end{figure*}

Beyond correlations between system parameters and substructure classes, we also examine correlations among $P_{j,k}$. In this case, we can simply substitute $D_{j,k}$ by $P_{j,k}$ in Equations (\ref{eq:r_ij}) and (\ref{eq:t_stat}) to compute $r_{i,j}$ and $p$. Figure \ref{fig:corr2} shows correlations among the stellar and disk parameters. There are strong correlations among stellar masses, stellar luminosities, and stellar temperatures, consistent with stellar evolution. The total flux is also correlated with the the dust disk size and weakly correlated with the stellar luminosities, as naturally expected. Unexpectedly, we identify a new correlation between $T_{\rm B, inner}$ and age, which is further examined in Section \ref{sec:age_bright}. 

\subsection{Discussion on substructures}
Here, we discuss the substructures in the DSHARP sample based on the correlation analysis. 
\subsubsection{Spiral}
Elias 27, IM Lup, and WaOph 6 show spirals. Figure \ref{fig:corr1} shows a \aizw{statistically significant} positive correlation with dust-disk radius with spiral structures, consistent with the argument in \cite{huang_spiral_2018} that the continuum emission in Elias 27 and IM Lup are extended to hundreds of au, larger than the average DSHARP disks. WaOph 6 is smaller than the two disks, but still larger than the average size. The correlation may reflect a simple detectability bias or the fact that extended disks offer a greater fraction of their area where the Toomre $Q$ parameter falls below unity. On the other hand, \citet{huang2025} report grand-design spirals in the compact disk around Haro 6-13 with a spiral radius of $\sim35$ au, providing evidence that spiral structures are not exclusively associated with large disks.

The possible common properties of the three systems with spirals are young ages, $<1$ Myr, as noted in \citet{huang_spiral_2018}, but the spiral structures is weakly correlated with age. There is also a weak positive correlation with the mass-accretion rate. The peak brightness temperatures is, on the other hand, moderately correlated with the spirals. Since $T_{\rm B, inner}$ is strongly correlated with both ages and the mass-accretion rates (Figure \ref{fig:corr2}),  $T_{\rm B, inner}$ may serve as a useful indicator for spiral formation.

\subsubsection{Geometrically-thick disk}
AS 209, DoAr 25, Elias 24, HD 142666, and MY Lup shows the possible vertical signals. In addition, we also include HD 163296, whose vertical structure in an outer ring is reported \citep{doi2021}, for the analysis. Only the inclination, which likely affects the detectability of the vertical thickness, exhibits \aizwrev{a weak correlation that does not reach statistical significance ($p>0.05$)}.

\subsubsection{Arc}
HD 143006 and HD 163296 exhibit bright arc structures \citep[e.g.][]{andrews2018}. Our correlation analysis reveals positive trends with stellar mass, age, luminosity and temperature, indicating that disks around more massive, luminous, hotter, and older stars are more likely to host bright arcs. Such arcs in millimeter continuum emission could trace the dust-trapping vortex \citep{perez2018} and/or gravitational interaction with yet-unseen planets \citep{isella2018}. \aizw{On the other hand, we do not identify a clear correlation with disk fluxes or sizes, indicating that the arc-hosting disks are not exceptionally large or bright disks in which arcs would be more easily detectable.} The correlation \aizw{might} imply that massive and evolved stars could provide preferential environments for arc formation. 

\subsection{Weak asymmetries}
Disks without $5\sigma$ residual beyond the innermost regions exhibit significant negative correlations with stellar mass and luminosity. Featureless disks thus tend to orbit lower-mass, lower-luminosity stars. Notably, we do not find strong correlations with disk total fluxes \aizw{or disk size}, suggesting that non-detection is not purely due to \aizw{the flux-related or size-related observational biases, but may reflect underlying stellar properties. }

\section{Correlations between peak brightness temperatures and ages} \label{sec:age_bright}

\begin{figure*} 

\begin{center}
\includegraphics[width=0.48\linewidth]{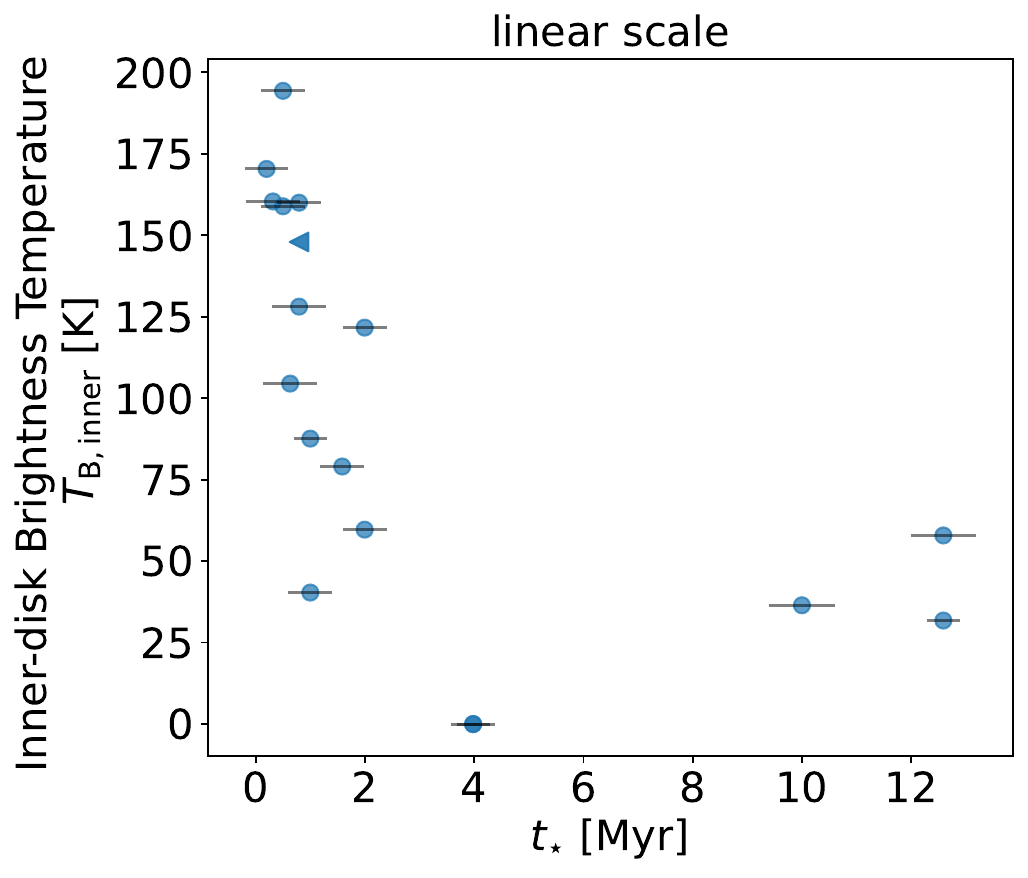}
\includegraphics[width=0.48\linewidth]{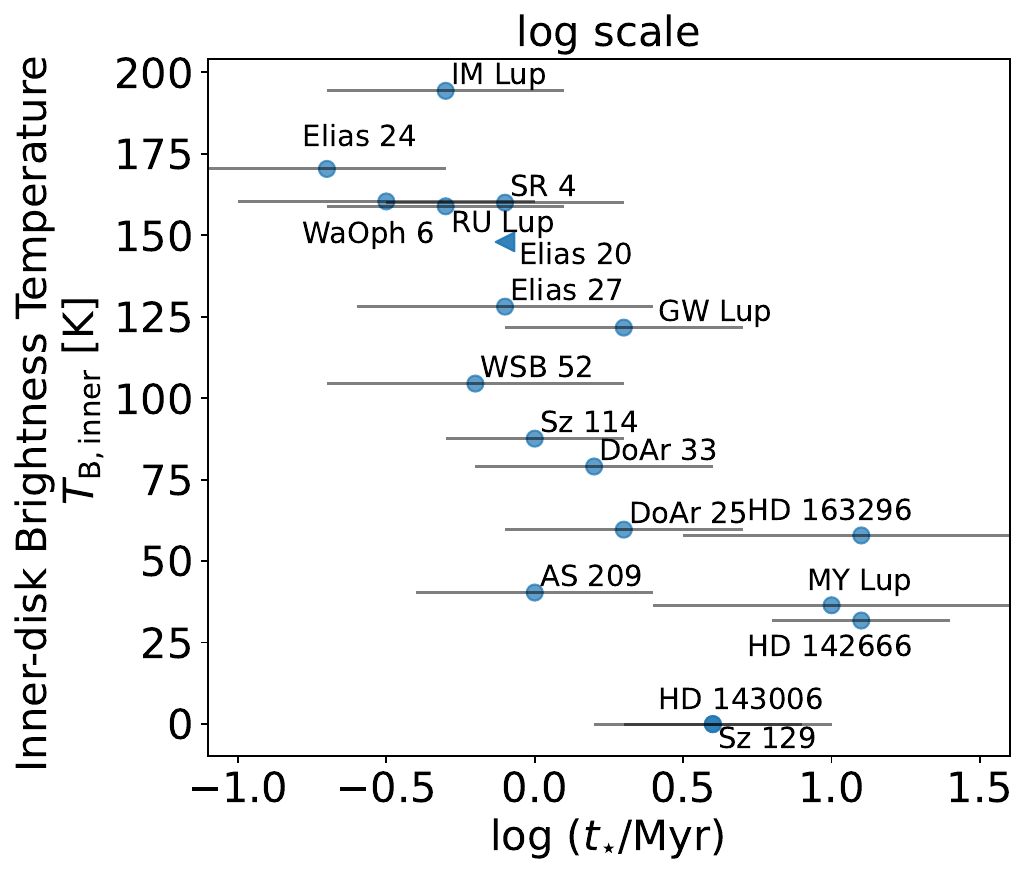}
\caption{Inner-disk brightness temperature $T_{\rm B, inner}$ and stellar age \aizwrev{$t_\star$} for the disk sample. In the right panel, the time is shown in the log scale, and each point is labeled by its source name. }
\label{fig:t_tb_r0}
\end{center}
\end{figure*}

There is a strong correlation between the \aizw{inner-disk} brightness temperature, $T_{\rm B, inner}$, and stellar age with $r=-0.80$ and $p=6.6 \times 10^{-5}$. A similar trend is found when using $T_{\rm B,inner}$ values derived from the brightness profiles of CLEAN-based images in \cite{huang_ring_2018} (Appendix~\ref{sec:huang_comp_tb}), although the significance is slightly weaker ($r=-0.67$, $p=0.0023$). \aizw{This difference is likely due to variations in effective angular resolution, with our modeling achieving higher resolution and therefore better capturing the hotter temperatures in the inner disk, which would strengthen the correlation.} Figure~\ref{fig:t_tb_r0} shows the scatter plot of $T_{\rm B, inner}$ versus stellar age, with time plotted on a linear scale (left) and a logarithmic scale (right). \aizw{Even with the large uncertainties in ages, the negative correlation is evident.} The plots indicate that $T_{\rm B, inner}$ saturates around $130-190$ K for $t_{\star}<0.5$ Myr, while it plateaus around $0$-$50$ K for $t_\star > 3$ Myr. We interpret the correlation as evidence for inner-disk dispersal in the DSHARP disks, as discussed below. 

\aizw{In addition to the dust thermal emission, free-free emission could contribute to the inner-region intensities. For instance, \cite{rota2024} measure spectral indices from cm to ALMA Band~3-7 for 11 transition disks, and find that the cm-3\,mm (Band 3)  indices are consistent with free-free emission.  Nevertheless, the contribution of the free-free emission at $\sim$1 mm (Band 6) is generally expected to be small, and is typically assumed to be negligible \citep[e.g.,][]{rota2024}. We therefore assume that $T_{\rm B, inner}$ is dominated by dust thermal emission. }

Brightness temperature ($T_{\rm B, inner}$) depends on both the local dust temperature and the optical depth (set by density and opacity).  Notably, several observed disks exhibit $T_{B,\mathrm{inner}}< 50$ K. This is more consistent with a drop in the density, because the dust temperature in inner regions is unlikely to decrease such drastically. Consistently $T_{B,\mathrm{inner}}$ shows no significant correlation with $L_\star$, which is expected to influence temperatures. By contrast, the mass accretion rate $\log \dot{M}$ shows a positive correlation with $T_{\rm B, inner}$ $(r=0.57, p=0.018)$, apparently suggesting a possible contribution to $T_{\rm B, inner}$ through accretion heating, which can vary mid-plane temperatures.  However, this correlation largely vanishes once stellar age is controlled for; the partial correlation coefficient is $r=0.07$ with $p=0.78$. Instead, the partial correlation analysis indicates the strong correlation between $T_{\rm B, inner}$ and age with $\log \dot{M}$ being controlled; we find $r=-0.68$ with $p=0.0037$. This indicates that the strong correlation is between age and $T_{\rm B, inner}$, while the apparent link with accretion rate is likely an age-related confounding effect. Summarizing, the decline in $T_{B,\mathrm{inner}}$ is more plausibly explained by a decrease in the dust density (or opacity), indicating progressive depletion of millimeter-sized dust. 

Next, we discuss what spatial scales in the disks are actually probed by $T_{B,\mathrm{inner}}$. It is defined from the intensity at the innermost radius $r=0.0076\arcsec$, and this value does not directly represent the brightness temperature at that radius, but rather reflects the temperature convolved with the effect angular resolution, which is characterized by $\gamma$. Assuming the same observational setup as the real data, \cite{aizawa2024} perform injection-recovery tests on the AS~209 radial profile compressed by a factor of 1/2 in radius. They find that emission within a half-width of $\sim0.01\arcsec$ can be recovered to within a few percent even with $\gamma\simeq 0.03\arcsec$ and a beam size of $0.04\arcsec$ (see Fig.~D.1 of \citealt{aizawa2024} in Appendix D). Thus, the central emission can be resolved at least three times finer than $\gamma$ in the simulations. 

Across our full sample, $\gamma$ spans $0.02\arcsec$-$0.1\arcsec$, implying that $T_{\rm B,inner}$ probes regions within $r<0.007$-$0.035\arcsec$, corresponding to $\sim1$-5 au at 150 pc. Performing an additional correlation analysis, we find no correlation between  $T_{\rm B,inner}$ and either the beam size or $\gamma$. \aizw{Although the comparison with the analysis with the CLEAN-based images suggests a potential link between $T_{\rm B,inner}$ and angular resolution, the lack of the correlation in our sample indicates} that $T_{\rm B,inner}$ is not strongly biased by angular resolution \aizw{in our modeling}. A more rigorous sets of injection-recovery tests for all the systems would be required to quantify the effective angular resolution for $T_{B,\mathrm{inner}}$, but such an analysis is beyond the scope of this paper. 

In summary, the decline in $T_{\rm B, inner}$ \aizw{with respect to stellar age} most likely traces the progressive depletion of millimeter-sized dust in the innermost disk regions ($\sim1$-5 au) of the DSHARP disks. In contrast, we find that neither the total flux nor the mean surface brightness (total flux divided by disk area) exhibits a clear correlation with age. This indicates that while the inner disk component probed by $T_{\rm B, inner}$ is dispersed more rapidly ($\sim$3 Myr) than the outer disk, which can persist for several Myr or longer. 

\cite{jennings2022} report that the oldest disks in the sample, HD 143006 and Sz 129, exhibit inner cavities, and note that other old systems such as HD 142666 and HD 163296 also show gaps inside 5 au, suggesting inner-disk dispersal at advanced ages. Our identified correlation provides a more general context for this hypothesis, further supporting the picture of inner-disk dispersal in the DSHARP disks. 

We consider the possibility that biases in age estimates might confound the observed correlation. Stellar ages are taken from \cite{andrews2018}, inferred by fitting isochrones to effective temperature and luminosity. An age estimation for a young star is generally challenging; episodic accretion can transiently modify both quantities, thereby biasing isochronal age estimates \citep[e.g.,][]{soderblom2014}. \citet{hosokawa2011} show that for stars \aizw{with effective temperatures} hotter than 3500 K isochronal ages tend to be overestimated: truly young objects can masquerade as older. All DSHARP targets are hotter than 3500 K, and the age estimates may be regarded as upper limits. Nevertheless, even under this assumption, it is striking that \aizw{we do not observe any object with high inner-disk temperatures and old ages (the upper-right region in Figure~\ref{fig:t_tb_r0}).} \aizw{It is therefore unlikely that age biases alone explain the observed correlation.}

Another possibility is that $T_{\rm B, inner}$ itself may influence stellar age estimates; high $T_{\rm B, inner}$ may signal a substantial amount of dust or accretion near the star, possibly complicating the isochrone fitting. Nevertheless, the weak direct correlation between  $T_{\rm B, inner}$ and mass accretion implies that the impact is limited. For quantifying the effect, a rigorous evaluation of such biases would require in-depth analysis for isochrone fitting, which is beyond the scope of this paper. 

\section{\aizw{Discussion}} \label{sec:discussion}
\subsection{\aizw{On the detection of circumplanetary disk emission and spirals in the DSHARP data}}
Annular structures in the DSHARP disks have been widely interpreted as potential signatures of embedded planets. However, despite the excellent sensitivity and angular resolution of the DSHARP data, we find no evidence for CPD emission or newly identified planetary spirals, except for one-armed spiral in Elias 27. Given that the DSHARP sample is among the most suitable ensembles for detecting faint structures owing to its long integration and large millimeter fluxes, our results suggest that detecting CPDs and planetary spirals in the millimeter continuum remains challenging with the current capabilities of ALMA. 

\cite{andrews2021} compare their CPD upper limits, typically 50-70$\mu$Jy, with the empirical relation between millimeter flux and object mass derived from nearby disks spanning planetary to stellar masses. The inferred planetary masses of most DSHARP disks are on the order of, or below, a few Jupiter masses \citep{zhang2018,andrews2021,wang2021}. The interpolated millimeter fluxes corresponding to such masses are typically  $\lesssim 10\mu$Jy, comparable to noise level in DSHARP images. Therefore, the non-detection of CPD emission is reasonable according to interpolations. We note, however, that the empirical relation is basically based on isolated objects, and CPDs embedded in circumstellar disk may exhibit higher luminosities.

\citet{speedie_sim_2022} investigate the detectability of planet-driven dust spirals across a range of disk properties, noise levels, and angular resolutions. \aizwrev{For} noise levels of \aizwrev{10-35$\mu$Jy with angular resolutions of 30-65 mas} (comparable to DSHARP \aizwrev{observations}), they find that \aizwrev{a planet with a mass of} $M_{p}=0.3 M_{\rm th} \simeq 2 M_{\rm Nep} \left(\frac{h}{0.07}\right)^3 \left(\frac{M_\star}{M_\odot}\right)$, where $M_{\rm th}$ is the thermal mass and $h$ is the disk aspect ratio, can generate detectable spiral signatures \aizwrev{under favorable disk conditions.} \aizwrev{In the current sample, the planet masses inferred from dust gaps are typically larger than Neptune \cite[e.g.,][]{zhang2018}, and the stellar masses span $\sim0.15$--$2 M_{\odot}$, with the majority clustered around $0.5$-$1 M_{\odot}$. Given these properties,} the lack of significant additional dust spirals in the DSHARP disks, aside from Elias 27 \aizwrev{and possibly HD 143006 in \cite{andrews2021}, may indicate a tension with predictions based on favorable disk conditions. } Additional searches for faint spirals, as well as tighter constraints, may benefit from forward modeling of spirals in visibility space \citep{stevenson2024}. A more quantitative investigation is deferred to future work.  

\subsection{\aizw{Implications and origins of the correlation between inner-disk brightness temperature and age}}
\aizw{We discuss implications of the identified relation between $T_{\rm B, inner}$ an age in Section \ref{sec:age_bright}}. It is well established that millimeter continuum fluxes for disks in various star-forming regions decline monotonically with age (Fig.~2 of \citealt{miotello2023}; Fig.~5 of \citealt{manara2023}).  In the Perseus region (mean age $<1$ Myr; \citealt{tychoniec2018}), a large fraction of disks retain substantial dust reservoirs.  At mean ages of $1$-$3$ Myr in Lupus \citep{ansdell2016} and Taurus \citep{andrews2013}, disk dust masses are reduced but still remain.  In Upper Scorpius (mean age $\sim10$ Myr  \citep{barenfeld2016}),  disks are largely dispersed. 

Infrared surveys likewise show a declining disk fraction with ages \citep[e.g.,][]{haisch2001,hernandez2007,ribas2014, ribas2015,ben2025}. Characteristic timescales for the disk dispersal are few to several Myr, while they depend on wavelength within the (near-)infrared band \citep{ribas2014,ben2025}. Because, (near-)infrared probes warmer, smaller dust than millimeter wavelengths, the inferred timescales can differ. 

Overall, both (near-)infrared and millimeter surveys suggest a monotonic decline of disk fluxes with with age. In contrast, we do not identify clear variations in SED shapes with ages in the DSHARP disks (see Fig 1. of \cite{andrews2018}), consistent with the weak age-flux correlations discussed in Section \ref{sec:corr}. This discrepancy likely reflects selection bias; the DSHARP sample are biased toward bright, extended disks, whereas demographic surveys are dominated by smaller, fainter disks, and their evolutionary tracks may thus differ \cite[e.g.,][]{marel2021}. Nevertheless, $T_{\rm B, inner}$  in the DSHARP sample still decline with age, in agreement with general trend in the fainter disks.  

Disk evolution is largely governed by viscous accretion, photoevaporation, and disk wind. Viscous accretion affects global evolution, implying that other processes may be required to facilitate inner-disk depletion or to sustain the outer disks. The disk winds mainly work at early phases of the disk resulting in the inner-disk dispersal \citep[e.g.,][]{suzuki2010}, while photoevaporation also possibly contributes to inner hole formation \citep[e.g.,][]{gorti2009,gorti2009b}. One widely-accepted hypothesis is that the inner disk are mainly evaporated by the disk-winds in the earlier phases and the outer disk are removed by photoevaporation in late phases \citep[e.g.,][]{alexander2014}. If magnetically driven winds were the primary driver of inner-disk clearing, one would expect a strong depletion of both gas and dust close to the star. However, once stellar age is controlled for, the correlation between  $T_{\rm B, inner}$ and $\log \dot{M}_{\star}$ is not significant. In addition, there is no apparent decline in infrared flux for stars with older ages. These likely indicate that gas and small dust particles could exist in inner disks at old ages in the current sample.

An alternative explanation is dust filtration: a local pressure maximum, possibly \aizw{carved} by a protoplanet, can trap large grains at outer radii while allowing gas and small dust to pass \citep[e.g.,][]{rice2006}. The DSHARP disks indeed exhibits rich annular structures, consistent with the possible pressure bumps trapping dust rings.  Analyzing a broad disk sample across ages, \cite{marel2021} find that structured (transition, ring, extended) disks can sustain large dust masses up to at least 10 Myr, while nonstructured compact disks show a steady decline in a shorter timescale. They attribute the difference to dust filtration. The inner-disk dispersal observed in the DSHARP disk is consistent with this picture: dust trapping suppresses the inward transport of large grains, leading to inner-disk depletion, likely driven by processes such as accretion, photoevaporation, and disk winds, while retaining dust in the outer disk. 

Another possible effect is dust growth over the time. The process is expected to proceed preferentially in the inner-regions of disks, where the shorter orbital periods lead to short growth timescales, thus facilitating segregation of dust into small and large grains \citep[e.g.,][]{tanaka2005}. Such growth could contribute to the observed decrease in millimeter flux in the present systems. However,  the characteristic timescale of  $\sim3$ Myr inferred here appears inconsistent with the theoretically predicted dust growth timescales of order $10^5$ years or less \citep[e.g.,][]{tanaka2005,ohashi2021}. Nevertheless, multi-frequency observations of inner regions would be helpful for constraining the dust population and testing this possibility. 

In summary, combining the correlations from this study with those from previous demographic surveys, we infer that both the inner disks of large, bright disks and compact, faint disks in millimeter flux could disperse in the similar timescales. Such rapid dispersal, which may signal solely the depletion of mm-sized dust or the population of larger dust grains, in inner disks could constrain planet formation and potentially limit the emergence of habitable planets, irrespective of disk size and flux.

\section{Summary and Conclusions}  \label{sec:summ}
We model the axisymmetric structures for 16 DSHARP disks by simultaneously fitting radial brightness profiles, geometric parameters, and hyperparameters. The residual images are then examined to explore deviations from axisymmetry. In addition, a systematic correlation analysis is performed for the DSHARP disks  except two binary systems.  Our main findings are:

\begin{itemize}
  
  \item None of the 16 disks exhibit compelling continuum-based signatures of circumplanetary disks, in agreement with the non-detections reported by \citet{andrews2021} for nine DSHARP systems.
  
  \item  In IM Lup and WaOph 6, both image-based and Fourier-based ($m=2$) analyses reveal regions where the spiral phase \aizwrev{sharply changes} and the amplitude is minimized; spiral structures appear to \aizwrev{be deflected} at these locations. In Elias 27, IM Lup and WaOph 6, we also identify possible bar-like structures near the disk centers, although limited angular resolution could introduce artificial features.
  
  \item  In Elias 27, we recover a prominent $m=1$ spiral whose morphology closely matches analytical planet-driven spiral models assuming a candidate planet inferred by kinematic deviations.

  \item No additional dust spirals beyond Elias 27 are detected. \aizwrev{The absence of such features may provide constrains on the nature of potential embedded planets associated with annular gaps. }
  
  \item \aizw{Five} disks show significant residuals attributable to the vertical extent of their dust layers. For the \aizwrev{three} disks, we mention that the vertical heights can be newly constrained.
  
  \item Inner-disk brightness temperature strongly correlates with age, suggesting the inner-disk dispersal over time in the DSHARP sample. The timescale is comparable to those found in compact, faint disks from demographic surveys. The dust filtration may explain both the inner-disk dispersal and little correlation between total flux and age in the present sample. 
\end{itemize}

The methodology developed for the residual and correlation analyses can be readily extended to other disk datasets. In addition, our study highlights the current limitation of detecting planetary signals in dust continuum emissions analyzing the asymmetric structures.  More qualitatively distinct constrains \aizw{solely from millimeter continuum} would require the development of future instruments with higher sensitivities and angular resolutions. Meanwhile, the observed correlation between $T_{\rm B, inner}$ and stellar age suggests that the inner disks can be dispersed even in bright systems. \aizw{The nature of the inner-disks, whose properties are essential for understanding planetary formation including the emergence of habitable planets, could be further constrained through multi-wavelength observations with ALMA and, in the future, facilities such as ngVLA.}

\section*{acknowledgments}
We thank Kazuhiro Kanagawa, Hongping Deng, Ruobing Dong, and Xuening-Bai for their insightful discussions. \aizw{We also thank the anonymous referee for constructive comments, which greatly improve the paper.}  This work was supported by JSPS KAKENHI grant No. 17H01103, 18H05441, 22H01274,  23K03463, and 25K17431. Data analysis was in part carried out on the Multi-wavelength Data Analysis System operated by the AstronomyData Center (ADC), National Astronomical Observatory of Japan. This research is based on the following ALMA data \#2013.1.00226.S, \#2015.1.00486.S, and \#2016.1.00484.L. ALMA is a partnership of ESO (representing its member states), NSF (USA) and NINS (Japan), together with NRC (Canada), MOST and ASIAA (Taiwan), and KASI (Republic of Korea), in cooperation with the Republic of Chile. The Joint ALMA Observatory is operated by ESO, AUI/NRAO and NAOJ. We acknowledge the use of ChatGPT (GPT-5; OpenAI) for grammatical and clarity improvements throughout the manuscript. 

{\it Software}: {\tt Astropy} \citep{astropy2013}, {\tt CASA} \citep{mcmullin2007}, {\tt corner} \citep{foreman2016}, {\tt emcee} \citep{foreman2013}, {\tt Jupyter Notebook} \citep{kluyver2016}, {\tt Matplotlib} \citep{hunter2007}, {\tt NumPy} \citep{walt2011}, {\tt Pandas} \citep{mckinney-proc-scipy-2010}, {\tt prodomidpy} \citep{aizawa2024}, {\tt SciPy}  \citep{virtanen2020}

\section*{Data Availability}
The data underlying this article are available in the DSHARP Data Release at \url{https://bulk.cv.nrao.edu/almadata/lp/DSHARP}.

\bibliographystyle{mnras}
\bibliography{ref}

\appendix

\section{Removal of outliers in data of DoAr 25} \label{sec:doar25_out}

Out of the 16 disks, we find that the data of DoAr 25 are significantly affected by correlated noises. The origins of these noises are not clear, but we attempt to remove them from the analysis.

The measurement sets for DoAr 25 consist of 230-233GHz and 244-247GHz data, each with 8 spectral windows. Using the geometric parameters in \cite{huang_ring_2018}, we compute deprojected spatial frequencies $q$. Then, we bin the real visibilities in the $q$-axis using a logarithmic grid with $N_{\rm bin}=200$ and interpolate them to construct a temporal model $f(q)$ for visibilities. 

For the data at each spectral window, we identify possible outliers deviated from $f(q)$ by more than $20 \sigma$. During this process, we find that real amplitudes of higher-frequency data are roughly $10\%$ larger than the lower-frequency data. \aizw{Under  the Rayleigh-Jeans approximation for thermal emission ($I_{\nu} \propto \nu^2$), the expected intensity ratio is $I_{245.5 \text{GHz}}/I_{231.5 \text{GHz}}\simeq 1.12$ in agreement with the observation.} To account for such frequency dependence, we consider a rescaled model $a_{\rm scale} f(q)$, and fit the model with $a_{\rm scale}$ to the data to obtain $a$ for each spectral window. After applying this rescaling, we repeat the outlier rejection with the $20\sigma$ criterion and refit the data to update $a_{\rm scale}$.

The visibilities are then rescaled using the fitted $a_{\rm scale}$, rebinned, and used to construct an updated model $f_{\rm updated}(q)$. With the updated model $f_{\rm updated}(q)$, we iterate the above procedure to determine outliers and $a_{\rm scale}$. Finally, we remove the identified outliers from the data, and rescale the visibilities and weights according to the fitted $a_{\rm scale}$. 

The processed DoAr 25 data, corrected for correlated noise and frequency scaling, are then used for the analyses presented in the main text.

\begin{figure*}
\begin{center}
\includegraphics[width=0.43\linewidth]{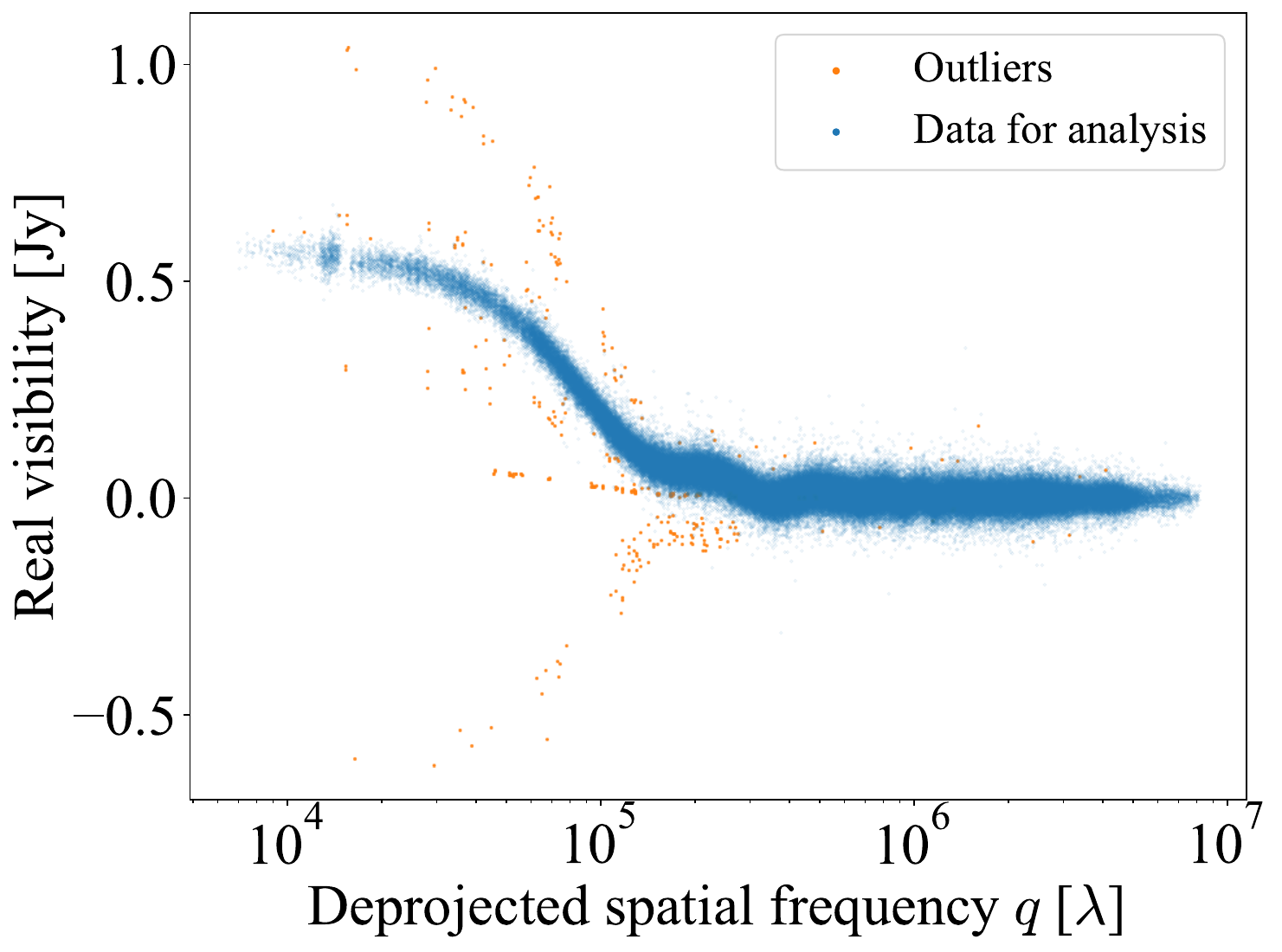}
\includegraphics[width=0.33\linewidth]{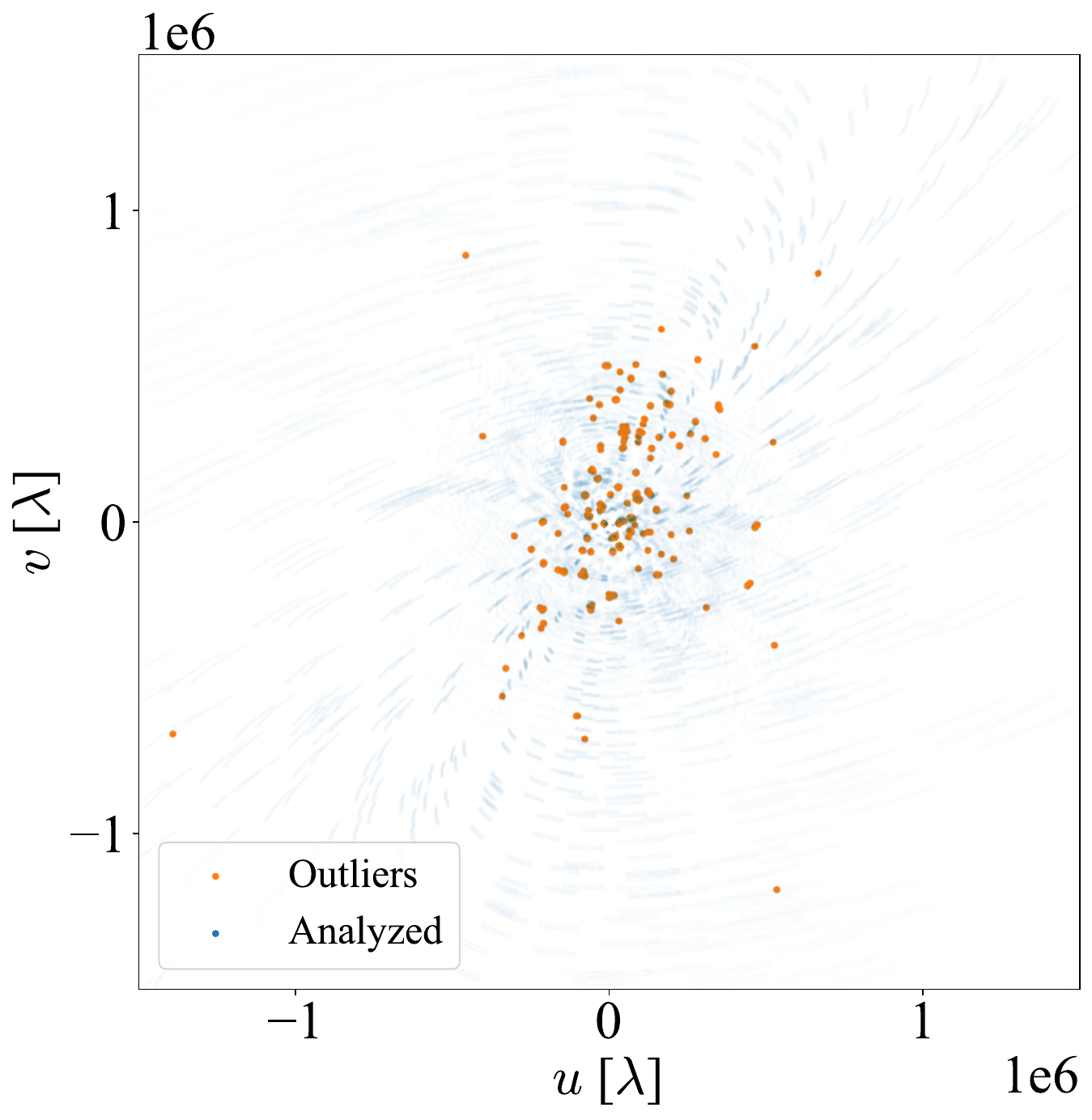}
\end{center}
\caption{Removal of outliers in the visibility data for DoAr~25. (Left) The real part of the visibilities (blue points) plotted against the deprojected spatial frequency $q$, with outliers indicated in orange. These outliers deviate significantly from the main distribution. (Right) The $(u,v)$ coverage, where the outliers are scattered rather than clustered in a specific region. 
} 
\label{fig:doar25_vis}
\end{figure*}

\section{Comparison of fitting results with previous studies } \label{sec:comparison_figure}

\subsection{Geometric Parameters}
\begin{figure*}
\begin{center}
\includegraphics[width=0.85\linewidth]{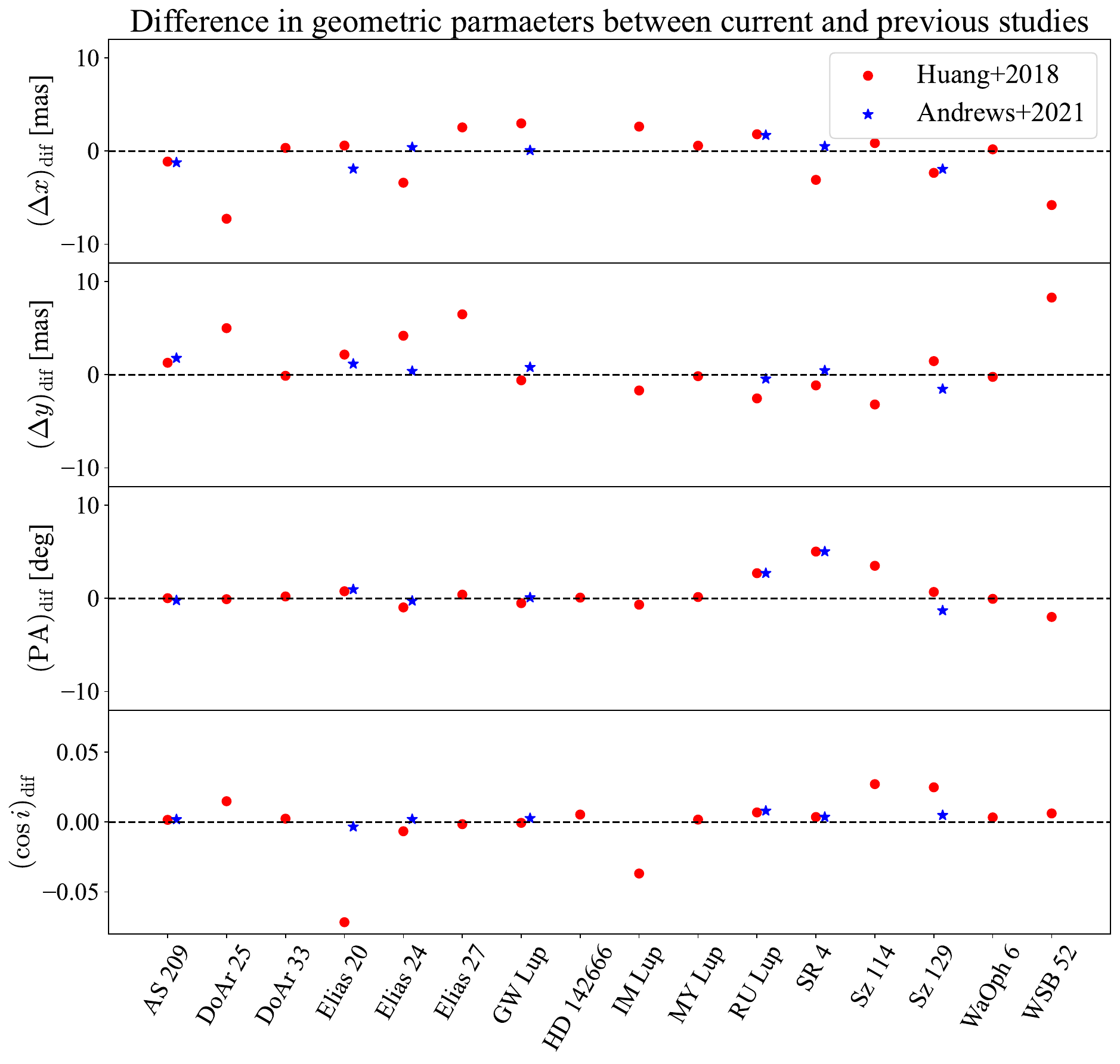}
\end{center}
\caption{Differences in geometric parameters between this work and two previous studies: \citet{huang_ring_2018} (circles) and \citet{andrews2021} (stars). The horizontal axis lists the disk names, and the vertical axes show differences in the fitted parameters (e.g., $\Delta x_{\rm{cen}}$, $\Delta y_{\rm{cen}}$, position angle, inclination). }
\label{fig:geometry_difference}
\end{figure*}

 Figure \ref{fig:geometry_difference} shows the differences in geometric parameters between our analysis and those reported in \cite{huang_ring_2018} and \cite{andrews2021}. We define $(\Delta x)_{\rm dif}$ as the difference of an estimated position in the right ascension between studies, and similarly define $(\Delta y)_{\rm dif}$, $({\rm PA})_{\rm dif}$, and $({\rm cos} i)_{\rm dif}$. 
 
For $(\Delta x)_{\rm dif}$, $(\Delta y)_{\rm dif}$, the absolute differences are less than $10$ mas, except for HD~142666 with $((\Delta x)_{\rm dif}, (\Delta y)_{\rm dif}) = (53.5{\rm mas}, -13.4{\rm mas})$, which is located outside the plotted range. For $({\rm PA} )_{\rm dif}$, the differences are at most $\sim 5$ deg. For $\cos i$, some of the differences are significant, especially for Elias 20 and IM Lup. 

The absolute values of differences look insignificant, but such small differences, e.g., $(\Delta x)_{\rm dif}=1$mas or $(\Delta {\rm PA})_{\rm dif}=1$ deg, still can introduce noticeable fake non-axisymmetric structures in the residual images. We note that geometric parameters remain biased in the presence of non-axisymmetric structure, though the accuracy achieved here is still superior to that from simple Gaussian fitting  \citep{aizawa2024}.

\subsection{Brightness profiles and visibilities}
\begin{figure*}
\begin{center}
\includegraphics[width=0.24 \linewidth]{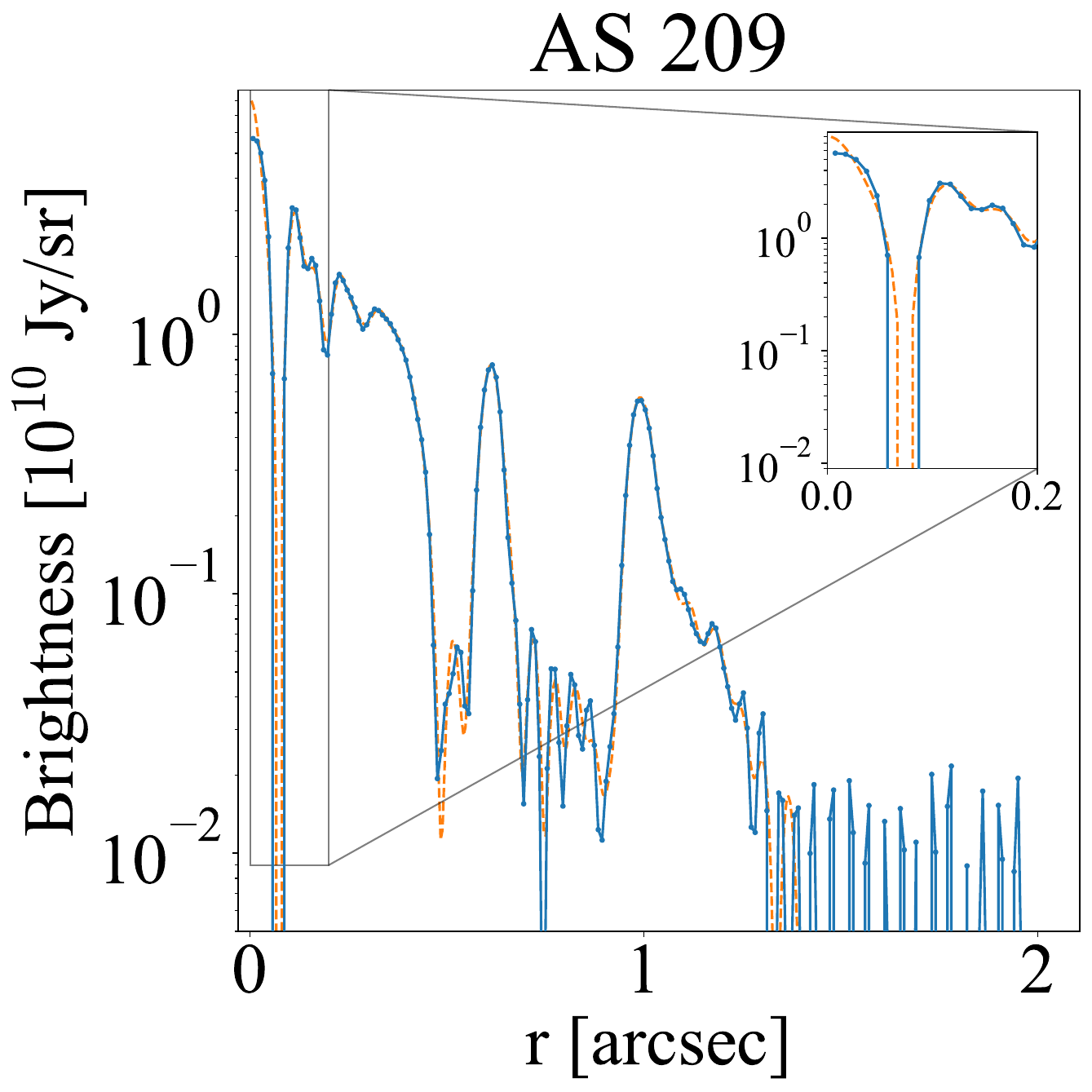}
\includegraphics[width=0.24 \linewidth]{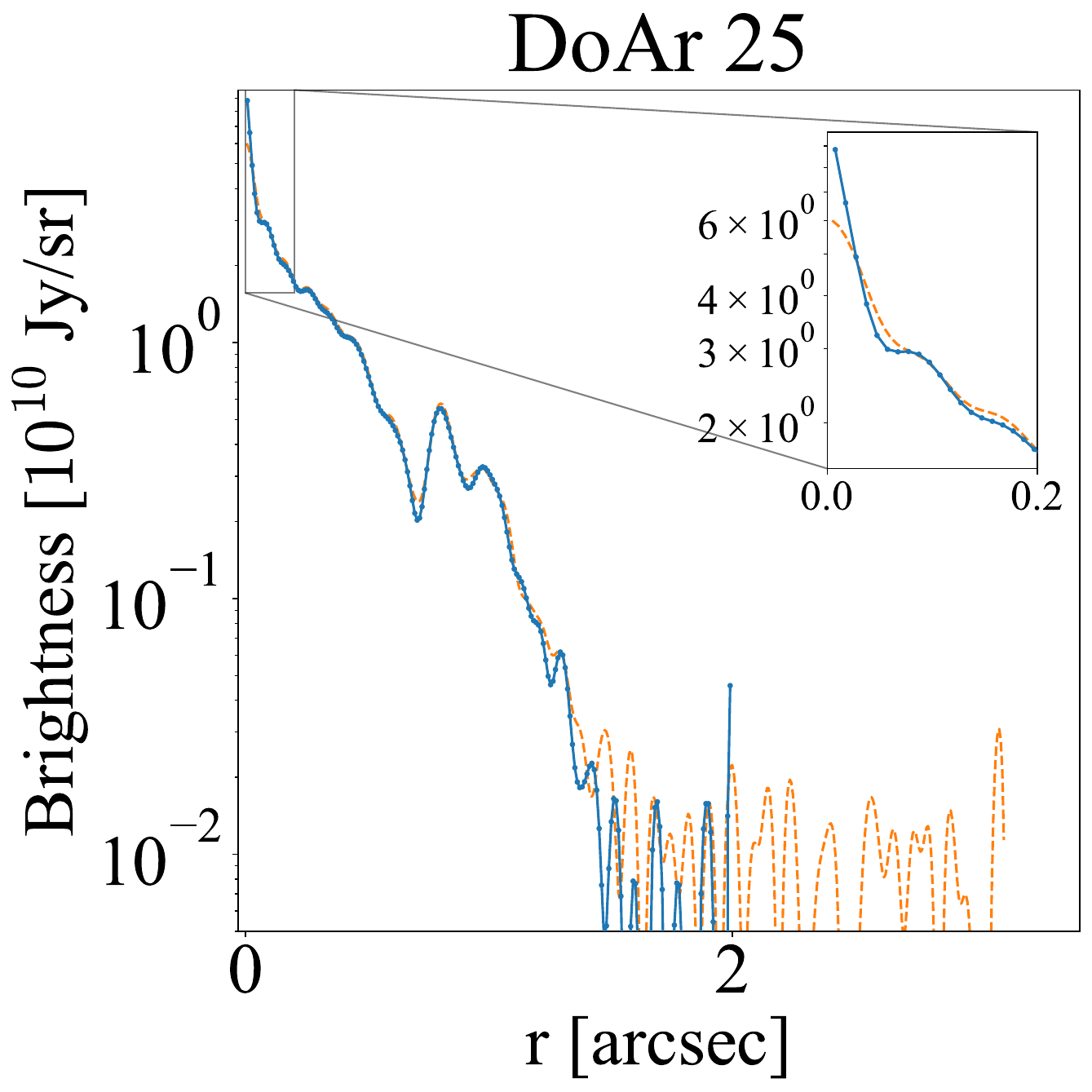}
\includegraphics[width=0.24 \linewidth]{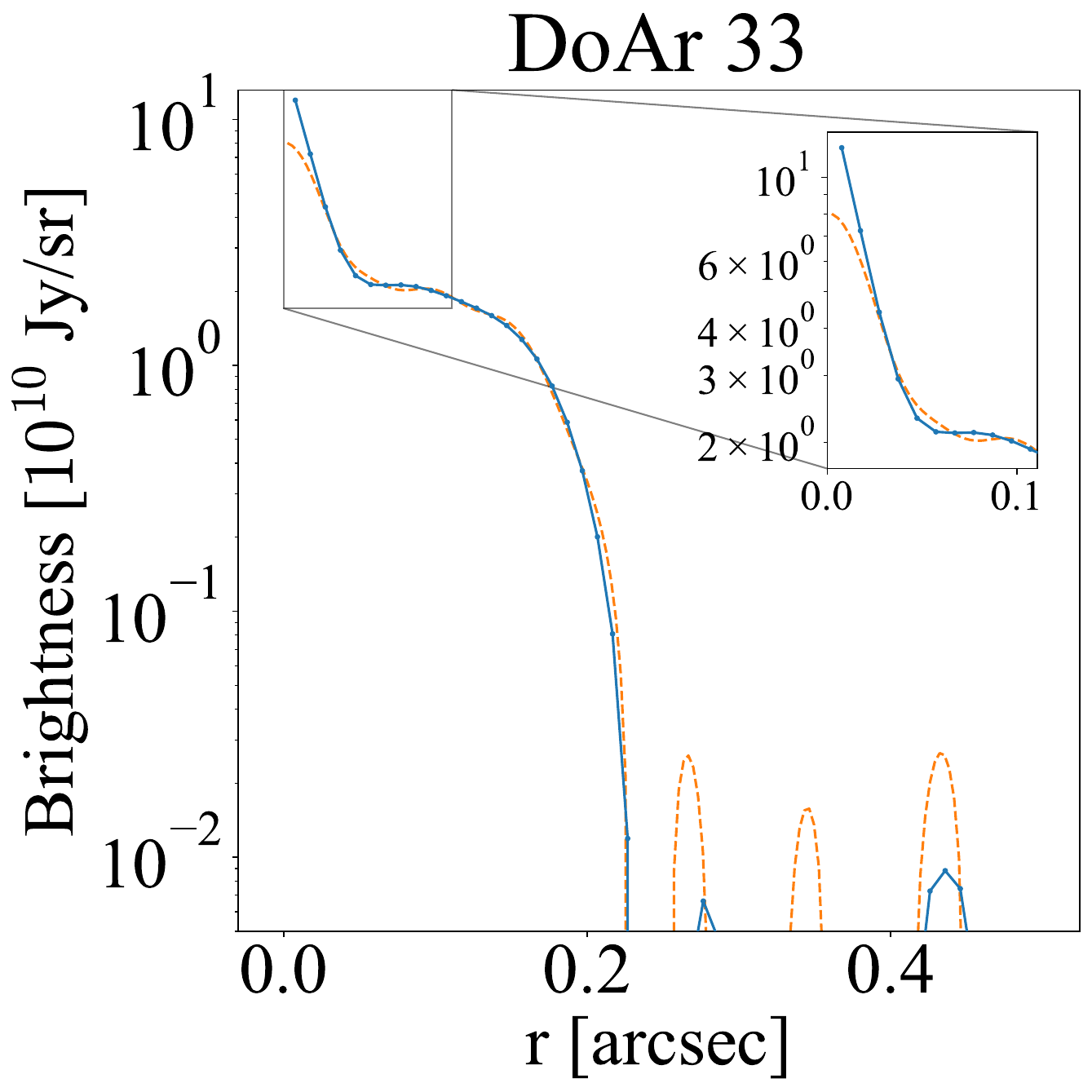}
\includegraphics[width=0.24 \linewidth]{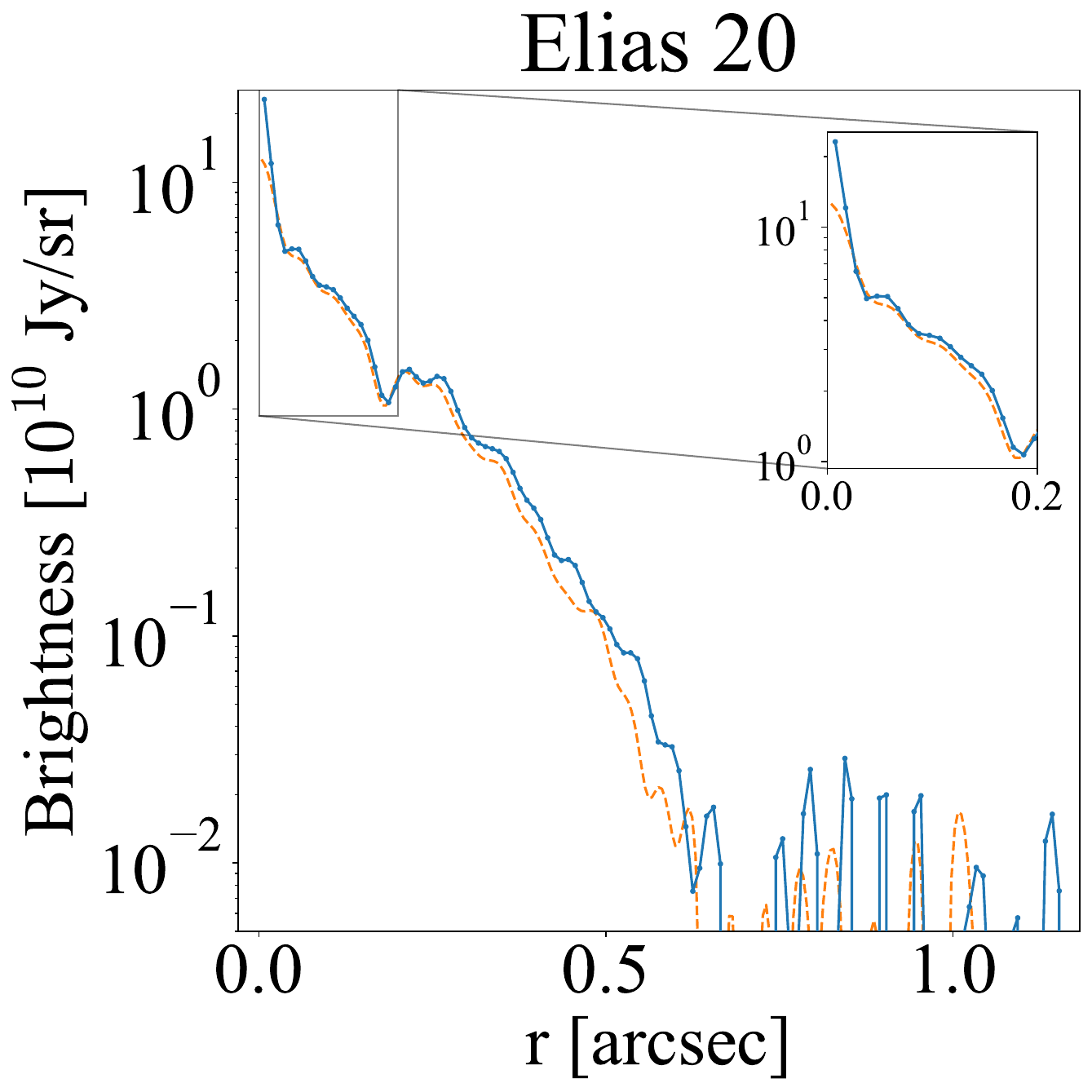}
\includegraphics[width=0.24 \linewidth]{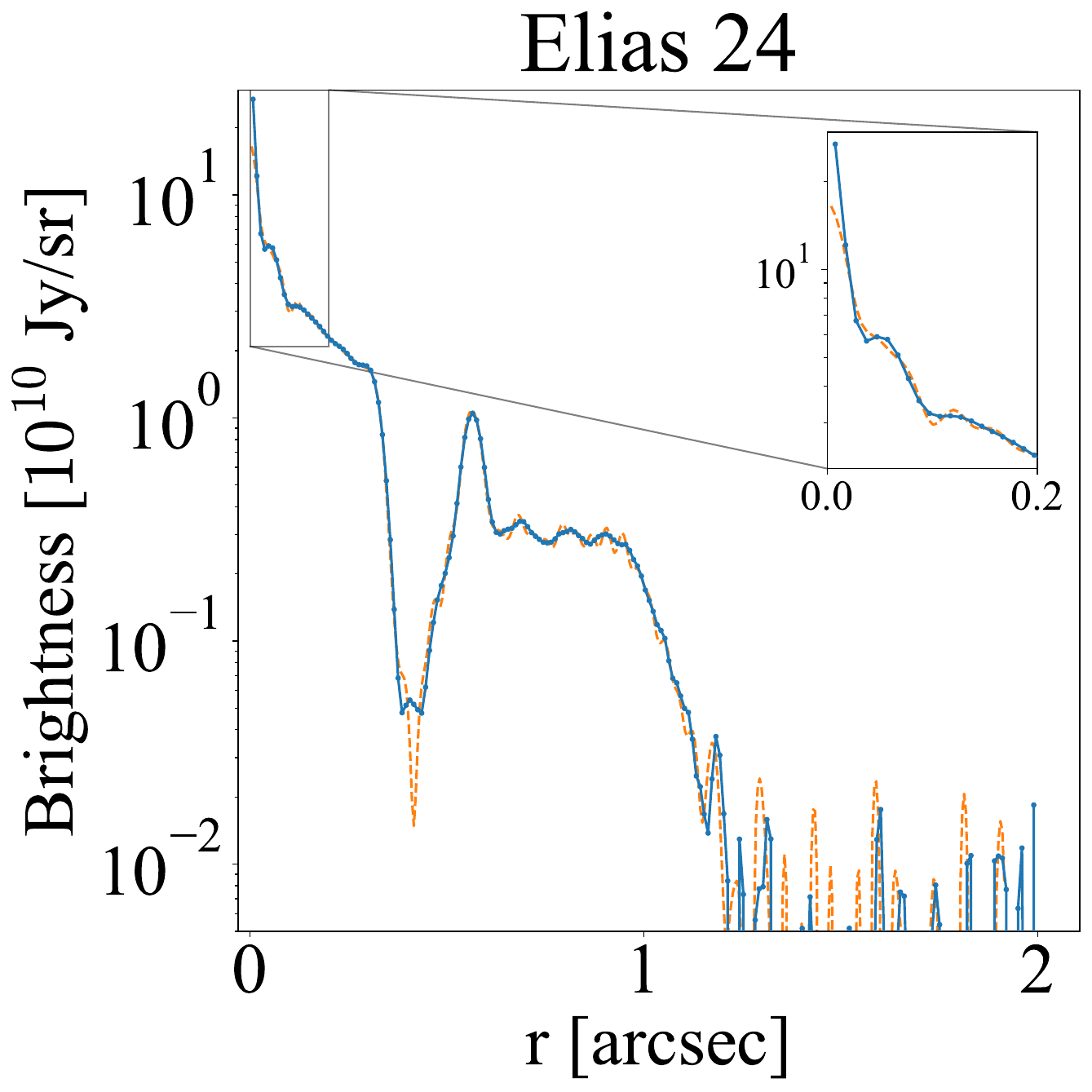}
\includegraphics[width=0.24 \linewidth]{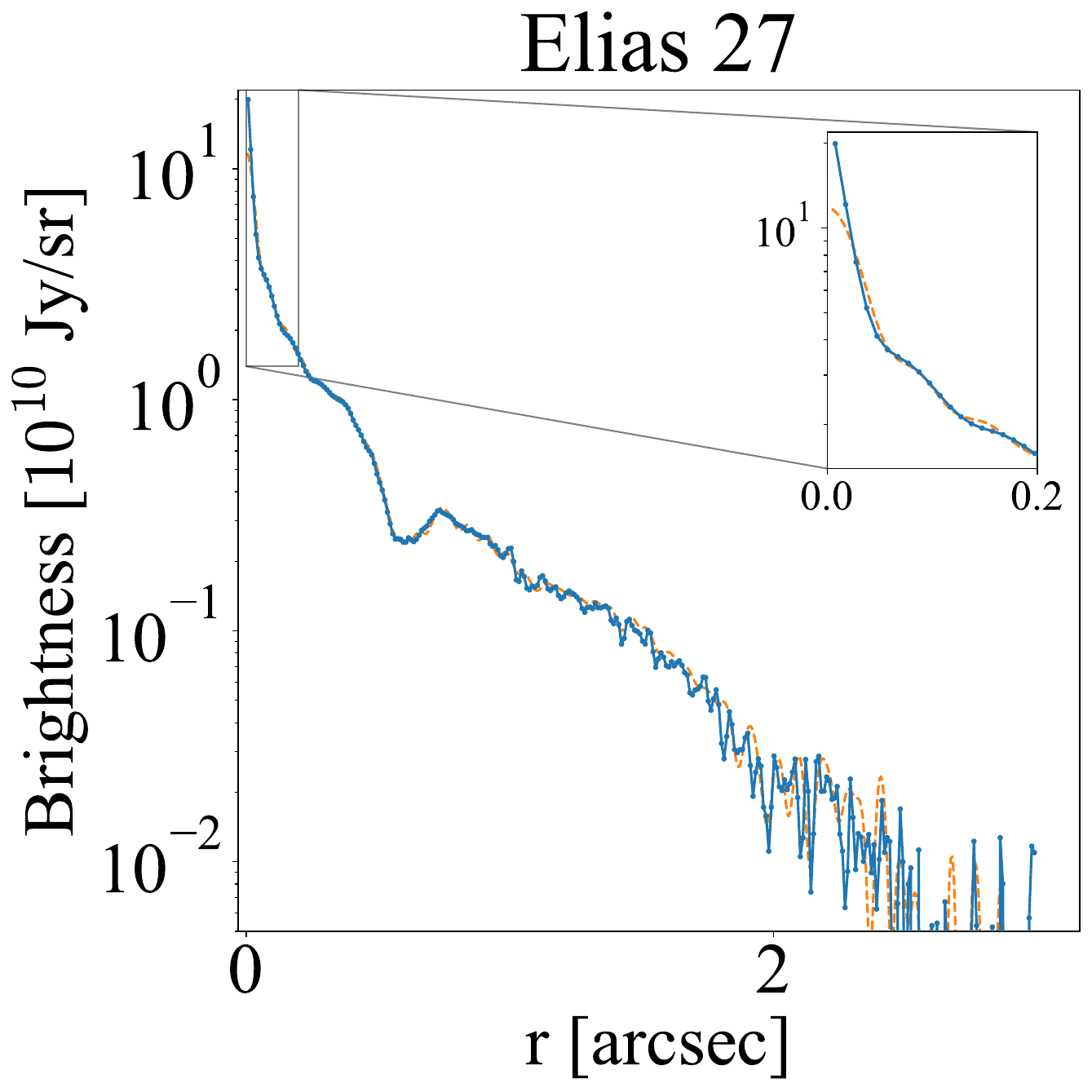}
\includegraphics[width=0.24 \linewidth]{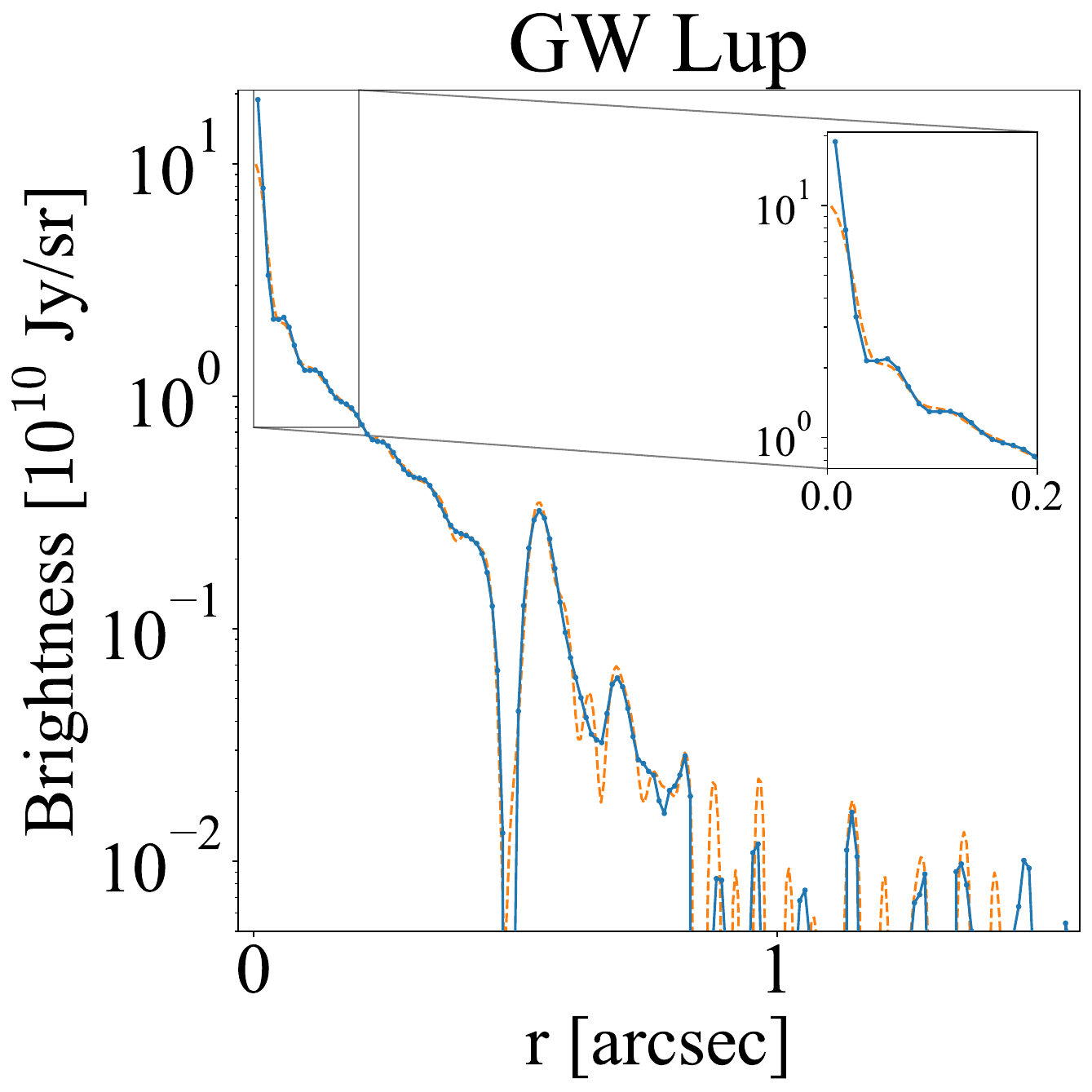}
\includegraphics[width=0.24 \linewidth]{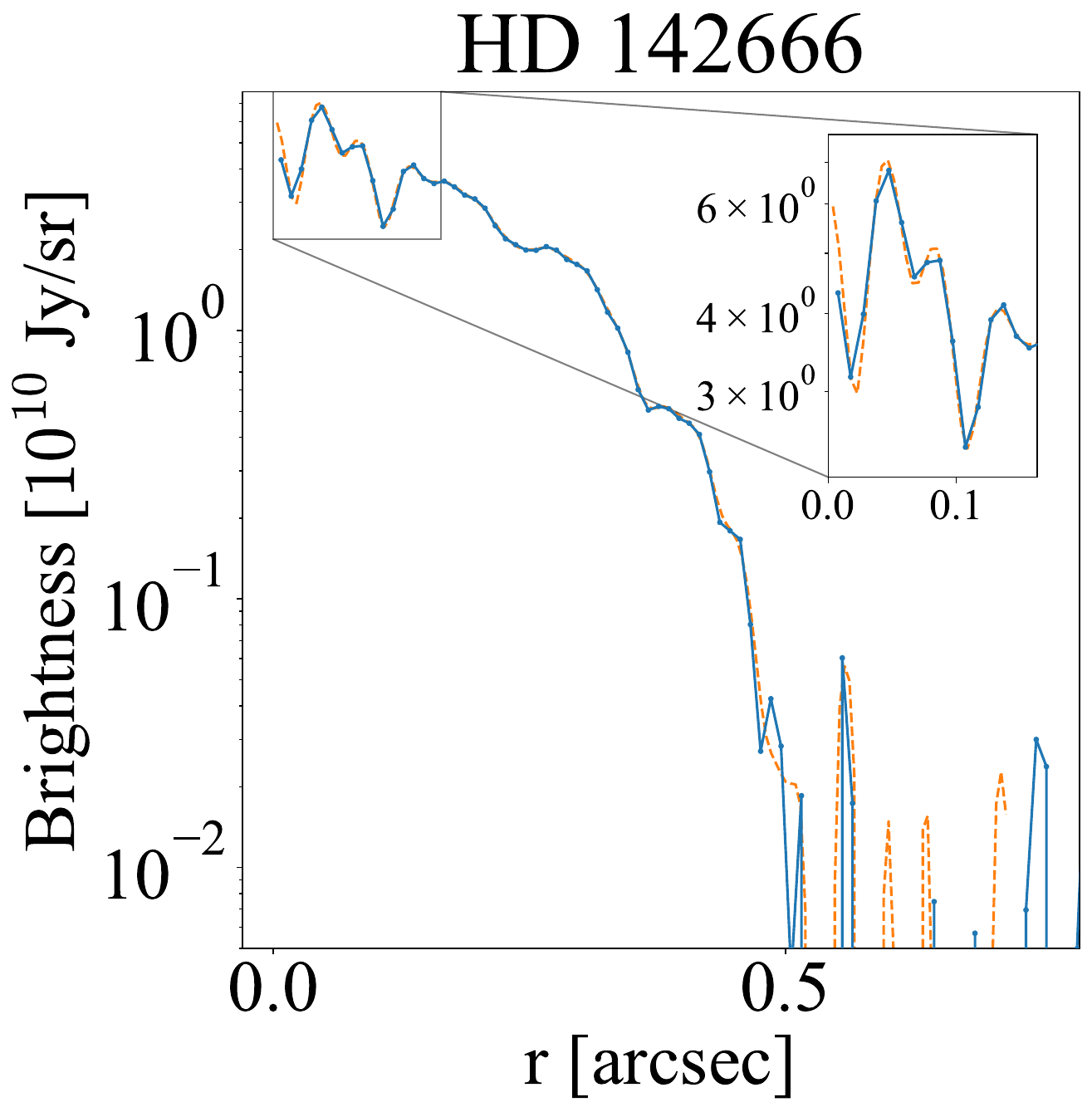}
\includegraphics[width=0.24 \linewidth]{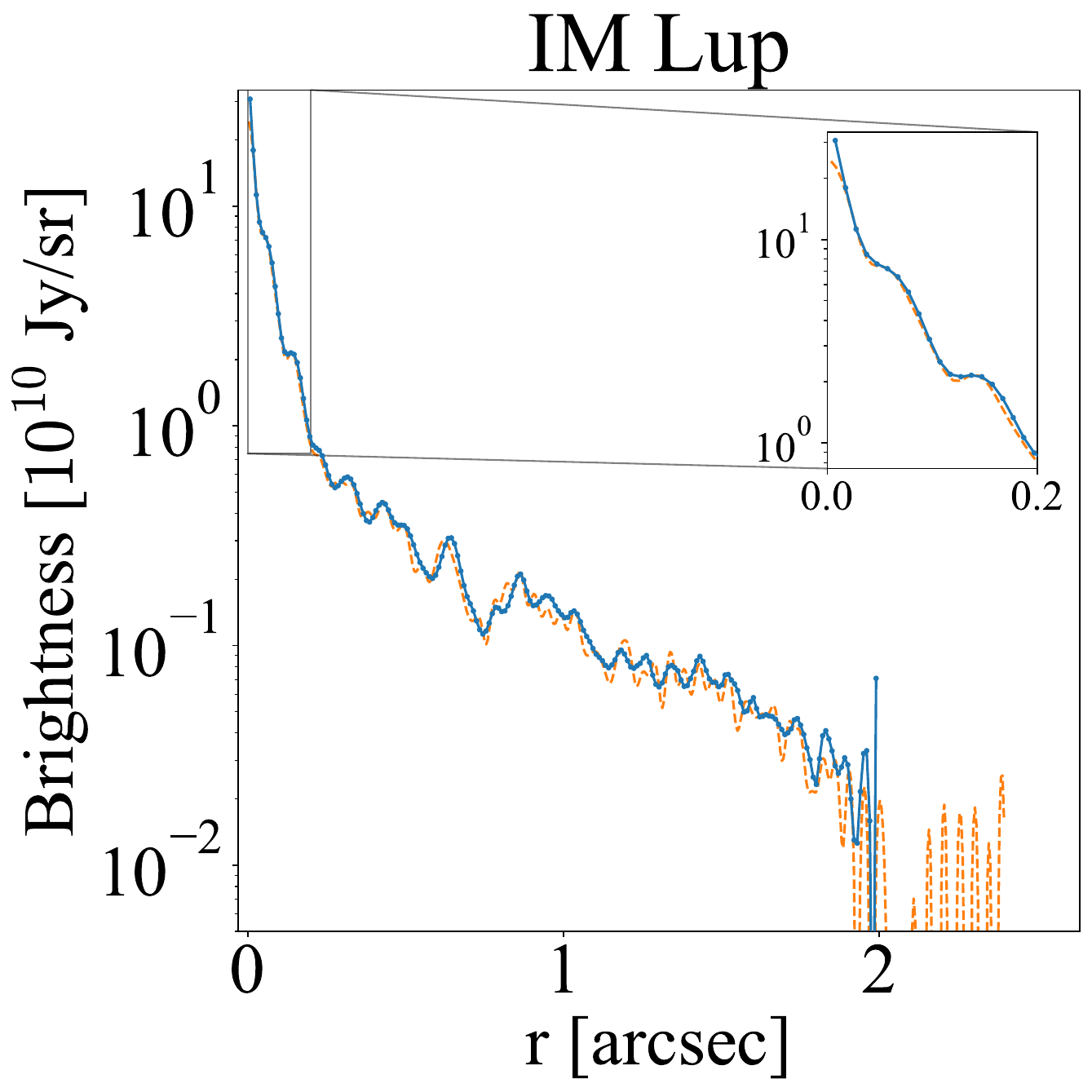}
\includegraphics[width=0.24 \linewidth]{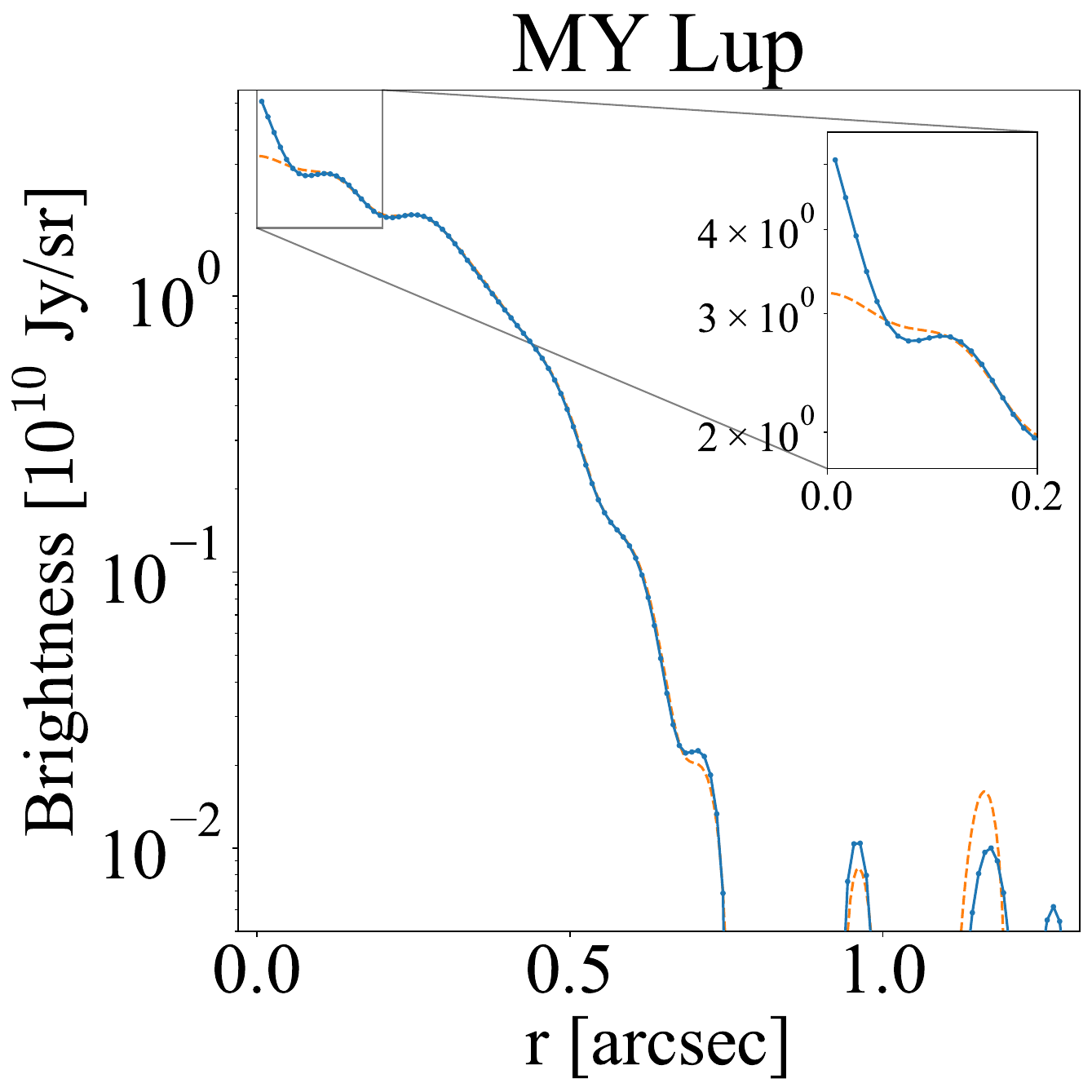}
\includegraphics[width=0.24 \linewidth]{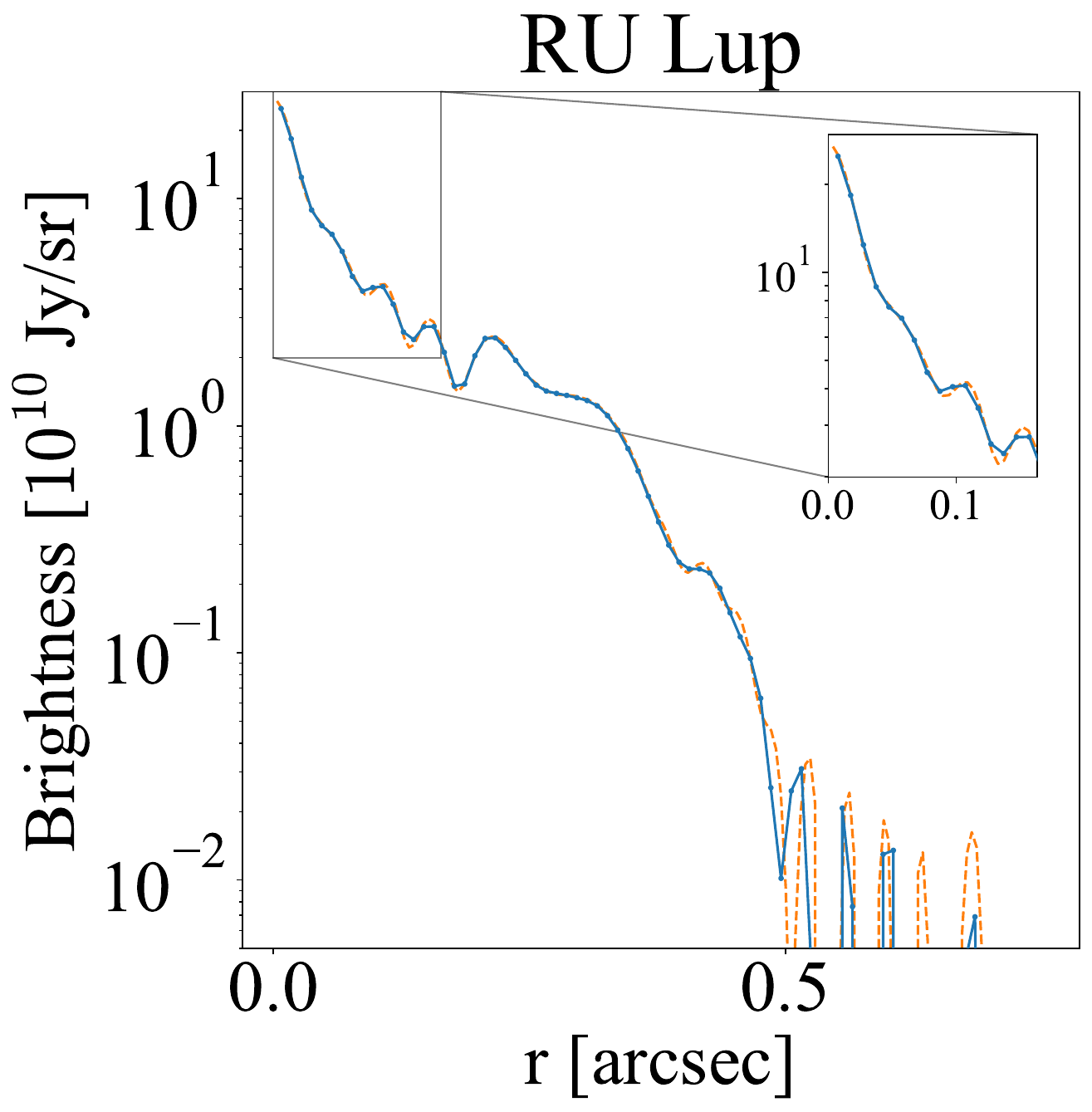}
\includegraphics[width=0.24 \linewidth]{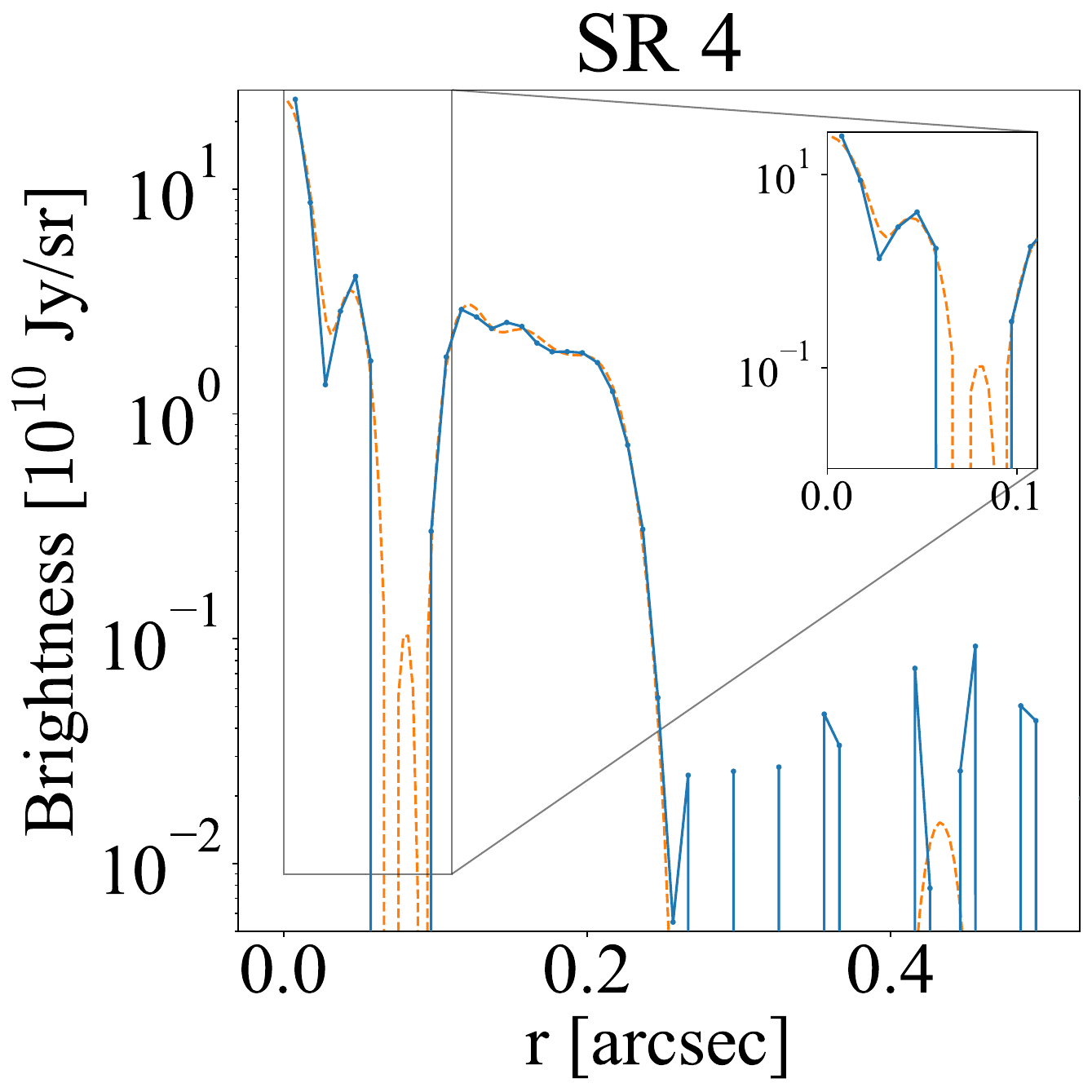}
\includegraphics[width=0.24 \linewidth]{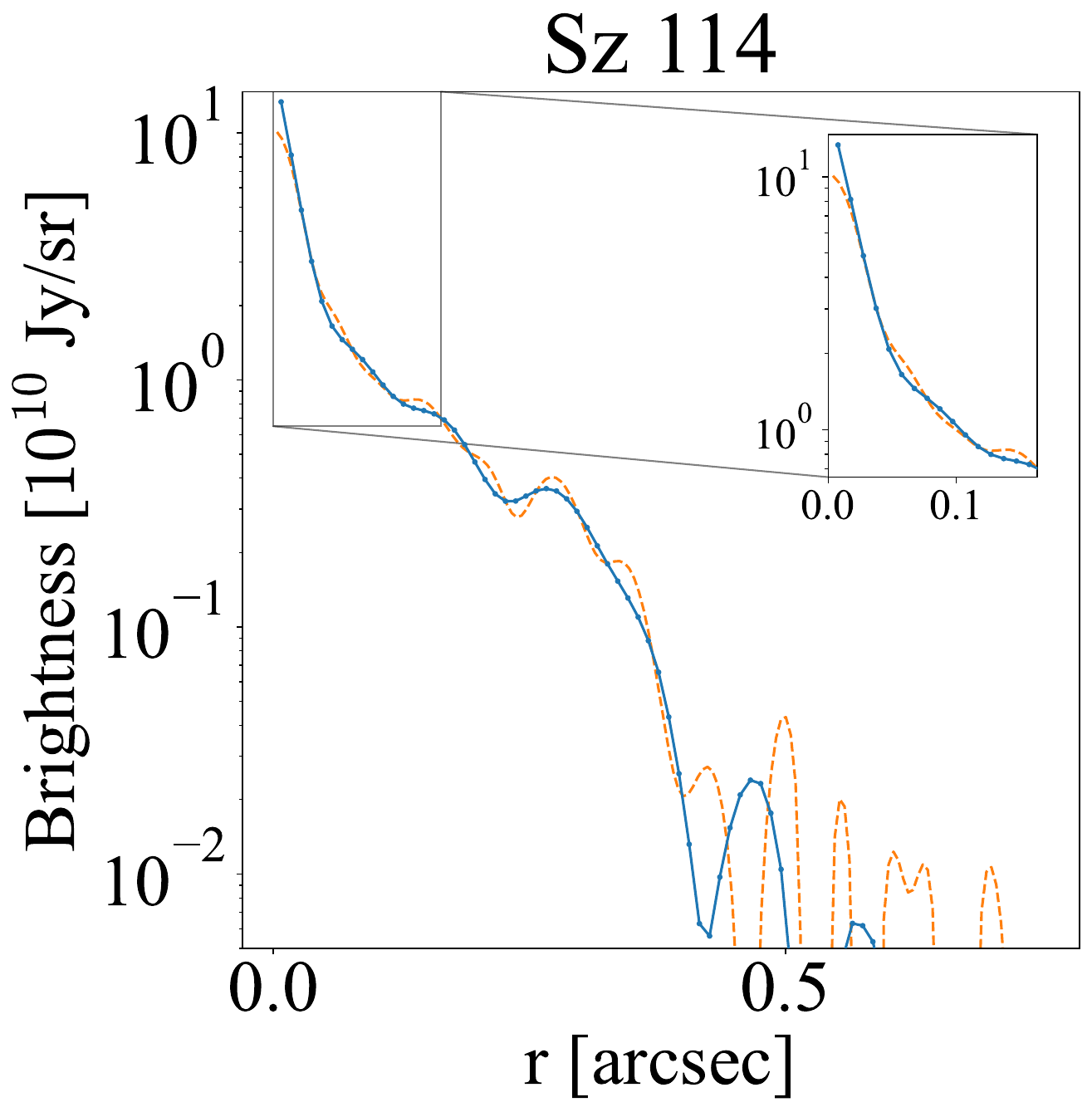}
\includegraphics[width=0.24 \linewidth]{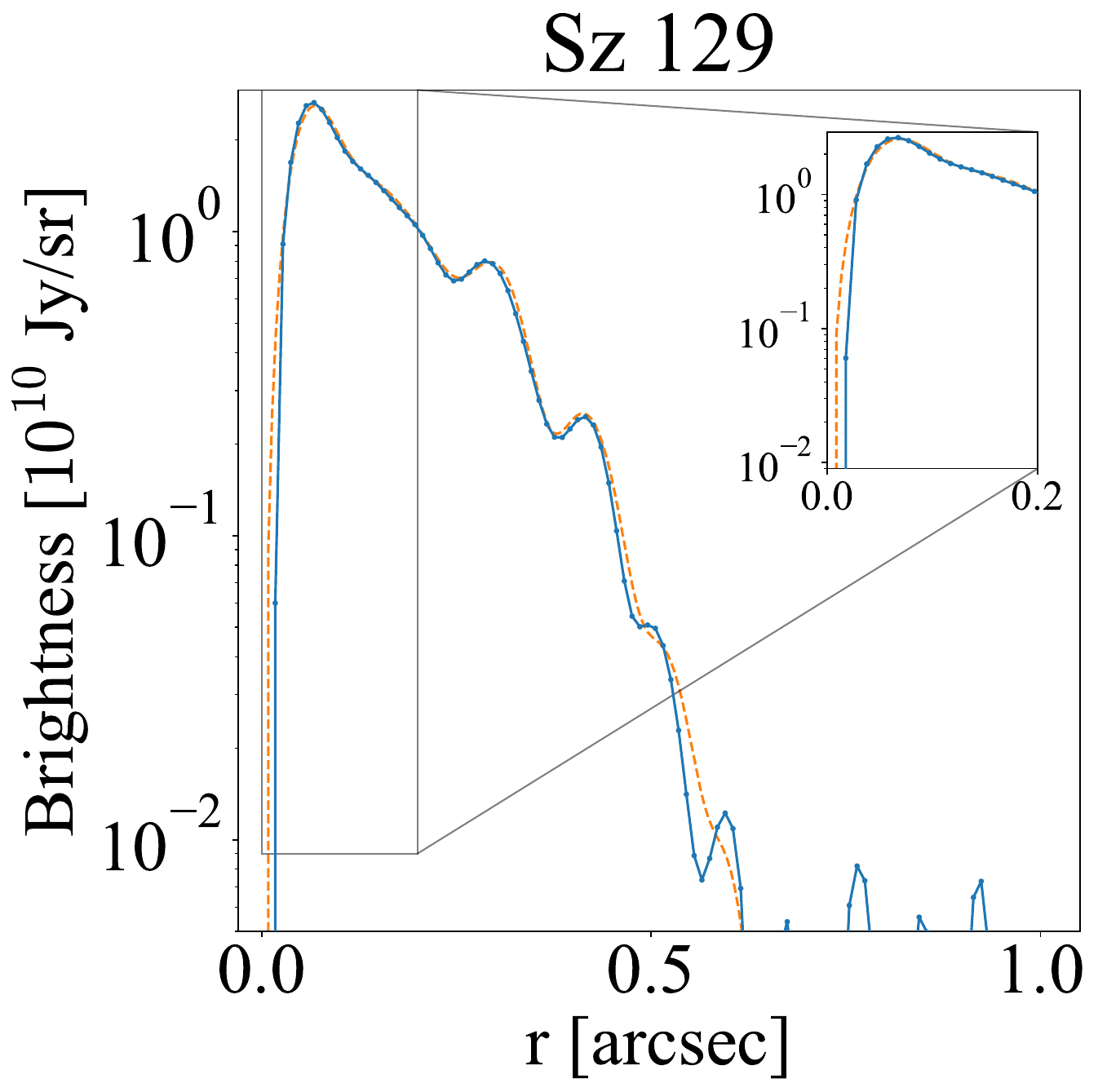}
\includegraphics[width=0.24 \linewidth]{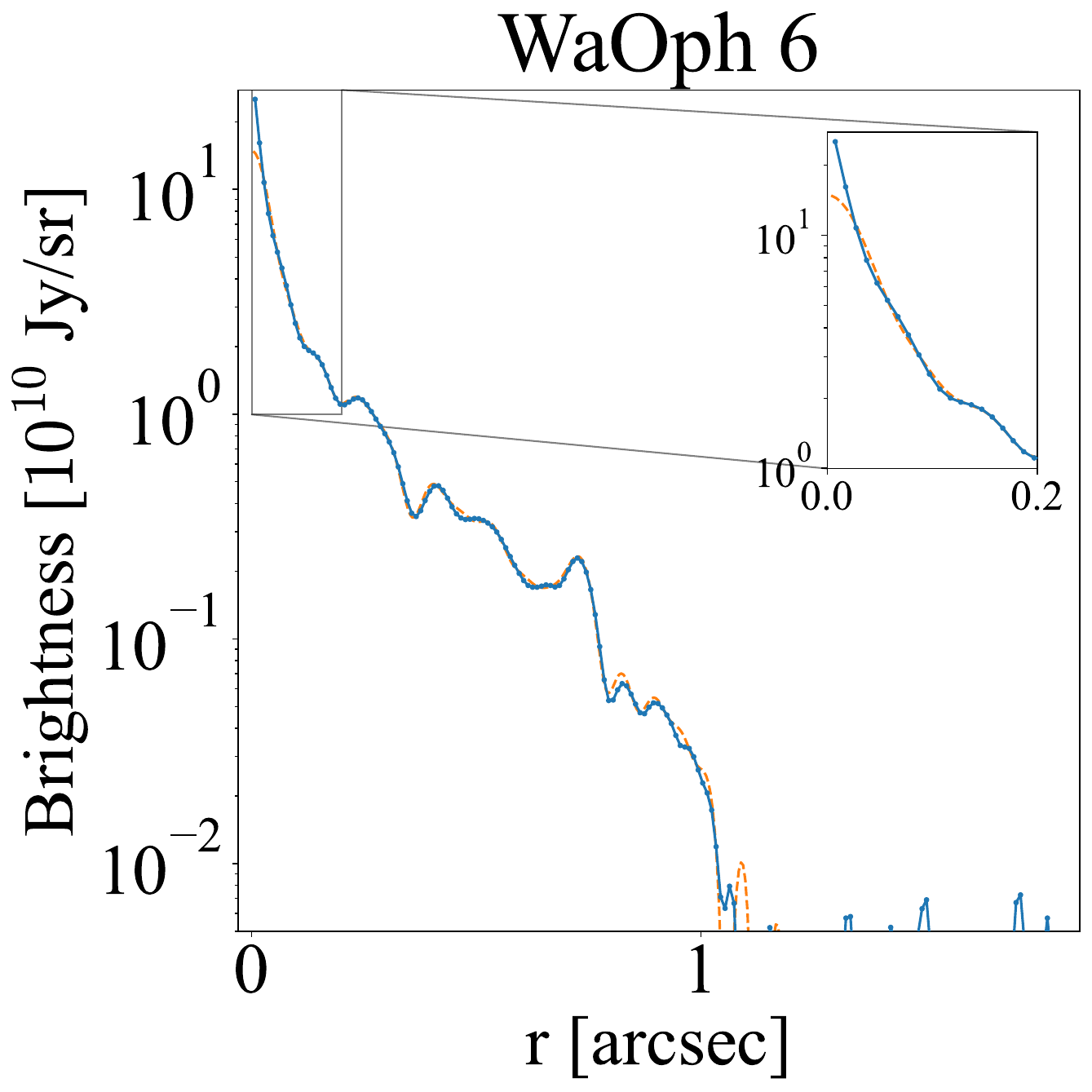}
\includegraphics[width=0.24 \linewidth]{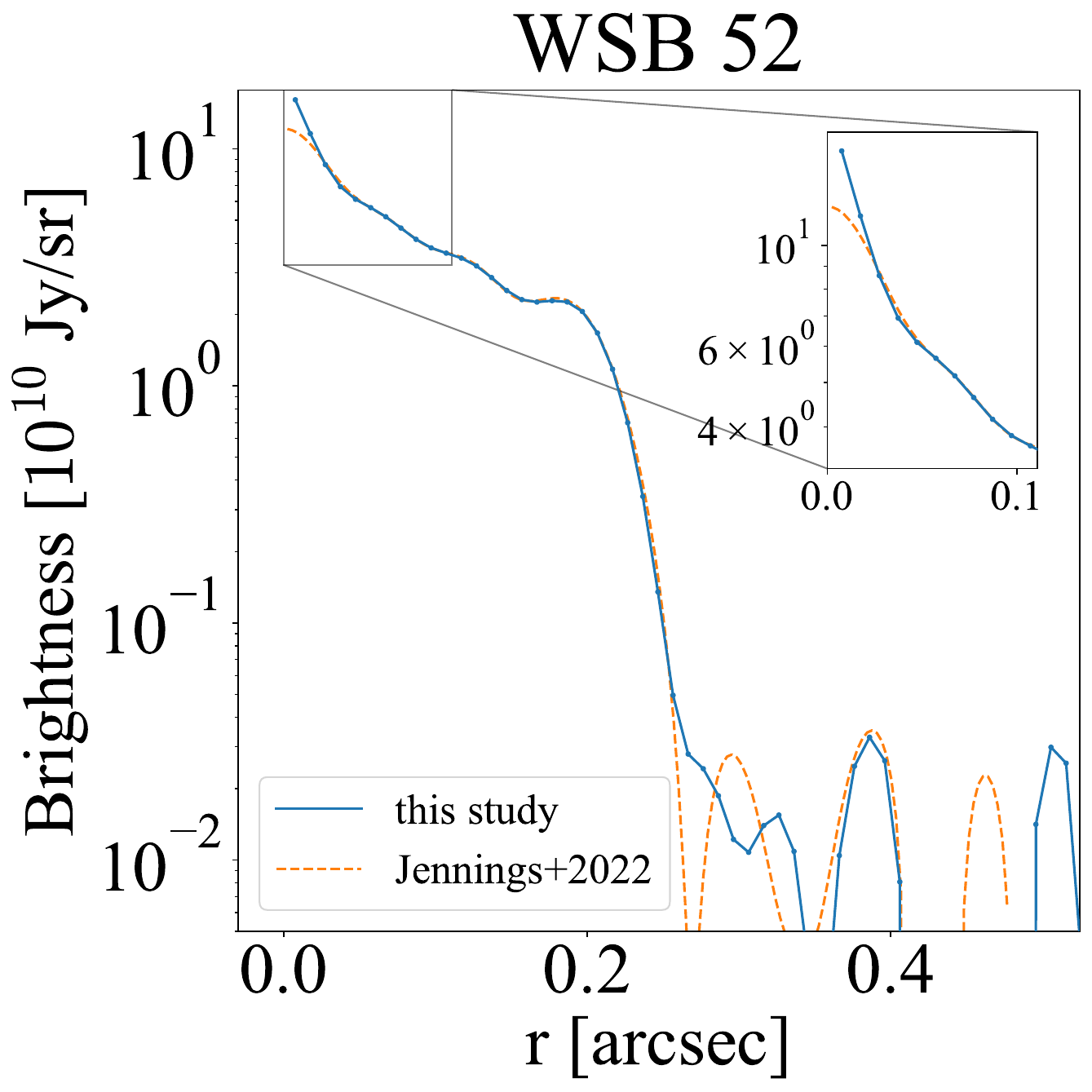}
\end{center}
\caption{Radial brightness profiles for 16 disks. \aizwrev{Blue} lines represent our models, while \aizwrev{orange} lines are adapted from \citet{jennings2022}. Insets provide zoomed views of the inner regions.  }

\label{fig:fitting_1d_bright}
\end{figure*}

\begin{figure*}
\begin{center}
\includegraphics[width=0.24 \linewidth]{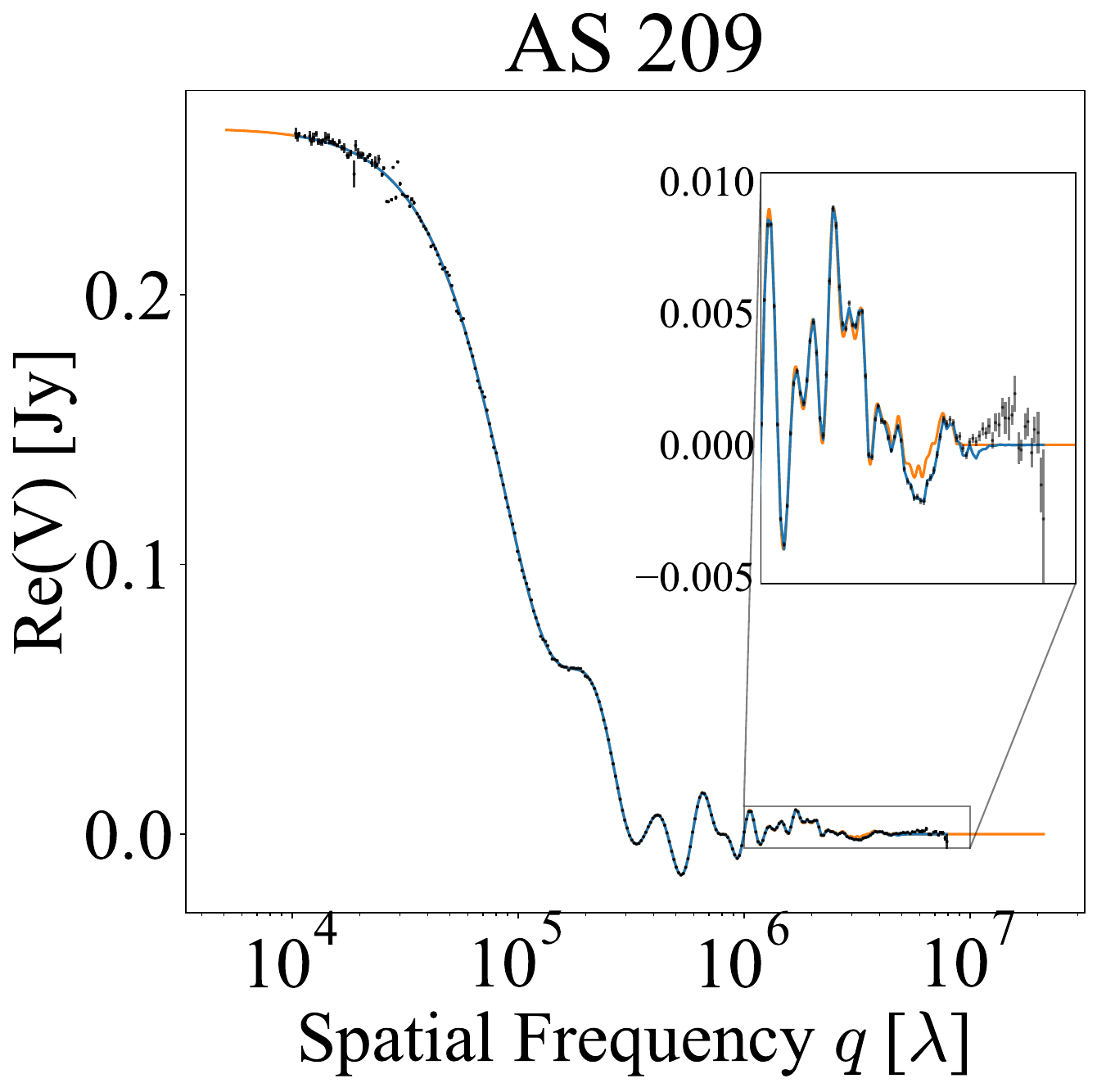}
\includegraphics[width=0.24 \linewidth]{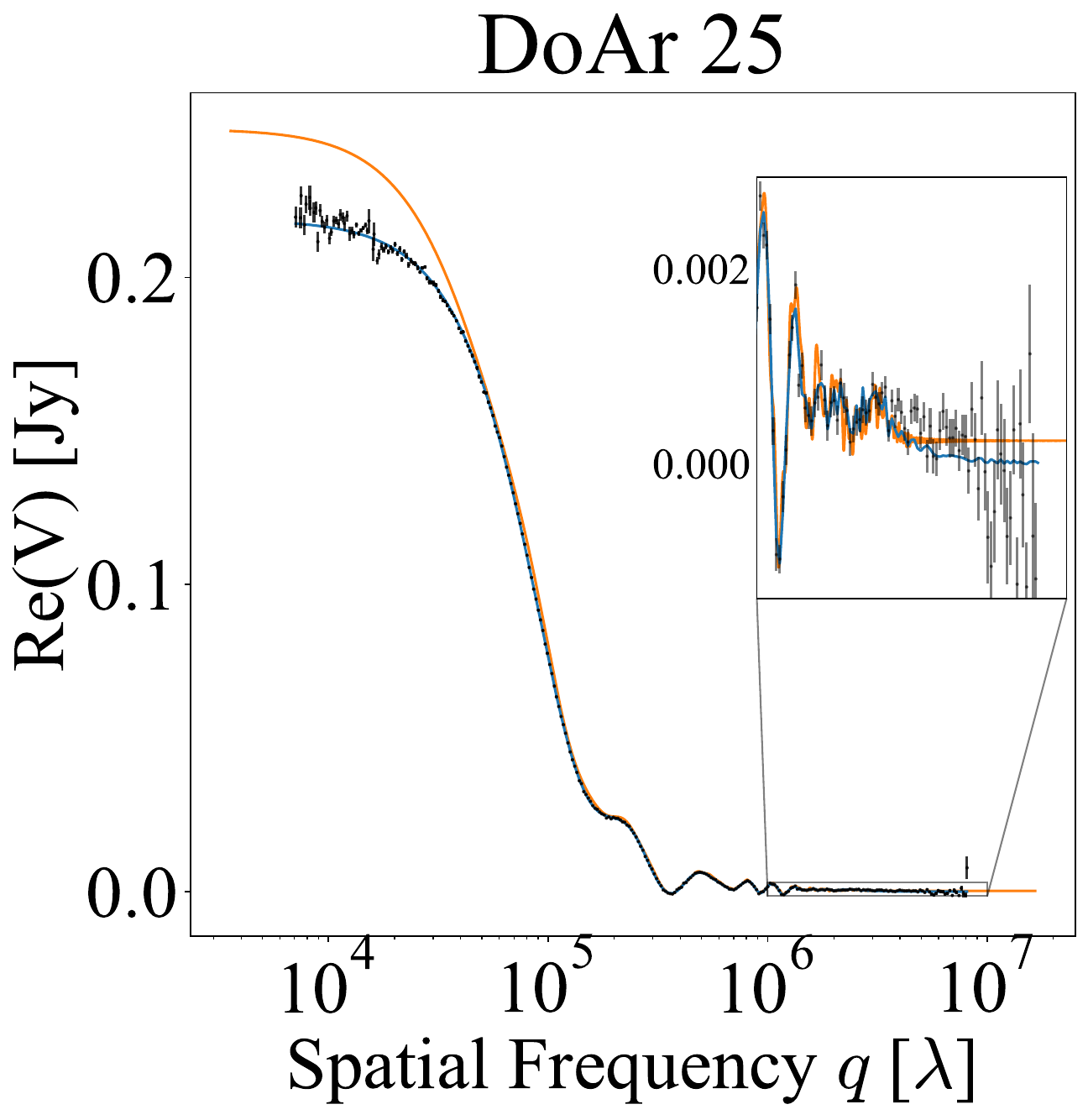}
\includegraphics[width=0.24 \linewidth]{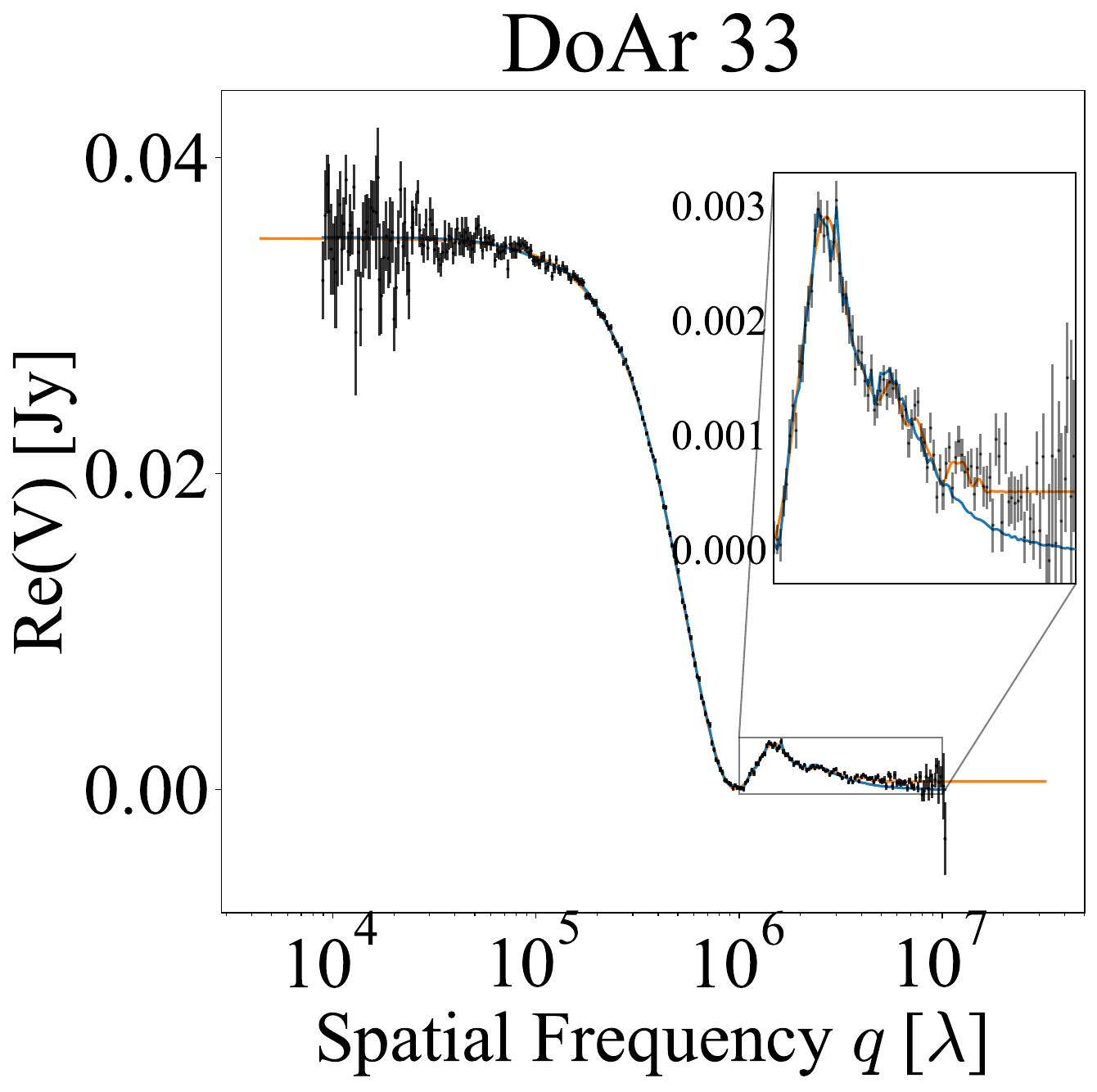}
\includegraphics[width=0.24 \linewidth]{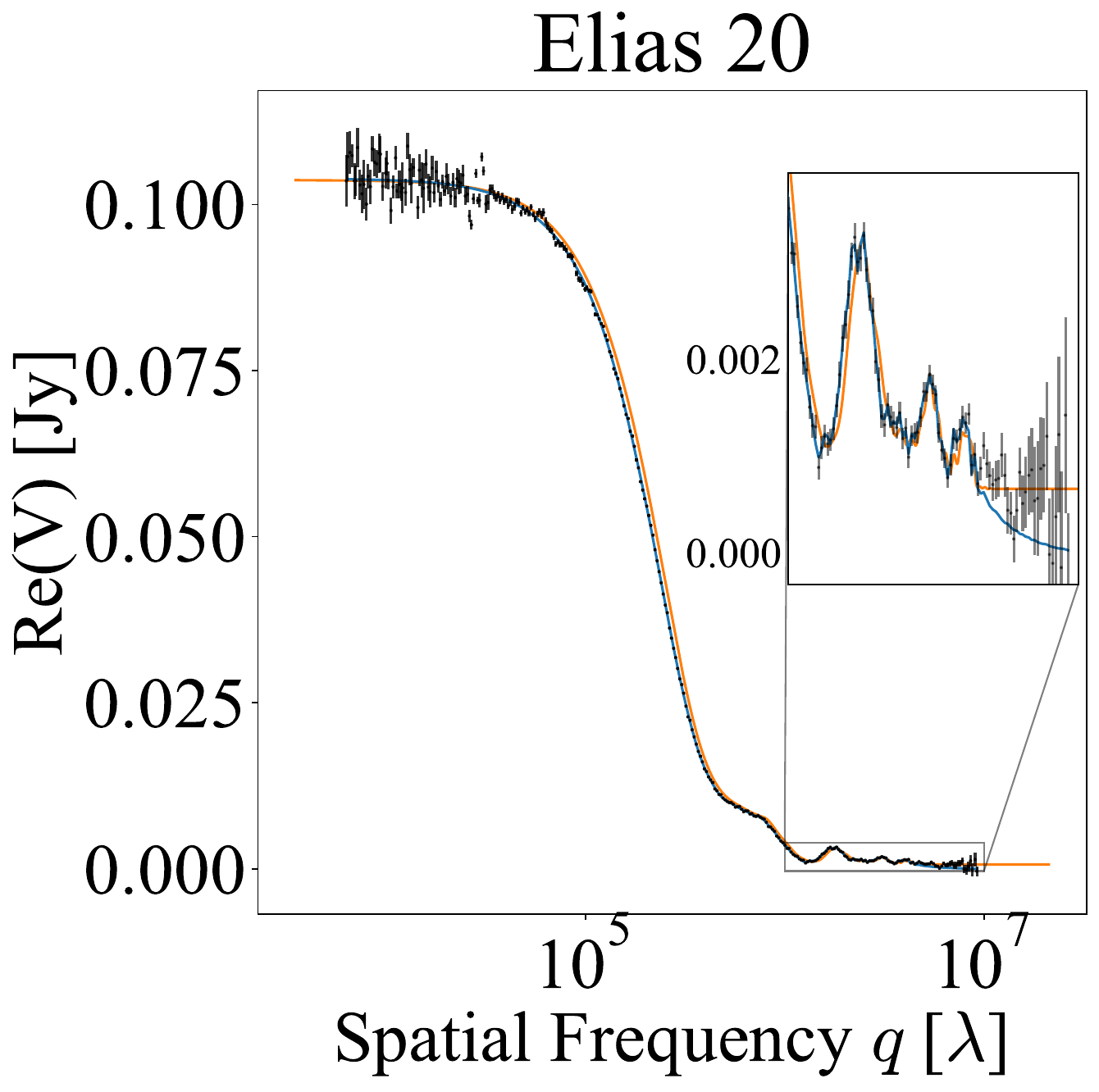}
\includegraphics[width=0.24 \linewidth]{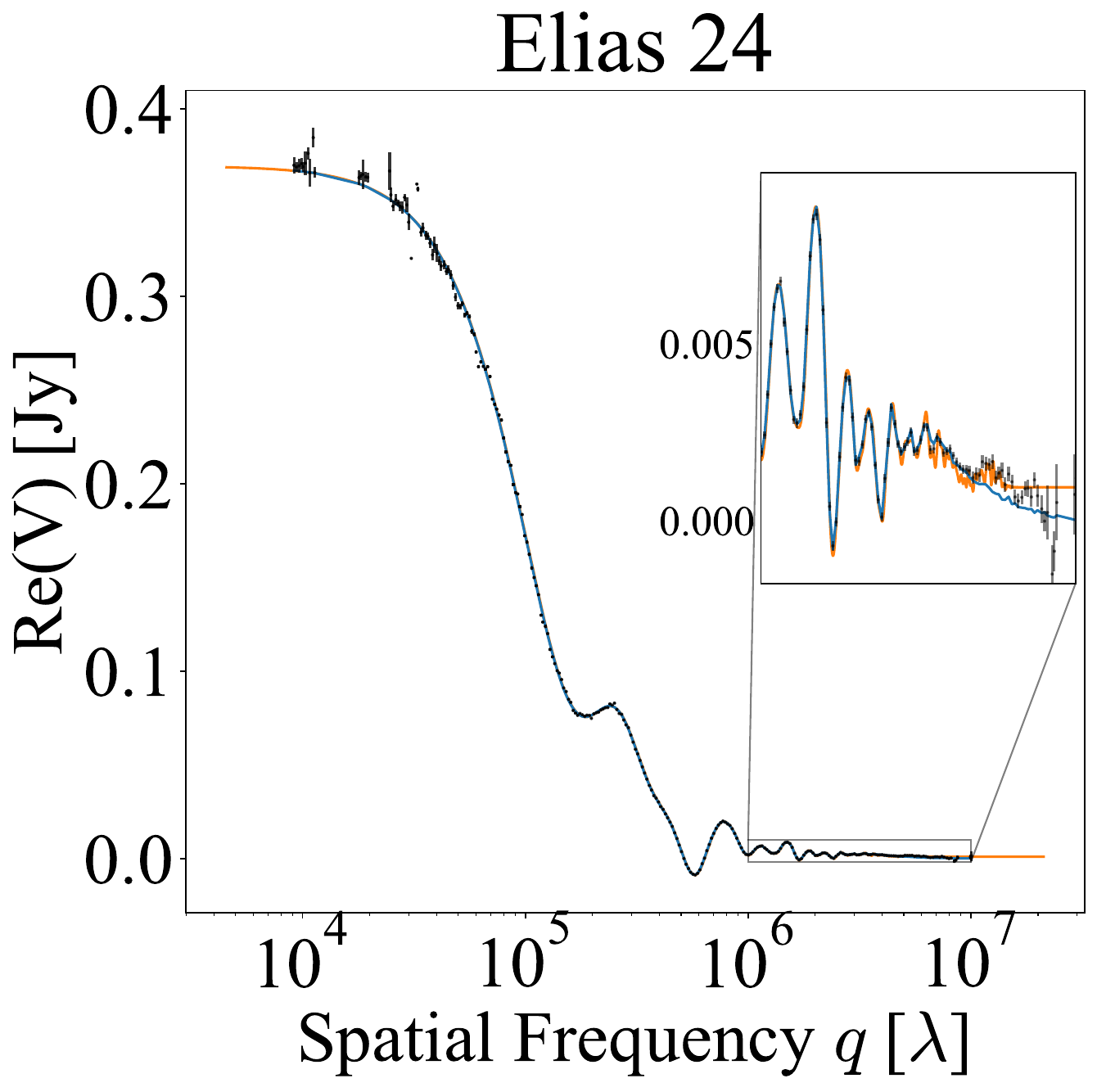}
\includegraphics[width=0.24 \linewidth]{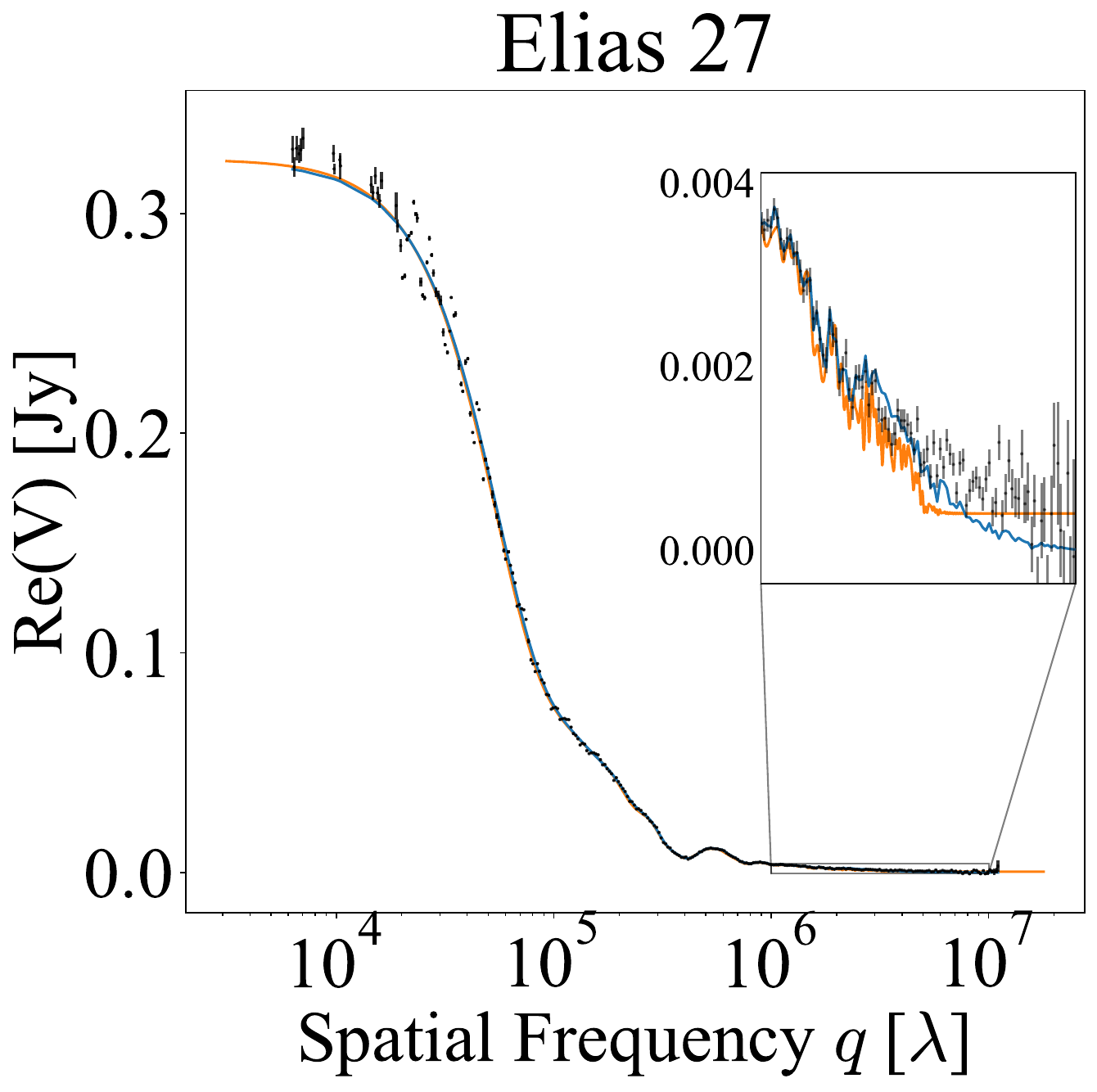}
\includegraphics[width=0.24 \linewidth]{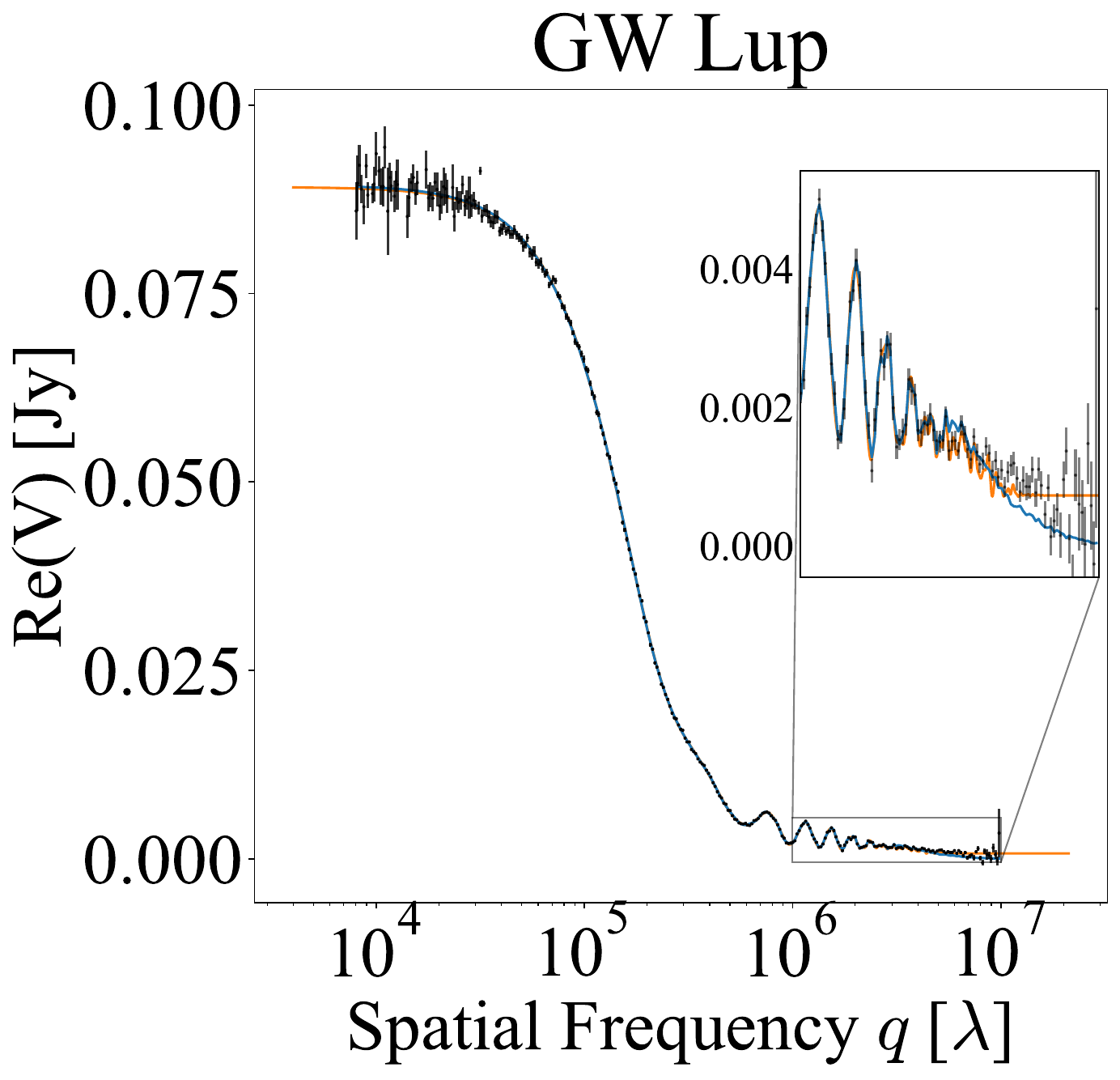}
\includegraphics[width=0.24 \linewidth]{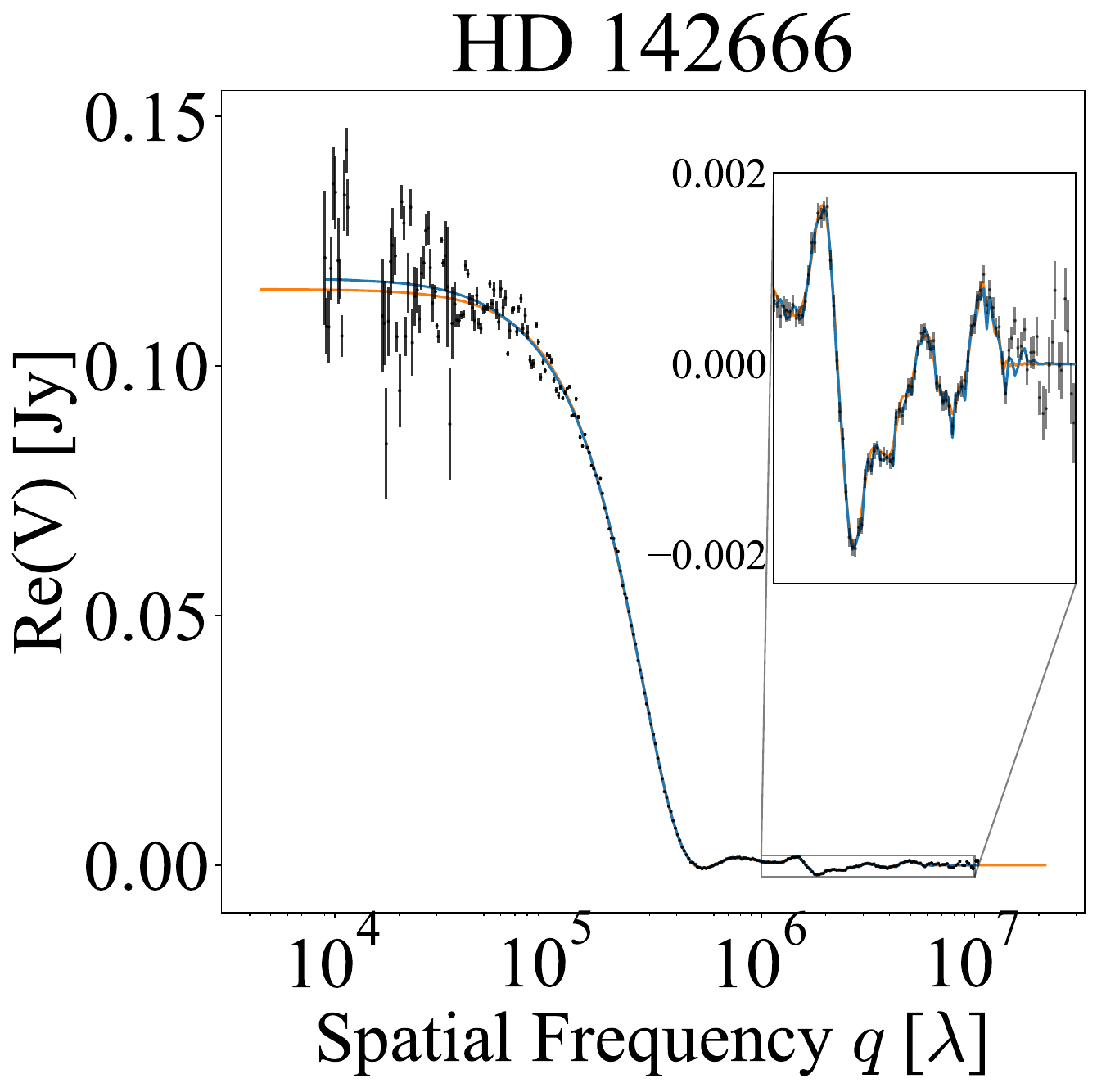}
\includegraphics[width=0.24 \linewidth]{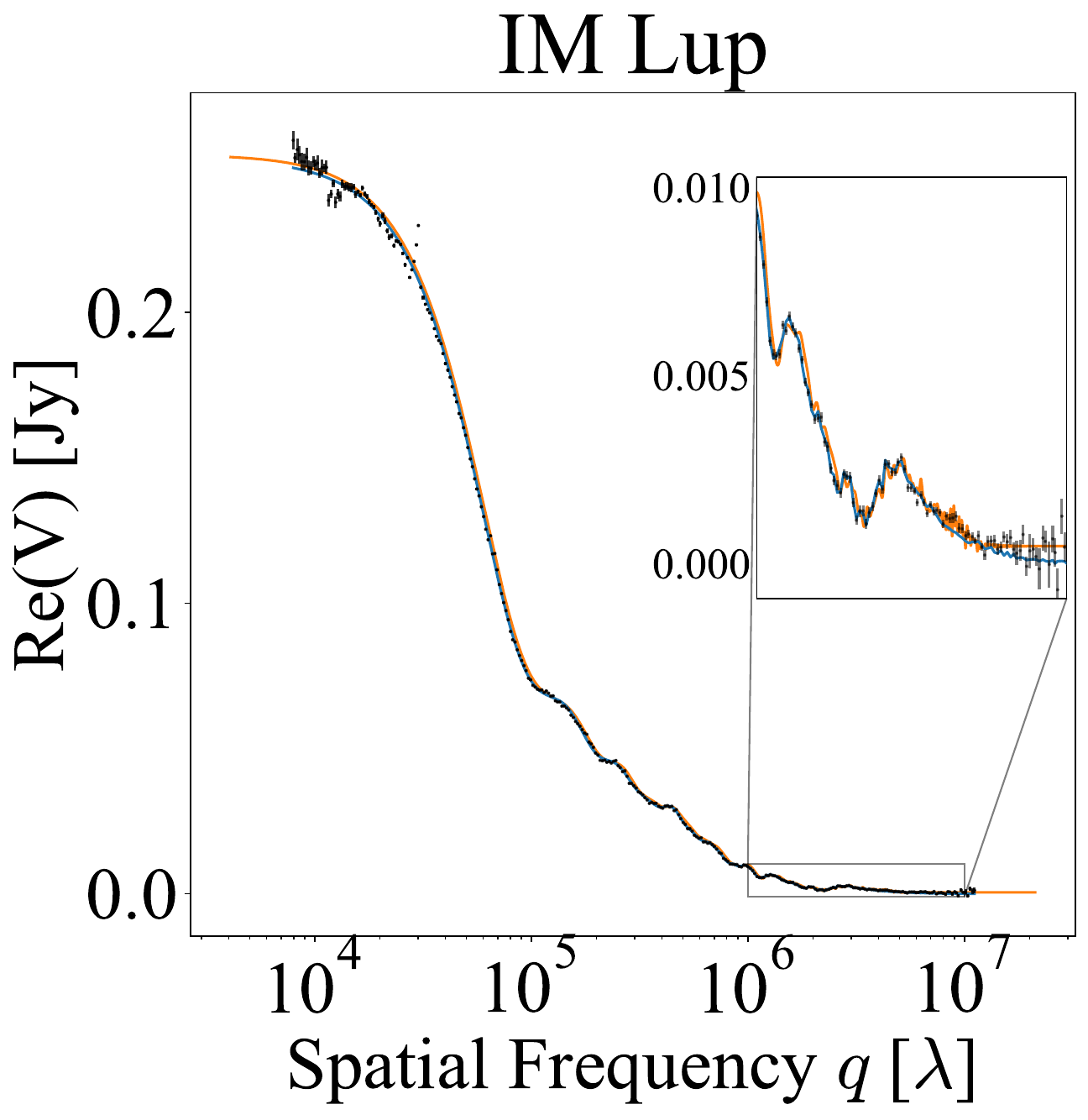}
\includegraphics[width=0.24 \linewidth]{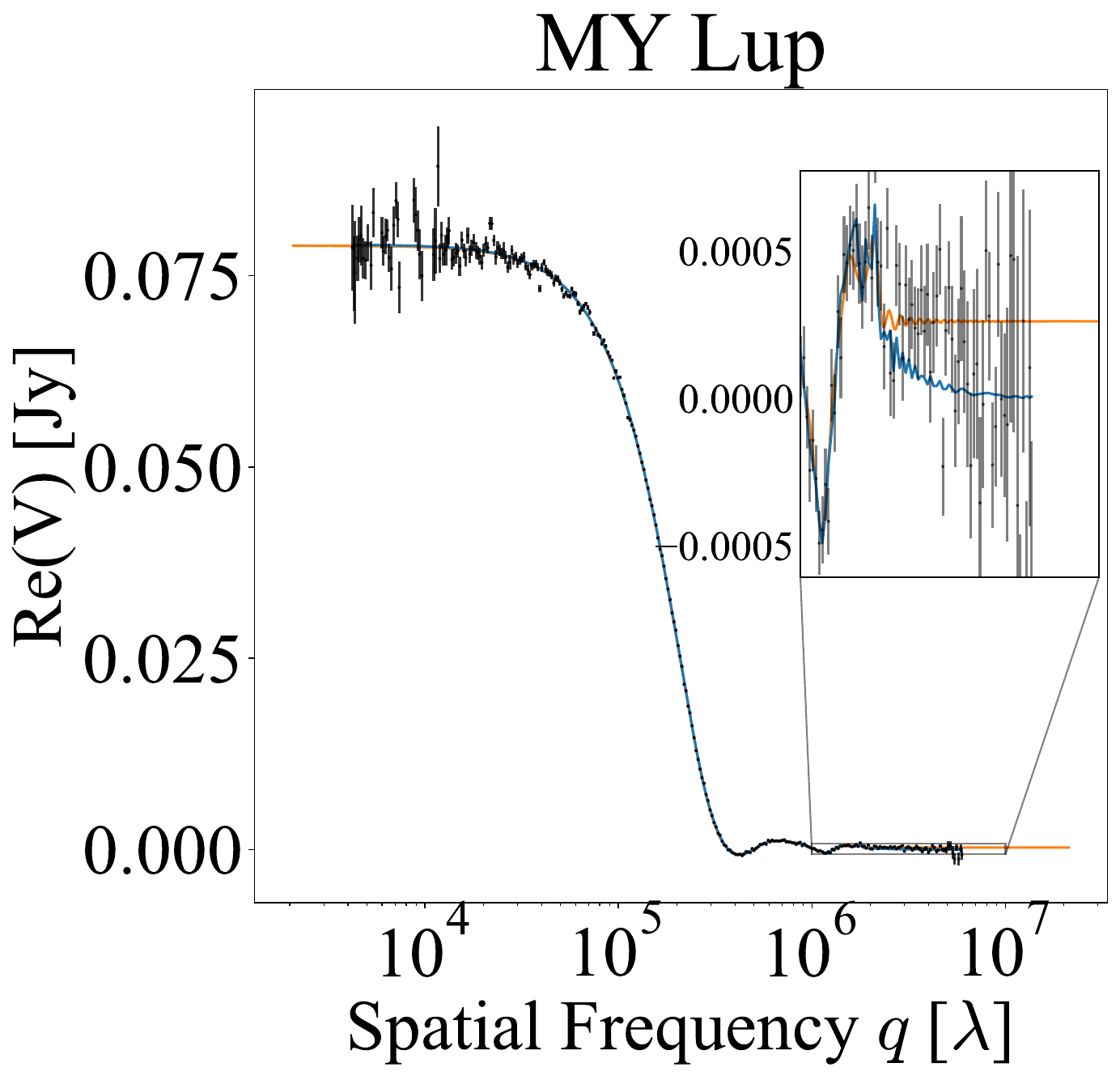}
\includegraphics[width=0.24 \linewidth]{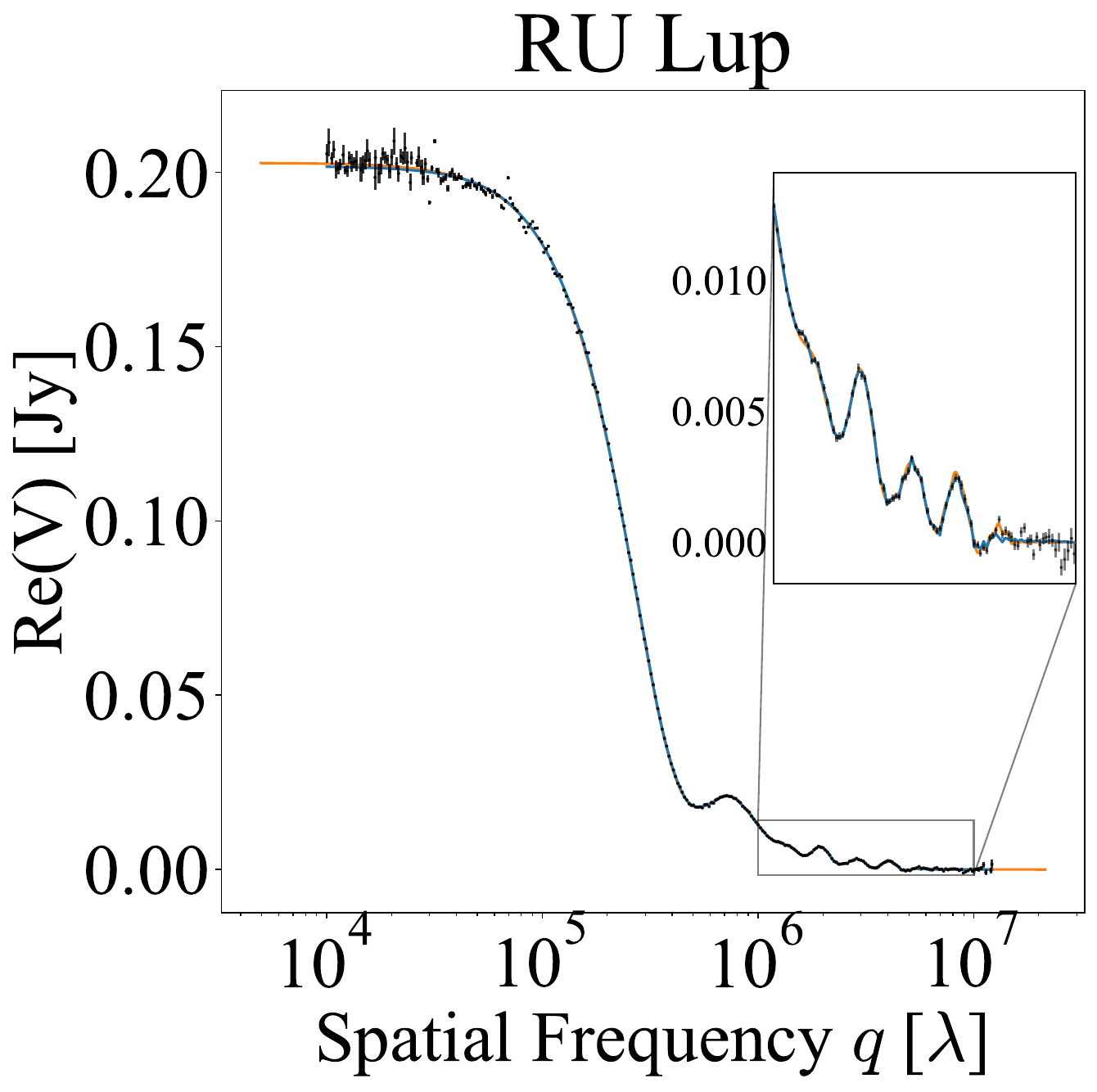}
\includegraphics[width=0.24 \linewidth]{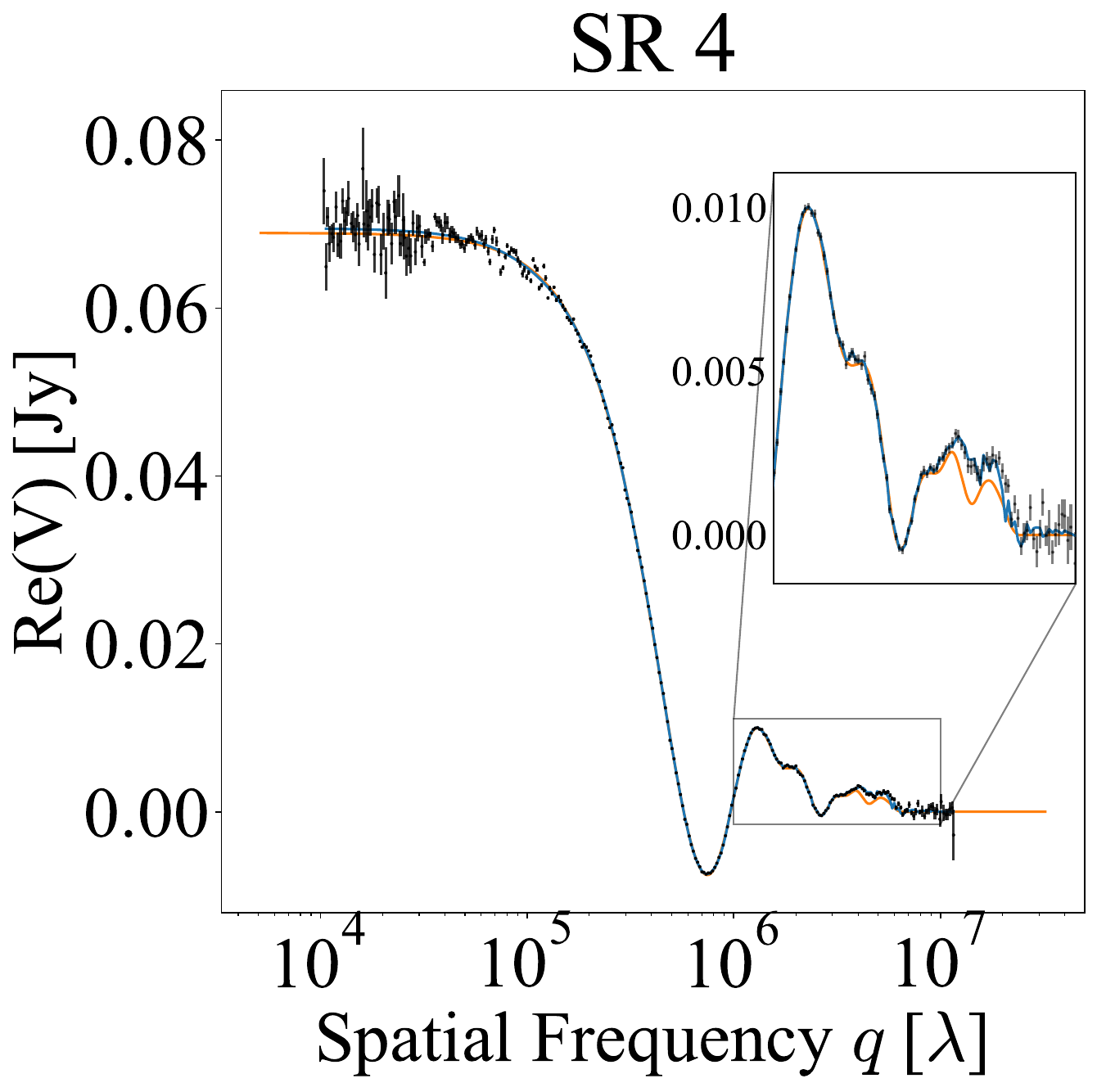}
\includegraphics[width=0.24 \linewidth]{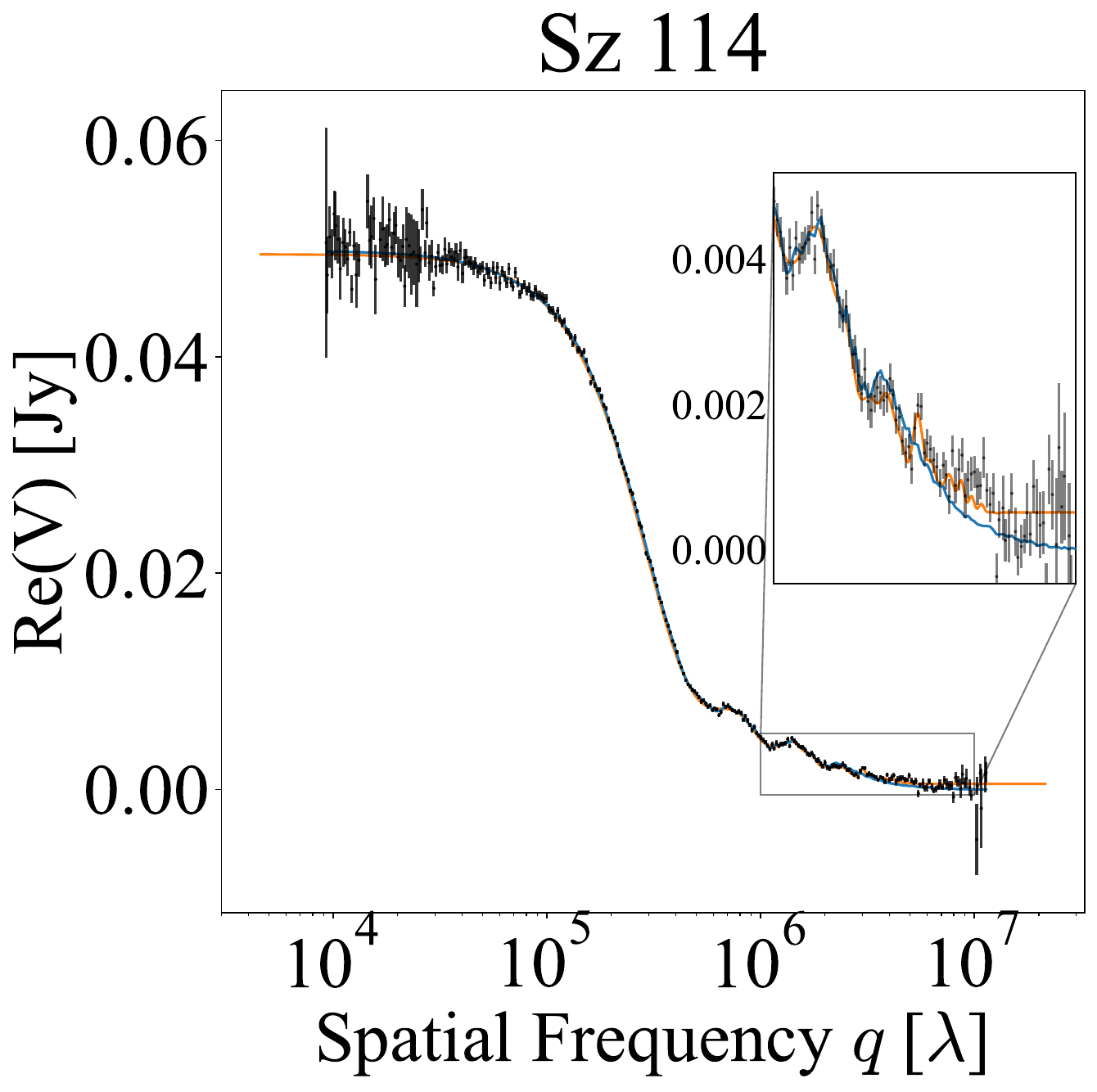}
\includegraphics[width=0.24 \linewidth]{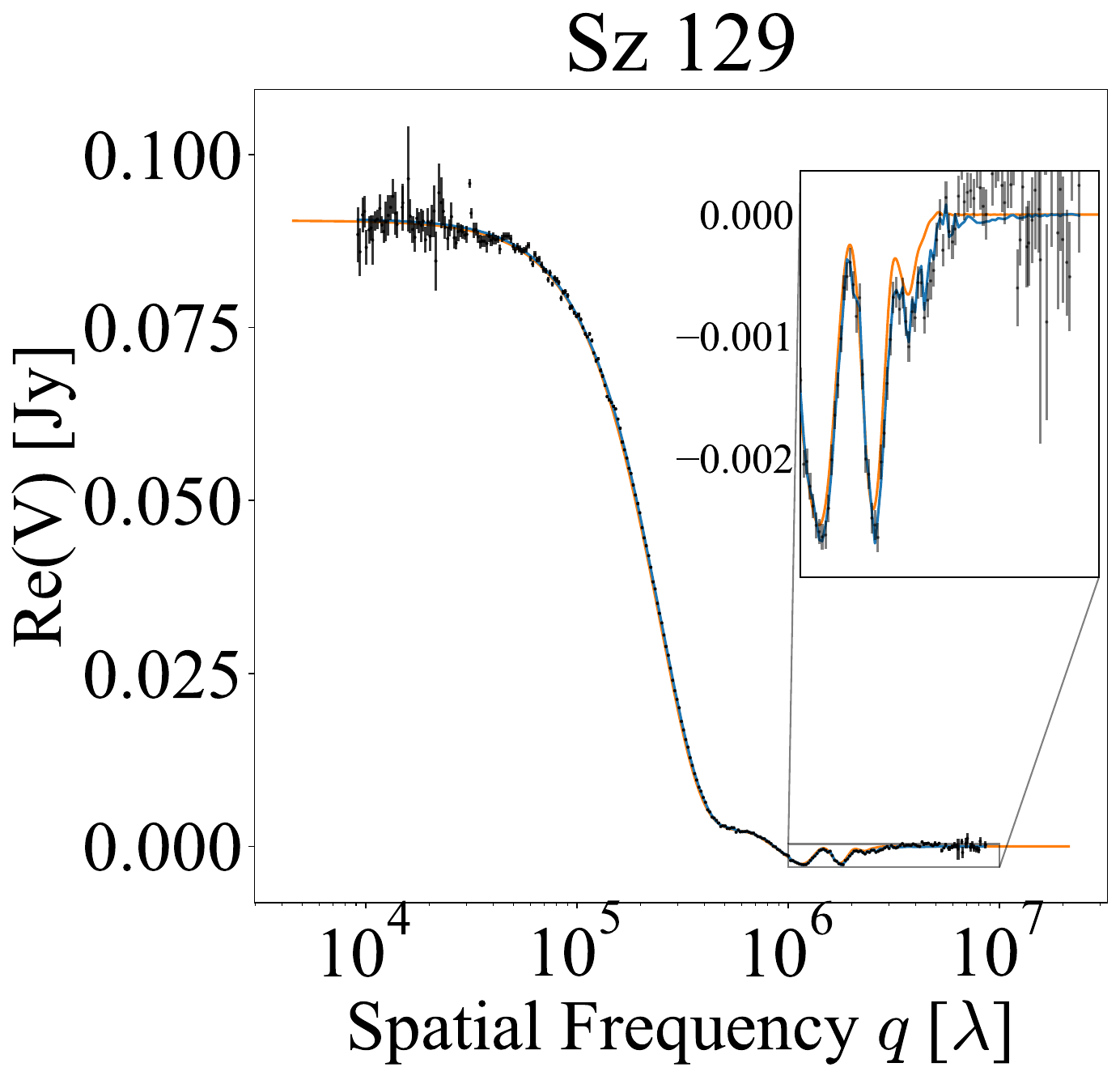}
\includegraphics[width=0.24 \linewidth]{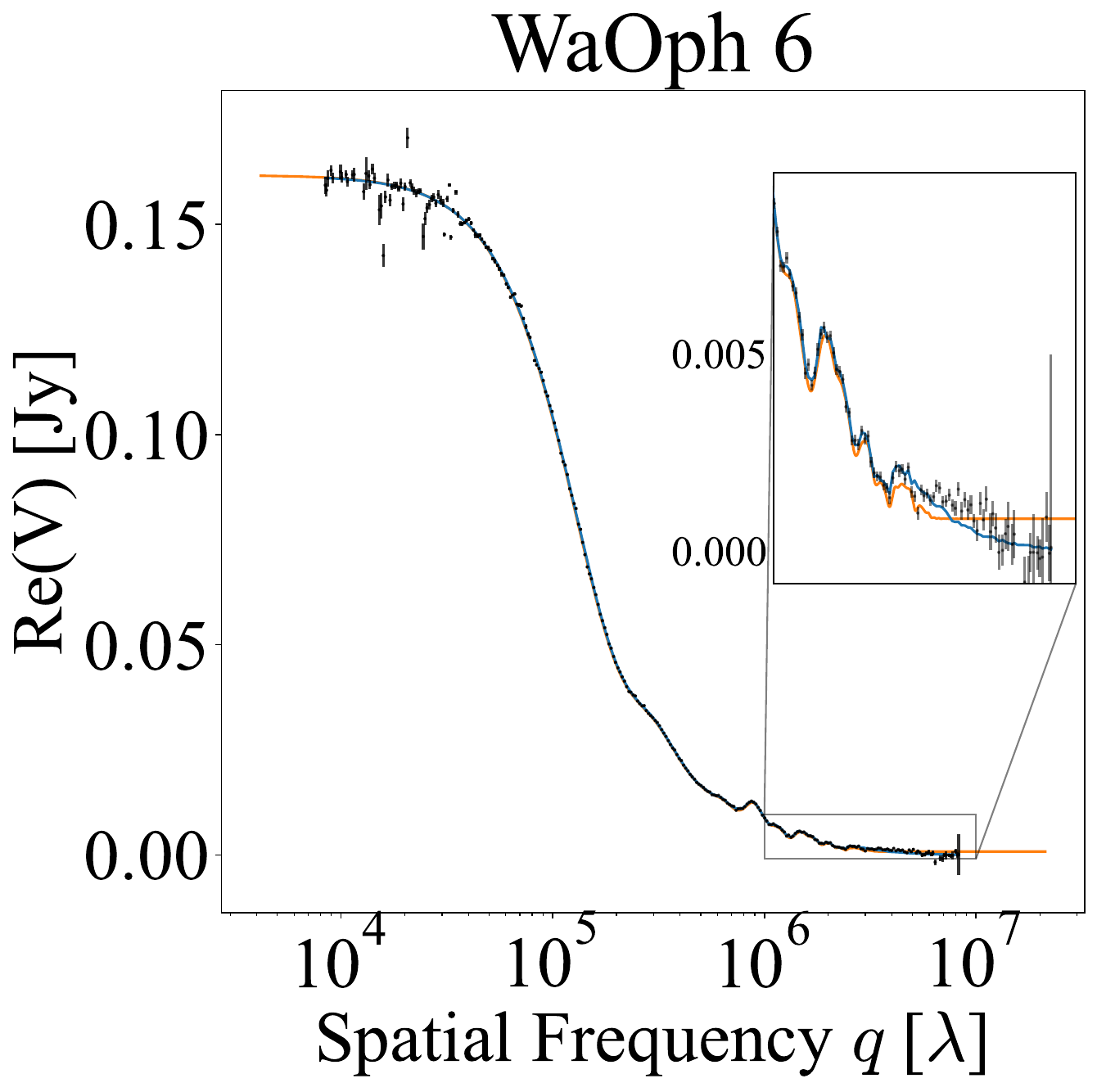}
\includegraphics[width=0.24 \linewidth]{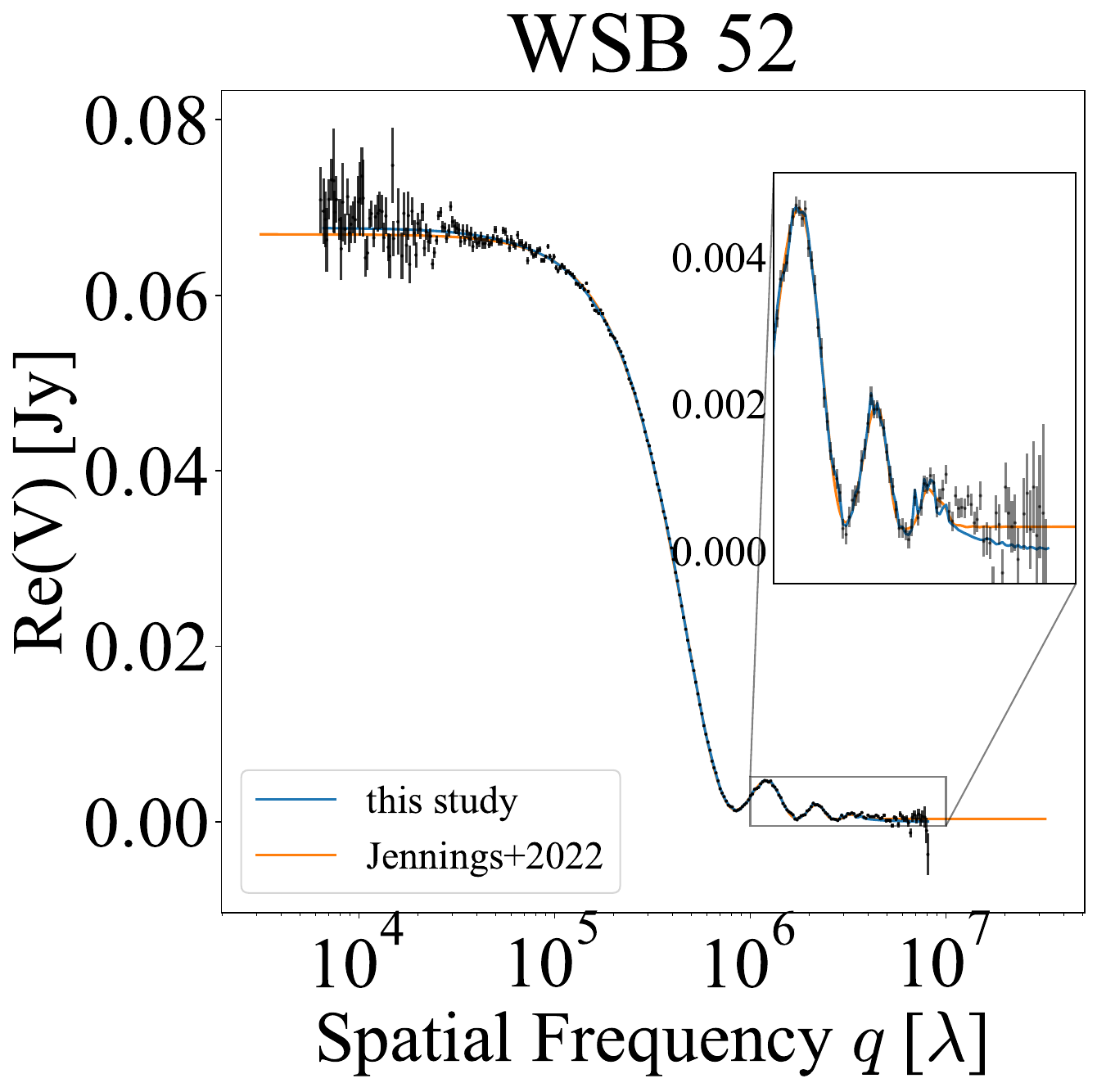}
\end{center}
\caption{Real part of the visibility as a function of deprojected spatial frequency for 16 disks. Blue lines represent our derived visibilities, while the orange lines are taken from \citet{jennings2022}. The insets highlight the data for long baselines. }
\label{fig:fitting_1d_vis}
\end{figure*}

 Figure \ref{fig:fitting_1d_bright} and \ref{fig:fitting_1d_vis} compare our recovered brightness profiles and corresponding model visibilities with those in \cite{jennings2022}. \aizw{We find overall good agreement, though we note some differences in the innermost brightness, which are addressed in the next section. }

\section{Comparison of disk central bright temperatures with previous literature} \label{sec:disk_central_intensity}
\subsection{Comparison with \cite{jennings2022}} \label{sec:jeff_comp_tb}
\begin{figure*}
\begin{center}
\includegraphics[width=0.48\linewidth]{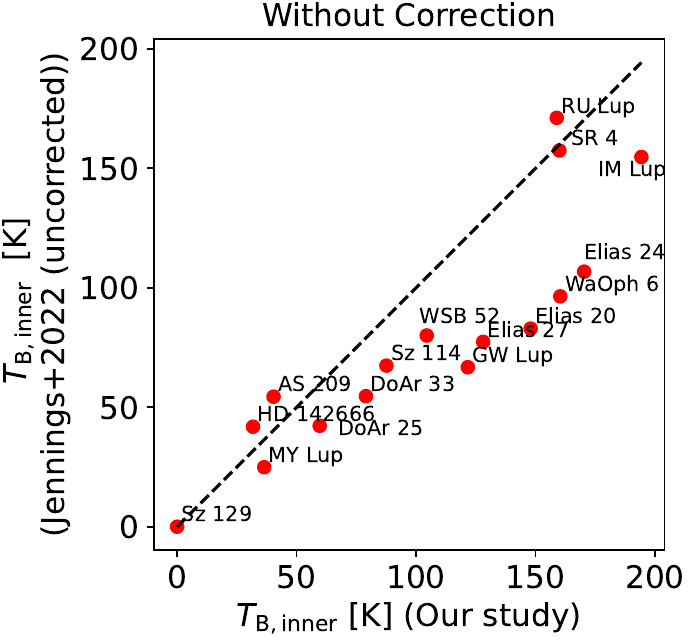}
\includegraphics[width=0.48\linewidth]{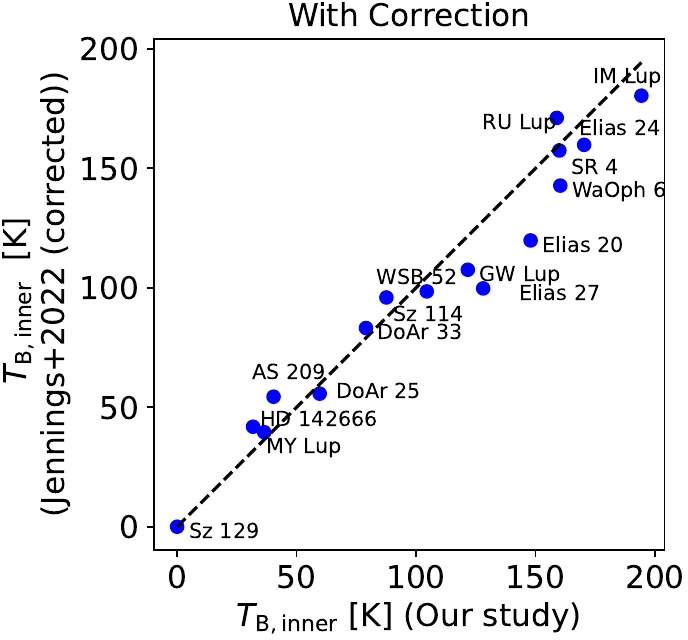}
\end{center}
\caption{Comparison of inner-disk brightness temperatures, $T_{B,\mathrm{inner}}$, between our study and \cite{jennings2022} (Left) Comparison with uncorrected values from \cite{jennings2022}. (Right) Comparison with corrected values after applying the point-source correction to fluxes in \cite{jennings2022}.  The dashed line indicates the $1:1$ relation, and disk names are labeled for reference.}
\label{fig:comp_tb}
\end{figure*}

The left in Figure \ref{fig:comp_tb} show the comparison of the disk central temperatures between our study and \cite{jennings2022}. We identify discrepancies for the the inner-disk intensities between our studies and \cite{jennings2022}. The innermost radius $r_1$ in the model for our study is different from \cite{jennings2022}; $r_1$ in \cite{jennings2022} is smaller than $r_1$ in our study by a factor of $\sim$1.5-3. This, however, tends to see brighter regions in the disk if $T_{\rm B,inner}$ is high in \cite{jennings2022}, in opposition to the observations, where our estimates are generally larger than \cite{jennings2022}. Indeed, the difference in $r_1$ is much smaller than the beam size and $\gamma$, suggesting that the inner disks probed by $T_{\rm B,inner}$ should be similar. 

We find out that prior-subtractions of point-source emission subtraction in \cite{jennings2022} largely explains the discrepancy. Specifically, in \cite{jennings2022}, they subtracted unresolved component, which is point source emission put at the disk center, before applying {\it frank} fitting. To study the effect of this subtraction, we correct the intensities in \cite{jennings2022} assuming that the subtracted flux are all compensated by the inner-disk intensities at $r_1$ in our study. 

Figure \ref{fig:comp_tb} shows the comparison of the disk central temperatures with and without the correction. We find that the temperatures in \cite{jennings2022} tend to be underestimated in comparison to our estimates, whereas both of the estimations mostly agree once the correction is taken into account. Though there are multiple differences in assumptions adopted in our study and \cite{jennings2022}, the consistency nevertheless demonstrates the robustness of estimation \aizw{of $T_{\rm B,inner}$}, as well as the trend in $T_{\rm B,inner}$ and $t_{\star}$

\subsection{Comparison with \cite{huang_ring_2018}} \label{sec:huang_comp_tb}
We compare our measurements of $T_{\rm B,inner}$ with those reported by \cite{huang_ring_2018}. For this purpose, we download the radial brightness profiles from the DSHARP website and extract the brightness temperature at the innermost radius. These profiles are derived from CLEANed images after beam smearing, so their effective angular resolution is lower than achieved by the one-dimensional modeling in our study  and in \cite{jennings2022}. 

\begin{figure*}
\begin{center}
\includegraphics[width=0.48\linewidth]{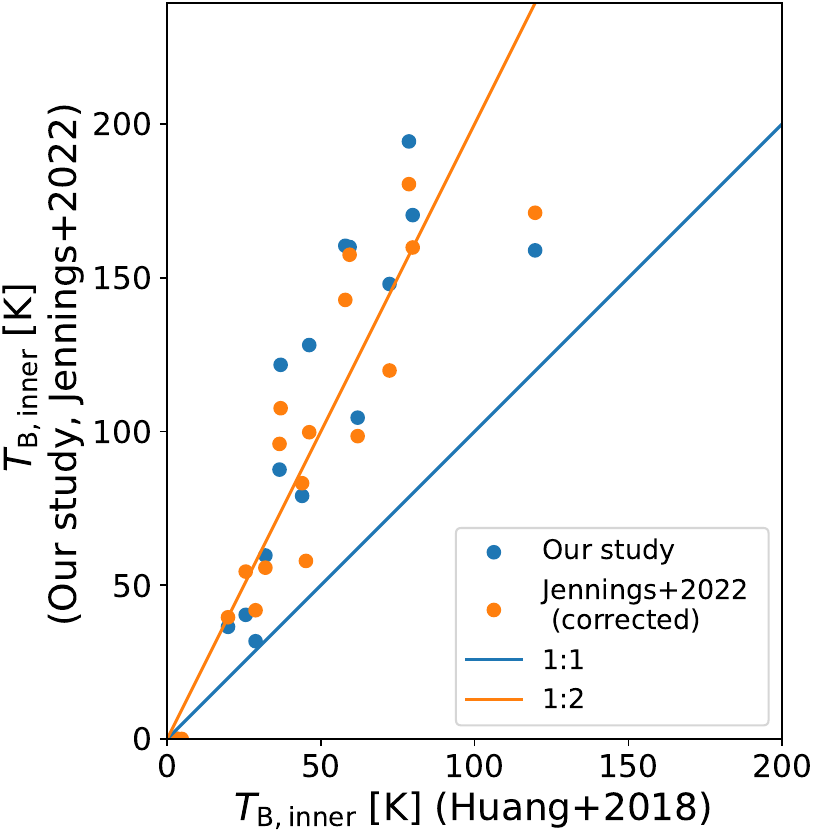}
\includegraphics[width=0.48\linewidth]{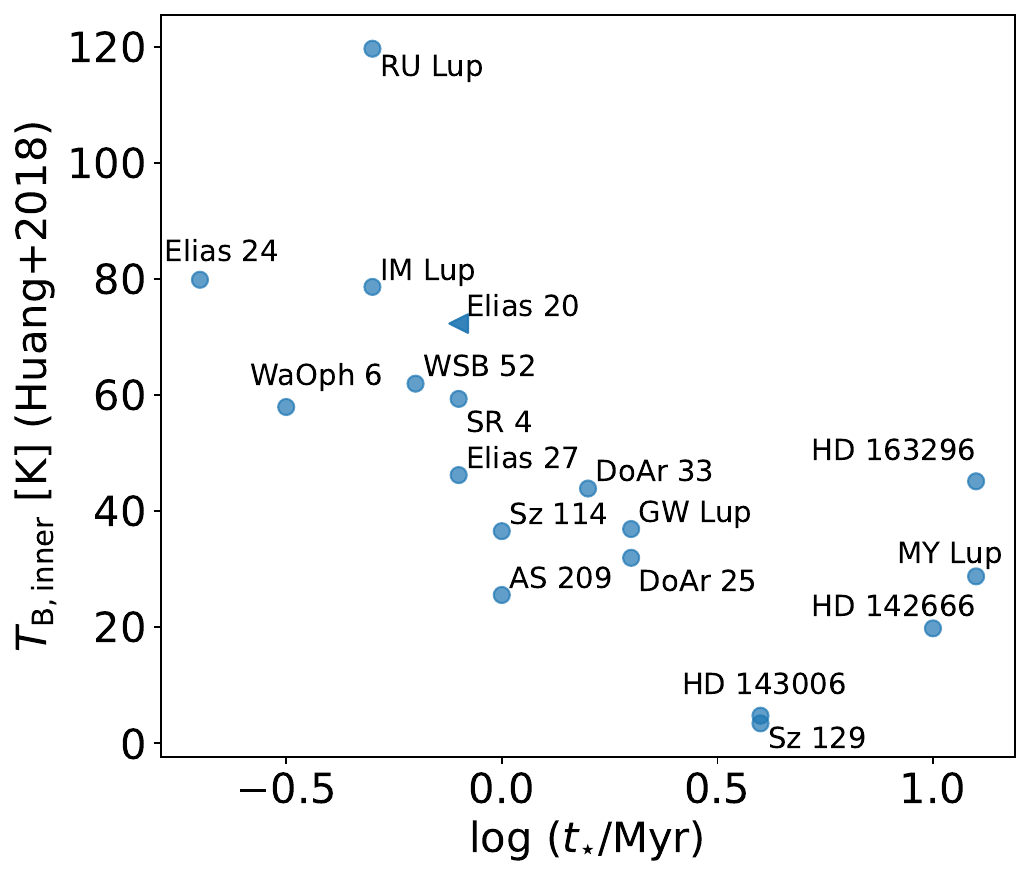}
\end{center}
\caption{Analysis of $T_{\rm B,inner}$ using the values from \cite{huang_ring_2018}. 
(Left) Comparison of $T_{\rm B,inner}$ between \cite{huang_ring_2018} and the one-dimensional analysis presented in this study and \cite{jennings2022}. Lines corresponding to $1:1$ and $1:2$ relations are shown for references.  
(Right) $T_{\rm B,inner}$ \citep{huang_ring_2018} and stellar age $\log_{10}(t_\star/\mathrm{Myr})$.}
\label{fig:comp_tb_huang}
\end{figure*}

The left panel of Figure~\ref{fig:comp_tb_huang} compares $T_{\rm B, inner}$ obtained from our one-dimensional modeling with those of \cite{huang_ring_2018}. Temperatures reported by \cite{huang_ring_2018} are systematically lower than in the other studies. Because one-dimensional modeling is more sensitive to the innermost disk regions than the CLEAN-based analysis, the higher temperatures found in our study and in \cite{jennings2022} are reasonable. The right panel of Figure~\ref{fig:comp_tb_huang} shows $T_{\rm B,inner}$ from \cite{huang_ring_2018} as a function of stellar age on a logarithmic scale. Although the significance of the correlation is weaker ($r=-0.67$, $p=0.0023$) than in our analysis, a declining trend is still apparent. The reduced significance is consistent with the lower effective angular resolution of the CLEANed images, whose $T_{\rm B, inner}$ probe a broader region beyond the inner disk.

\section{Spiral model} \label{sec:spiral_model}

Here, we describe a spiral model used in Section \ref{sec:odd_sym_spiral}. To compute the spiral phase, we adopt the analytic phase equation \citep{ogilvie2002,bae_2018}:

\begin{align}
\phi_{m,n} (R) &= - \phi_{p} - {\rm sgn}(R-R_{\rm p})\frac{\pi}{4m} + 2\pi \frac{n}{m} - \int^{R}_{R^{\pm}_{m}} \frac{1}{H(R')} \left| \left(1 - \frac{R'^{3/2}}{R_{\rm p}^{3/2}}\right)^{2}  - \frac{1}{m^{2}}\right| ^{1/2} dR',
\label{eq:analytical_phase}
\end{align}

Here, $\phi_{p}$ denotes the azimuthal phase of the planetary position, $R_{\rm p}$ is its radial location, $m$ is the azimuthal mode index, and $n$ is an integer specifying the spiral mode (with $n=0,1,\ldots,m-1$). In this expression, $H(R')=c_{s}(R')/\Omega(R')$ is the scale height of the disk, where $\Omega(R')$ represents the angular velocity of the disk, and $c_{s}(R')$ is the local sound speed. The quantities $R^{\pm}_{m} = (1 \pm 1/m)^2 R_{\rm p}$ indicate the radial locations of the Lindblad resonances, where density waves are launched. Consequently, the phase $\phi_{m,n}(R)$ traces a planetary wake, which appears as a positive trail in a residual map.

We adopt the same assumptions as \cite{speedie_sim_2022}. For the inner spiral ($R < R_{\rm p}$), we consider mode numbers in the range $2\leq m \leq 10$, while for the outer spiral ($R > R_{\rm p}$) we restrict $m$ to $1\leq m \leq 10$. We assume $n=0$ to isolate the primary spiral mode. The disk aspect ratio $H(R')/R'$ is assumed to follow the power law: $H(R')/R' = c_{s}(R')/(R' \Omega(R')) \propto R'^{1/4}$, assuming the passively irradiated flaring disk. Specifically, the Keplerian angular velocity is assumed to be determined solely by stellar gravity, so that $\Omega(R')\propto M_{\star}^{1/2} R'^{-3/2}$. To calculate the sound speed $c_{s}(R')$, we adopt the temperature profile of an irradiated flaring disk model, $T(R')\propto \varphi^{1/4} L_{\star}^{1/4} R'^{-1/2}$ \citep{dullemond_2018}, where $\varphi$ is the disk flaring parameter and $L_{\star}$ is the stellar luminosity. The sound speed is then given by $c_{s}(R')=\sqrt{k_{B}T(R')/M}$, where $k_{B}$ is the Boltzmann constant and $M$ is the mean molecular weight. In our calculations, we assume $\varphi=0.02$ and $M=2.37 m_{p}$, where $m_{p}$ is the proton mass. The direction of the planetary motion is assumed to follow the disk rotational direction; we adopt it to be clockwise for Elias 27 according to \cite{huang_spiral_2018}, which determin the rotation direction from $^{12}$CO emission. For the planetary location, we visually determine the values of $R_p = 0.6\arcsec$ and $\phi_{p}=-120^{\circ}$, to match the observed spiral shape.

\end{CJK*}
\end{document}